\documentclass[twocolumn]{aa}  

\usepackage{graphicx}
\usepackage{subcaption}
\usepackage{xcolor}
\usepackage[symbol]{footmisc}
\usepackage{booktabs}
\usepackage{physics}
\usepackage{multirow}
\usepackage{amsmath}
\usepackage{hyperref}
\usepackage{url}
\usepackage{academicons}
\usepackage{svg}
\hypersetup{breaklinks=true,colorlinks=true,urlcolor=blue, linkcolor=blue,  citecolor=blue, bookmarksopen=true}

\usepackage{txfonts}
\begin{document}

\title{Forbidden emission lines in protostellar outflows and jets with MUSE}

\author{Lizxandra Flores-Rivera
          \inst{1}
          \and
          Mario Flock \inst{1} 
          Nicol\'{a}s Kurtovic \inst{1}
          \and
          Bernd Husemann \inst{1}
          \and
          Andrea Banzatti \inst{7}
          \and
          Simon C. Ringqvist \inst{2}
          \and
          Sebastian Kamann \inst{9}
          \and
          André Müller\inst{1}
          \and
          Christian Fendt \inst{1}
          \and
          Rebeca Garc\'{i}a Lopez \inst{3}
          \and
          Gabriel-Dominique Marleau \inst{5,6,1}
          \and
          Thomas Henning \inst{1}
          \and
          Carlos Carrasco-Gonz\'{a}lez \inst{4}
          \and 
          Roy van Boekel \inst{1}
          \and
          Miriam Keppler \inst{1}
          \and
          Ralf Launhardt \inst{1}
          \and
          Yuhiko Aoyama \inst{8}
           }

\institute{Max Planck Institut f\"ur Astronomie, K\"onigstuhl 17, 69117 Heidelberg, Germany\\
              \email{flores@mpia.de}
              \and
               Simon C. Ringqvist, n\'{e} Eriksson; Institutionen f\"or astronomi, Stockholms universitet, AlbaNova universitetscentrum, 106 Stockholm, Sweden 
              \and
              School of Physics, University College, Belfield, Dublin 4, Ireland
              \and
              Instituto de Radioastronomía y Astrofísica(IRyA), Universidad Nacional Autónoma de México (UNAM), Mexico
             \and
              Institut f\"ur Astronomie und Astrophysik,
              Universit\"at T\"ubingen,
              Auf der Morgenstelle 10,
              72076 T\"ubingen, Germany
              \label{Tue}
              \and
              Physikalisches Institut,
              Universit\"{a}t Bern,
              Gesellschaftsstr.~6, 
              3012 Bern, Switzerland
              \label{Bern}%
              \and
              Department of Physics,
              Texas State University,
              San Marcos, 
              601 University Dr, 
              San Marcos, TX 78666, United States
              \label{Texas}%
              \and
              Institute for Advanced Study,
              Tsinghua University,
              Beijing, 
              100084, China
              \label{China}%
              \and
              Astrophysics Research Institute,
              Liverpool John Moores University,
              IC2, Liverpool Science Park,
              146 Brownlow Hill,
              Liverpool,
              L3 5RF,
              United Kingdom
             }

             \date{Accepted on December 19, 2022}

\abstract
   {Forbidden emission lines in protoplanetary disks are a key diagnostic in studies  of the evolution of the disk and the host star. They signal potential disk accretion or wind, outflow, or jet ejection processes of the material that affects the angular momentum transport of the disk as a result.}
   {We report spatially resolved emission lines, namely,  [\ion{O}{i}]~$\lambda\lambda$6300, 6363, [\ion{N}{ii}]~$\lambda\lambda$6548, 6583, H$\mathrm{\alpha}$, and [\ion{S}{ii}]\,$\lambda\lambda$6716, 6730 that are believed to be associated with jets and magnetically driven winds in the inner disks, due to the proximity to the star, as suggested in previous works from the literature. 
   With a resolution of 0.025$\times$0.025~$\mathrm{arcsec}^{2}$, we aim to derive the position angle of the outflow/jet (PA$_\mathrm{outflow/jet}$) that is connected with the inner disk. We then compare it with the position angle of the dust (PA$_\mathrm{dust}$) obtained from previous constraints for the outer disk. We also carry out a simple analysis of the kinematics and width of the lines and we estimate the mass-loss rate based on the [\ion{O}{i}]~$\lambda$6300 line for five T Tauri stars.}
   {Observations were carried out with the optical integral field spectrograph of the Multi Unit Spectroscopic Explorer (MUSE), at the Very Large Telescope (VLT). 
    The instrument spatially resolves the forbidden lines, providing a unique capability to access the spatial extension of the outflows/jets that make the estimate of the PA$_\mathrm{outflow/jet}$ possible from a geometrical point of view.}
   {The forbidden emission lines analyzed here have their origin at the inner parts of the protoplanetary disk. From the maximum intensity emission along the outflow/jet in DL Tau, CI Tau, DS Tau, IP Tau, and IM Lup, we were able to reliably measure the PA$_\mathrm{outflow/jet}$ for most of the identified lines. We found that our estimates agree with PA$_\mathrm{dust}$ for most of the disks. These estimates depend on the signal-to-noise level and the collimation of the outflow (jet). The outflows/jets in CIDA 9, GO Tau, and GW Lup are too compact for a PA$_\mathrm{outflow/jet}$ to be estimated. Based on our kinematics analysis, we confirm that DL Tau and CI Tau host a strong outflow/jet with line-of-sight velocities much greater than 100 km~s$^{-1}$, whereas DS Tau, IP Tau, and IM Lup velocities are lower and their structures encompass low-velocity components to be more associated with winds. Our estimates for the mass-loss rate, $\dot{M}_{\mathrm{loss}}$, range between (1.1-6.5)~$\times$10$^{-7}$-10$^{-8}$~$M_{\odot}~yr^{-1}$ for the disk-outflow/jet systems analyzed here.
   
  } 
  {The outflow/jet systems analyzed here are aligned within around 1$^{\circ}$ between the inner and outer disk. Further observations are needed to confirm a potential misalignment in IM Lup.}


\maketitle
%

\section{Introduction}

The detection of forbidden emission lines in different samples of T Tauri stars has provided access to different physical conditions depending on their line-of-sight (LoS) velocity information. Line decomposition methods have led to the classification of different velocity components that originate from different physical mechanisms \citep{Hamann_1994, Hartigan_1995, Ercolano_2017, Fang2018, Banzatti2019, Pascucci_2020}. Outflows, jets, magnetospheric accretion, and disk winds influence the angular momentum transport (i.e., \citet{Casse_2000, Casse_2002, Nezami_2012,Stepanovs_2016}). 
In most of the disk regions, the magneto-rotational instability (MRI; \cite{Balbus_1991}) is almost entirely suppressed by non-ideal magnetohydrodynamic (MHD) effects \citep{Bai_2013} because of the low level of coupling between the gas and the magnetic field. At the same time, strong magnetic fields give rise to magnetically driven winds \citep{Blandford_1982}, as in the uppermost layers of the disk atmosphere, the gas and magnetic field are coupled again. On the other hand, thermal photoevaporative winds can co-exist with magnetically driven winds \citep{Rodenkirch_2020}. In the regions where thermal disk winds are generated, the surrounding gas must be highly ionized at least enough for electrons to be thermally excited; temperatures can be greater than 5,000 K with gas densities between 10$^{5}$-10$^{6}$ cm$^{-3}$ \citep{2Simon_2016}. Therefore, studying forbidden emission lines can provide clues about the main processes taking place in the inner disk.

\begin{figure*}[!htbp]
    \centering
    \includegraphics[width=18cm]{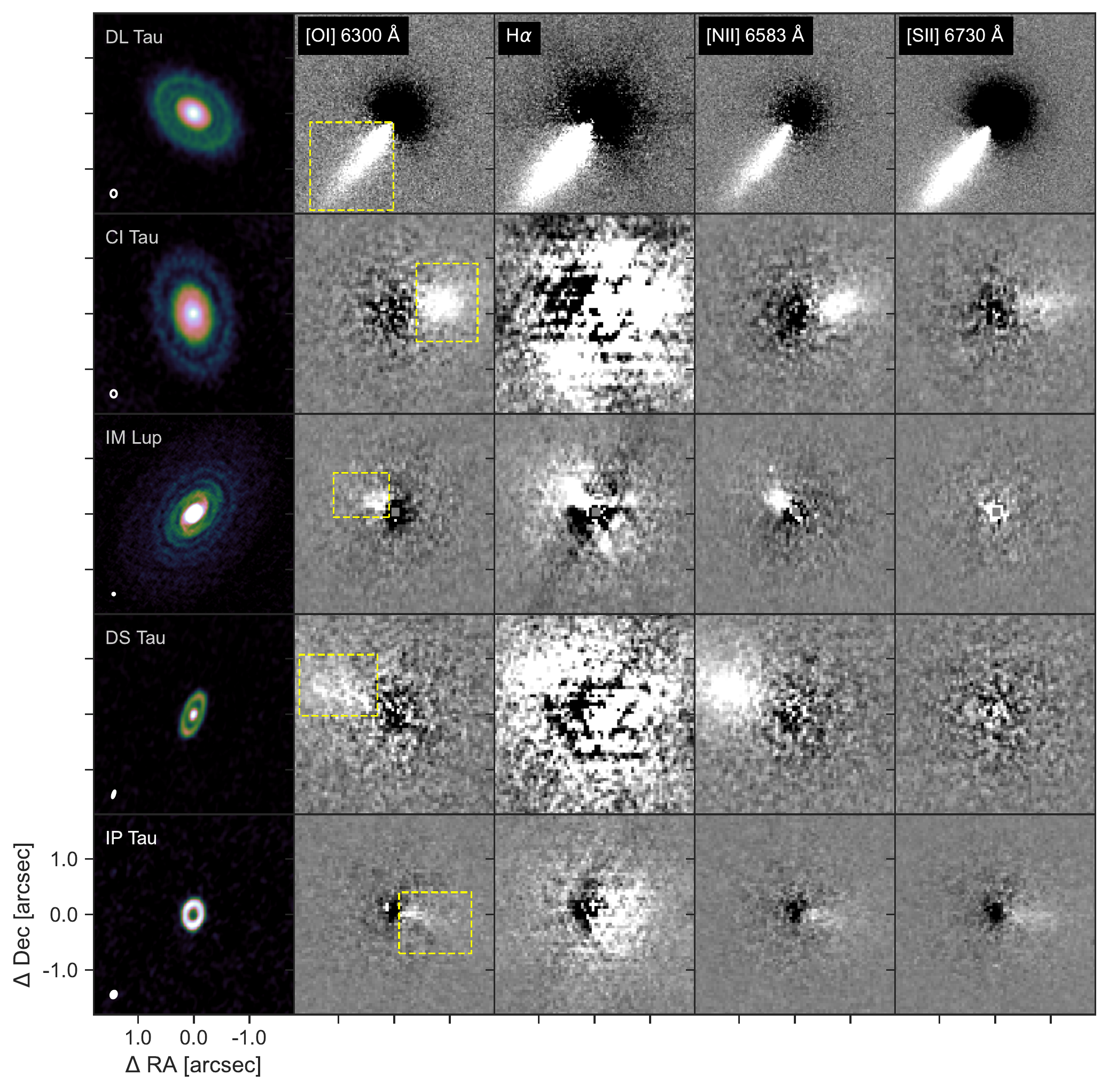} %
    \caption{Composite image of different protoplanetary disks at four different forbidden emission lines. First column shows the dust continuum emission at 1.3 mm from ALMA cycle 4 \citep{Long_2018, Long_2019}. Other columns show the different forbidden emission lines from MUSE as labeled in the top left  of the first row. Dashed yellow rectangles mark the area from where we gather the spectrum of the jet by summing over the spatial axes within the dashed yellow frame of the image.}
    \label{fig:supra}
    \end{figure*}
\begin{table*}[pht!]
    \caption{Disk properties (geometric parameters).} 
    \label{tab:disk_properties}
    \small 
    \centering 
    \begin{tabular}{lccccccccr} 
    \hline \hline
    \rule{0pt}{1.0\normalbaselineskip}
        Name & $\alpha$(J2000)  & $\delta$(J2000) & d$_\mathrm{pc}$  & incl  & PA$_{\mathrm{dust}}$  & $\overline{\mathrm{PA}}_{\mathrm{outflow/jet}}$  & difference & ref.\\
        &   &  & (pc) &  ($^{\circ}$) & ($^{\circ}$) & ($^{\circ}$) & ($^{\circ}$) \\
   \hline
    DL Tau & 04 33 39.07 & +25 20 38.09  &159.9 & -45.0$\pm$0.2  &  52.1$\pm0.4$ &  143.25$\pm$0.61 & 1.5$\pm$0.72 & $^{a}$   \\ 
    CIDA 9 & 05 05 22.86 & +25 31 31.23 &165.6 & 45.6$\pm0.5$  & 102.7$\pm0.7$ & -- &-- &$^{a}$\\ 
    CI Tau  & 04 33 52.01 & +22 50 30.09 & 160.3 & 50.0$\pm0.3$  &  11.2$\pm$0.13 & 79.12$\pm$1.36 & 0.32$\pm$1.37 & $^{a}$ \\ 
    DS Tau  & 04 47 48.59 & +29 25 11.18 & 158.4 & -65.2$\pm0.3$ & 159.6$\pm0.4$& 70.69$\pm$0.78 & 1.09$\pm$0.88  & $^{a}$ \\ 
    GO Tau  & 04 43 03.07 & +25 20 18.70 & 142.4 & 53.9$\pm0.5$ & 110.9$\pm0.24$& -- & -- & $^{a}$\\ 
    IP Tau  & 04 24 57.08 & +27 11 56.54 & 129.4 & -45.2$_{-0.8}^{+0.9}$ & 173.0$\pm1.1$ & 81.35$\pm$1.53 & 1.65$\pm$1.88  &$^{a}$ \\ 
    IM Lup  & 15 56 09.17 & -34 30 35.67 & 155.8 & -47.5$\pm$0.5 &  143.9$\pm0.6$ & 56.02$\pm$1.90  & 2.12$\pm$1.99  &$^{b}$ \\ 
    GW Lup  & 14 46 44.72 & -34 30 35.67 & 155.2 & 38.7$\pm$0.3 & 127.6$\pm0.5$ & -- & -- & $^{b}$\\ 
    \hline
    \multicolumn{8}{@{}l@{}}{\scriptsize \textbf{Notes.} The inclination, and PA$_{\mathrm{dust}}$ are obtained from $^{a}$\citet{Long_2019}, and $^{b}$\citet{Huang_2018}} \\ 
    \multicolumn{8}{@{}l@{}}{\scriptsize The distances, d$_\mathrm{pc}$, are from the third release of the \citet{Gaia_2021}} \\
    \multicolumn{8}{@{}l@{}}{\scriptsize The $\overline{\mathrm{PA}}_{\mathrm{outflow/jet}}$ represents the average among the lines we measure a PA$_{\mathrm{outflow/jet}}$ in Table \ref{tab:jets_properties} }\\
    \multicolumn{8}{@{}l@{}}{\scriptsize Symbol '--' means no detection} \\
    \end{tabular}
    \end{table*}

T Tauri sources show complex velocity line profiles. The low-velocity component (LVC) can be classified in two components: the narrow component (NC) and the broad component (BC); both presumably tracing photoevaporative or magnetohydrodynamic winds. \citet{Whelan_2021} reported the first detection of MHD winds, which spatially resolved the forbidden emission lines using spectroastrometry analysis. On the other hand, a small percentage \citep[$\sim$30$\%$;][]{Nisini_2018} shows a high-velocity component (HVC) associated with jets. Though different mechanisms form winds and outflows and jets, collimated disk winds could form the base of an outflow-like or jet-like structure. Many surveys \citep[i.e.,][]{Hamann_1994, Hartigan_1995, Hirth_1997, Antoniucci_2011, Rigliaco_2013, Natta_2014, 2Simon_2016, Nisini_2018, Fang2018, Banzatti2019} have shown that the HVC of the [\ion{O}{i}]~$\lambda$6300 excitation line is common and likely probing outflows, jets, or MHD winds \citep{Edwards_1987, Hartigan_1995, Bacciotti_2000, Lavalley_2000, Woitas_2002}.  Moreover, the HVC in T Tauri disks has been spatially confirmed to be associated with jets. Their emission reaches a greater extension from the star than those lines with the LVC \citep{Dougados_2000}. Other lines, such as [\ion{N}{ii}]~$\lambda$6583 and [\ion{S}{ii}]~$\lambda$6731, have also been reported to probe outflows and jets \citep{Solf_1993, Dougados_2000, Lavalley_2000, Woitas_2002, Natta_2014, Nisini_2018}, in which the detailed kinematic structure has been determined from the analysis of such intrinsic line profiles. Another important line is H$\mathrm{\alpha,}$ which also traces the velocity fields of outflows and jets \citep{Edwards_1987, Bacciotti_2000, Woitas_2002}, as well as accretion processes in the accretion flow onto a planet and in the circumplanetary disk \citep[i.e.,][]{Haffert_2019, Hashimoto_2020, Marleau_2022}.

Studying the connection between the inner and outer disk has attracted much attention lately. Despite the rings and gaps seen in protoplanetary disks, shadows could indicate some degree of misalignment between the inner and outer disk \citep{Marino_2015, stolker_2016, Min_2017, Debes_2017, Benisty_2017, Casassus_2018, Benisty_2018, Pinilla_2019, Muro-Arena_2020}. Resolving the inner parts of the disk is quite challenging. A recent study by \citet{Bohn_2022} showed that it is possible to assess the inner disk using near-infrared and submillimeter observations in T Tauri and Herbig Ae/Be disks. They emphasize the difficulty of measuring a possible misalignment between inner and outer disks in T Tauri disks, mainly due to the limited angular resolution of VLT/GRAVITY or the specific geometry orientation of the disks. 

    In this paper, we introduce the detection of forbidden emission lines such as: [\ion{O}{i}]~$\lambda\lambda$6300, 6363, [\ion{N}{ii}]~$\lambda\lambda$6548, 6583, H$\mathrm{\alpha}$, and [\ion{S}{ii}]\,$\lambda\lambda$6716, 6730 for the first time in some T Tauri sources with the MUSE/VLT instrument in the narrow-field mode (NFM). These forbidden emission lines originate from the innermost part of the disk as outflows/jets and we estimate the position angle (PA$_\mathrm{outflow/jet}$) that is connected with the inner disk and compare it with the dust position angle (PA$_\mathrm{dust}$), derived from previous work. In \S\ref{sec:sect_2}, we describe the physical properties of the disk using the Atacama Large Millimeter/submillimeter Array (ALMA)  and the MUSE observations. In \S\ref{sect_3}, we present our estimates of the PA$_\mathrm{outflow/jet}$ and analyze the velocity components from the line profiles. In \S\ref{sect_4} and \S\ref{sect:sect_5}, we present our summary and conclusions, respectively.
\section{Disk parameters and observations}
\label{sec:sect_2}
We briefly summarize the disk observational description of the dust continuum emission. The physical properties we use to compare the properties of the outflows/jets are originally constrained by previous studies. Lastly, we introduce the MUSE observations. 

\subsection{Dust continuum parameters}
    
    The disks presented here have been well studied and are known to have dust substructures detected at millimeter wavelengths. We adopted some additional disk properties, such as the inclination and PA$_\mathrm{dust}$, from previous work that performed the disk model fitting based on the dust morphology and the origin of the substructures. The distance of each source is obtained from the \citet{Gaia_2021}.

   The disk properties were obtained and analyzed from three different datasets. The disk properties from IM Lup and GW Lup were obtained from the Disk Substructures at High Angular Resolution Project (DSHARP) at 1.25 mm continuum observations with a high spatial resolution of 0$\farcs$04 \citep{Andrews_2018, Huang_2018}. For the disks in the Taurus star-forming region, we considered the ALMA Cycle 4 observations (ID: 2016.1.01164.S; PI: Herczeg) that were observed at 1.33 mm and a resolution of $\sim$0$\farcs$12 \citep{Long_2018, Long_2019}. 
   
  \subsection{MUSE observations and data reduction}
    
    The observations were carried out between November 2019 and January 2020 with VLT/MUSE in the NFM under program-ID 0104.C-0919 (PI: A. M\"uller). The MUSE optical integral field spectrograph covers the spectral range from 4650--9300~\AA~ at a resolving power of $\sim$2000 at 4600~\AA~ and $\sim$4000 at 9300~\AA~  \citep{2010SPIE.7735E..08B}, with a spectral sampling of 0.125 nm per pixel. The adaptive-optics assisted NFM has a field of view (FOV) of 7.4$\times$7.4~$\mathrm{arcsec}^{2}$ with a spatial scale of 0.025~$\mathrm{arcsec}$ per pixel, which is a pixel scale that is about ten times smaller  than the wide-field mode. With this spatial resolution, we can recover the spatial extension, by means of x and y pixels, of the emission coming from the outflows/jets. The VLT adaptive optics system uses four laser guide stars whose wavelengths range from 5781~\AA~to 6048~\AA. This spectral range is removed from the MUSE spectra to avoid the presence of additional lines that occur when the laser interacts with air molecules.
    
    The MUSE-NFM observations were reduced with the standard ESO pipeline v2.8.1 \citep{2012SPIE.8451E..0BW, 2014ASPC..485..451W, 2016ascl.soft10004W}, which includes bias subtraction, spectral extraction, flat-fielding, wavelength calibration, and flux calibration. Although the NFM includes an atmospheric dispersion corrector (ADC), the centroid variations can still be a few spatial pixels along the entire wavelength range. Therefore, we corrected the calibrated pixel tables for each exposure based on the centroid shifts with wavelengths measured on intermediate cube reconstructions. The corrected pixel tables were then used to create the combined cube of all calibrated exposures. 
    
    The wavelength calibration was done using a helium-argon ARC lamp from the day before each night of observations. Due to temperature variations, the wavelength solution for a given spectrum can be offset and an empirical wavelength correction using bright skylines should be applied \citep{Weilbacher_2020}. Skylines are, however, faint in the NFM, so this wavelength correction is hard to apply. These offsets can vary from target to target and are on the order of fractions of a pixel, up to a full pixel in extreme cases \citep[5-50 km~s$^{-1}$; see also][]{Xie_2020}. We report eight data cubes that were obtained over 0.5~h each.

   \begin{figure*} [!htbp]
    \centering
    \begin{subfigure}[]{.45\textwidth}
    \includegraphics[width=7cm]{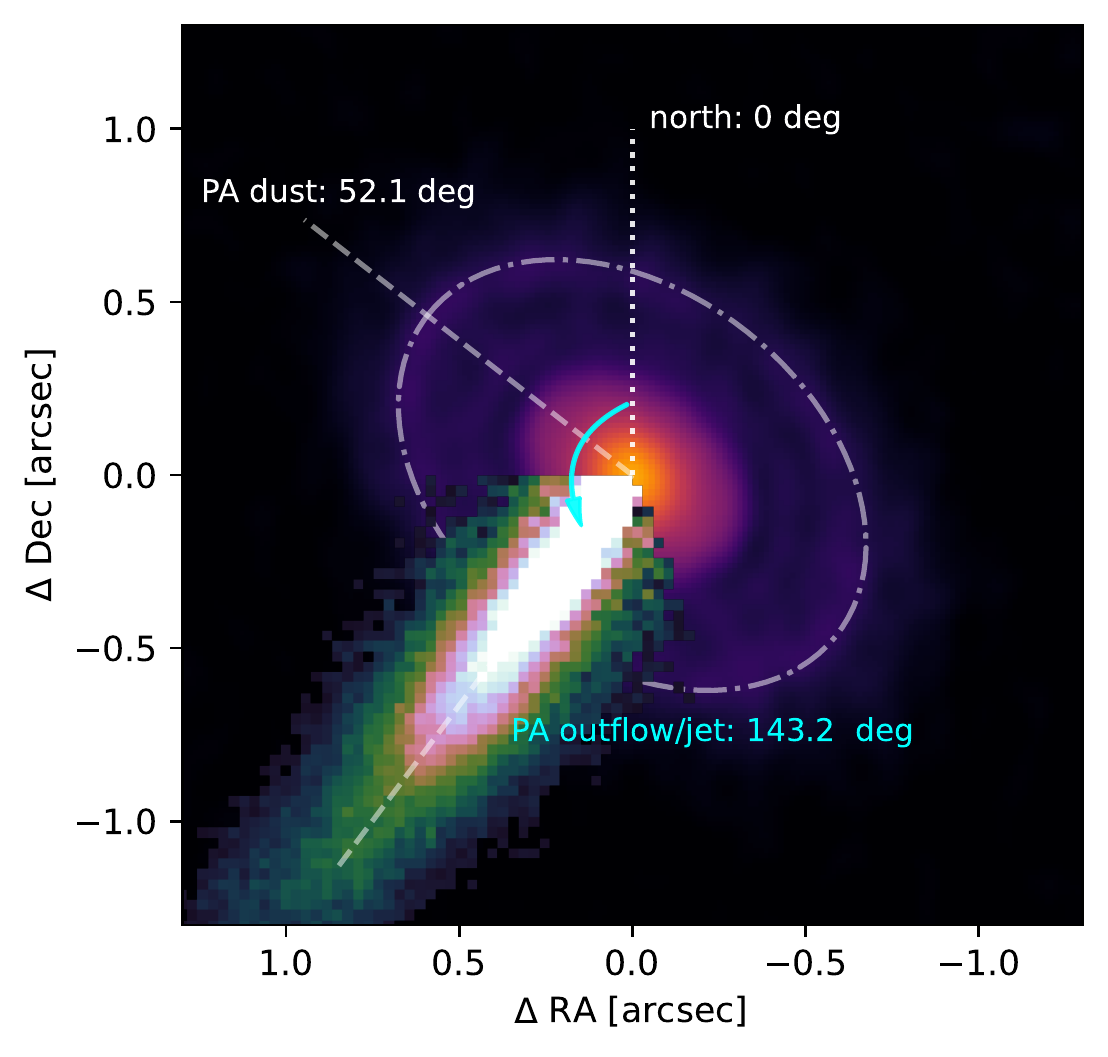} 
    \end{subfigure}
    \begin{subfigure}[]{0.45\textwidth}
    \includegraphics[width=7cm]{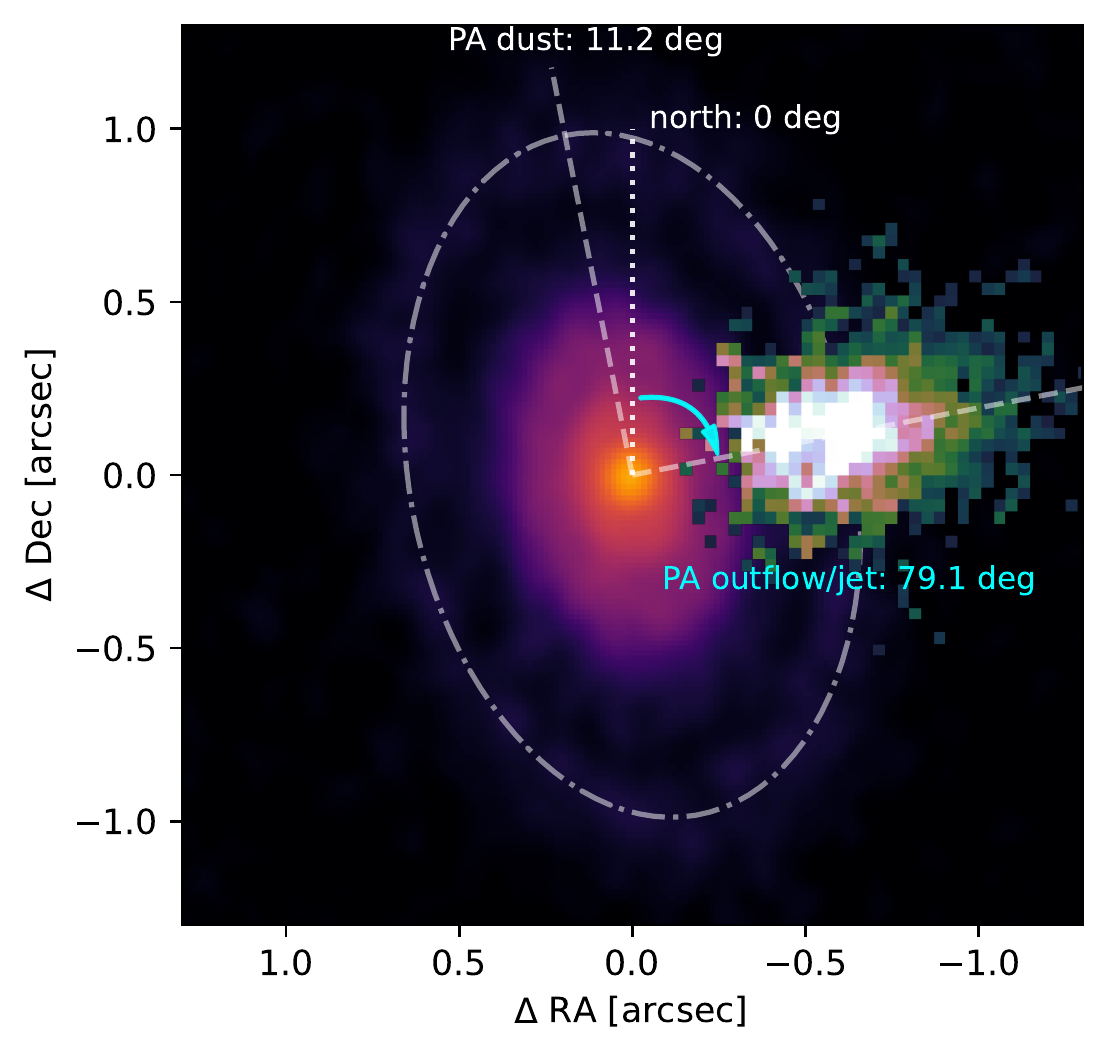} 
    \end{subfigure}
    \begin{subfigure}[]{.45\textwidth}
    \includegraphics[width=7cm]{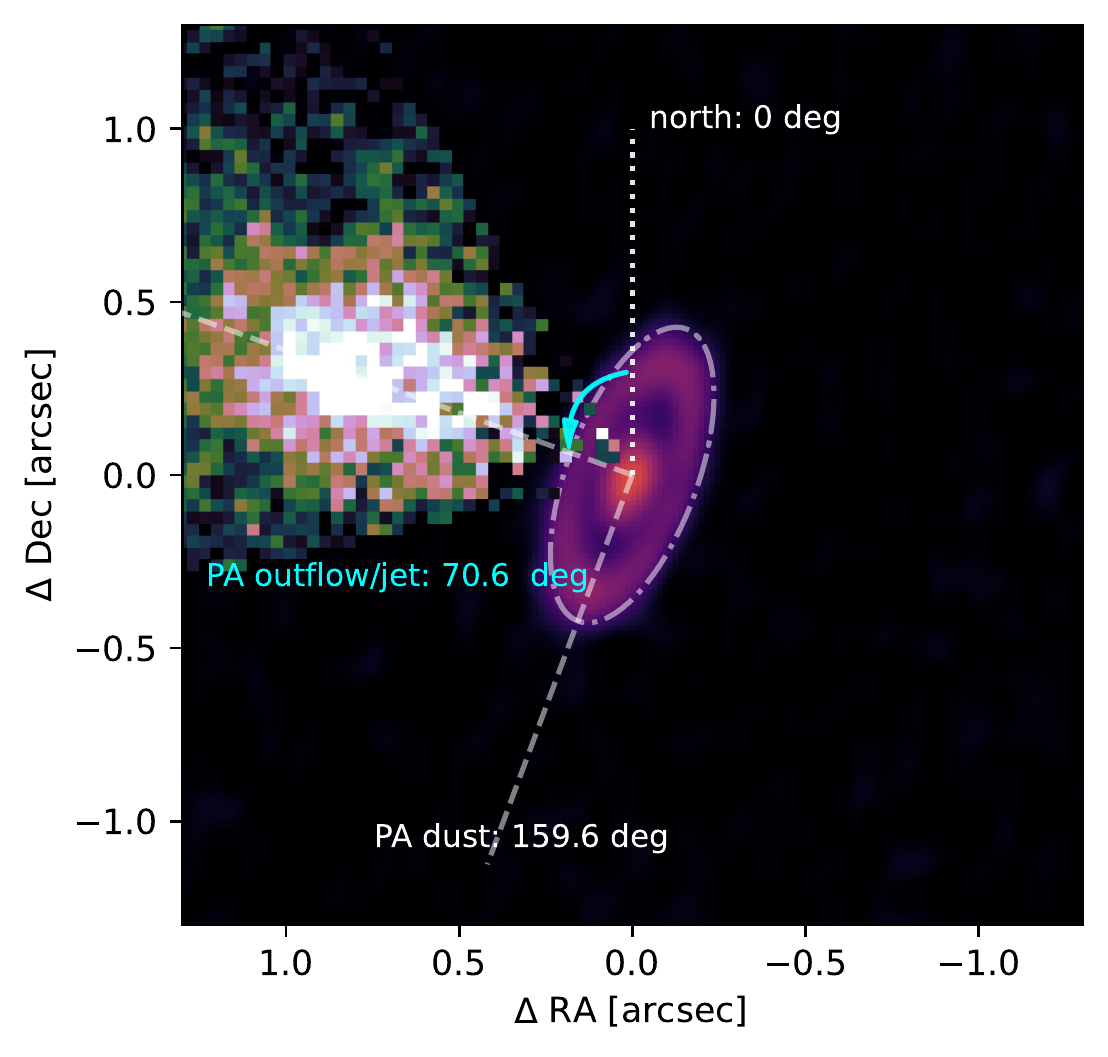}
    \end{subfigure}
    \begin{subfigure}[]{.45\textwidth}
    \includegraphics[width=7cm]{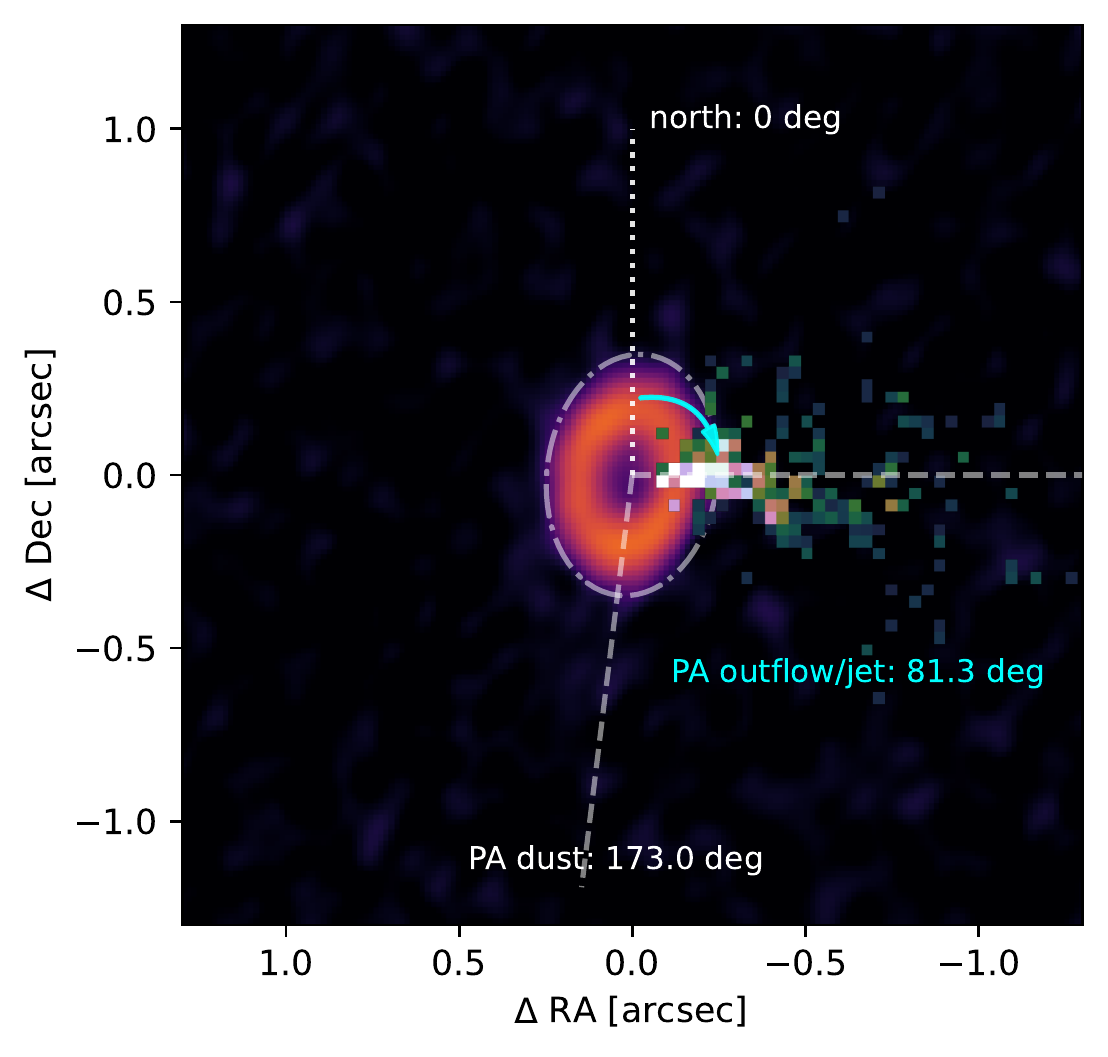}
    \end{subfigure}
    \begin{subfigure}[]{.45\textwidth}
    \includegraphics[width=7cm]{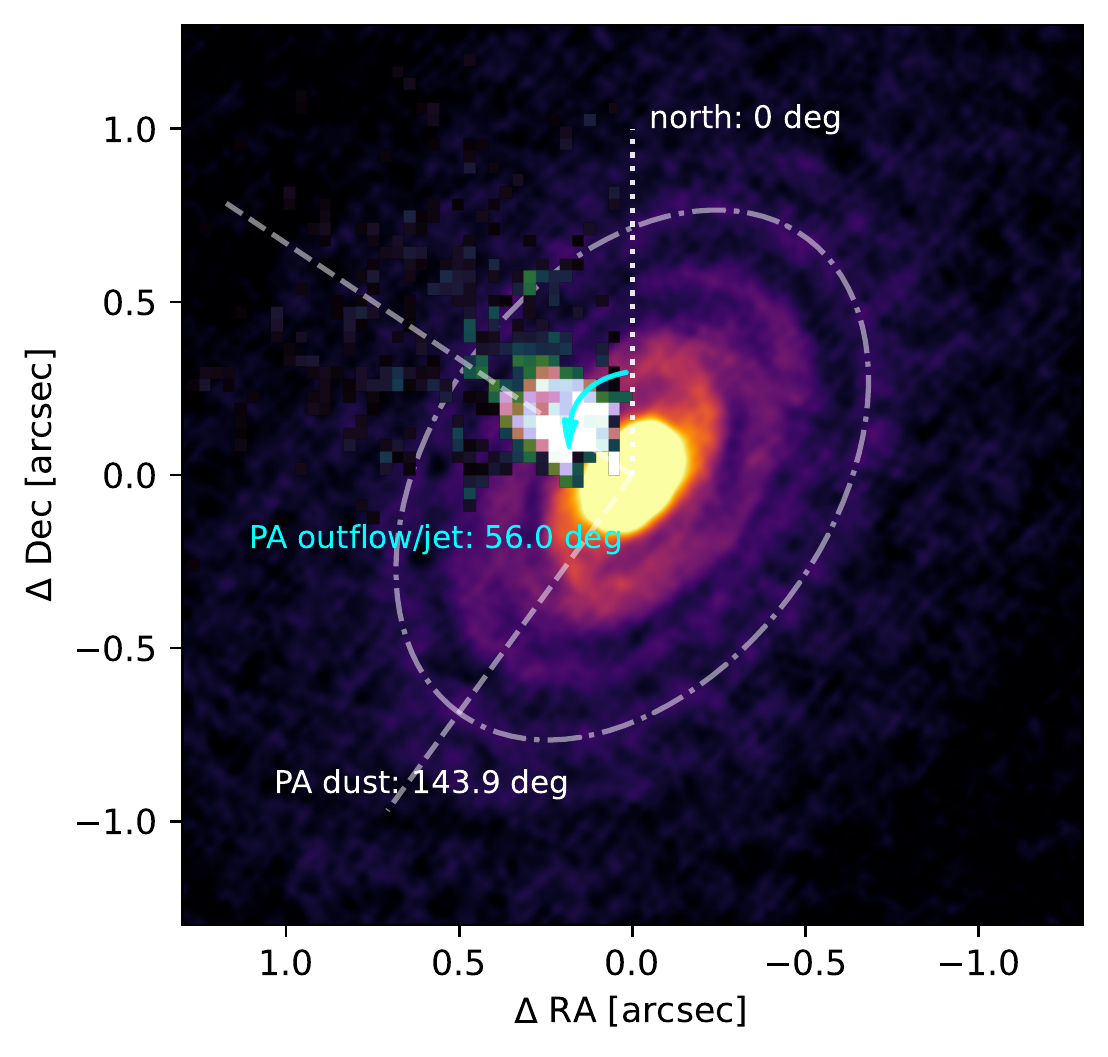}
    \end{subfigure}
    \caption{Visualization of the estimated $\overline{\mathrm{PA}}_{\mathrm{outflow/jet}}$ compared to the PA$_{\mathrm{dust}}$. Top panels show DL Tau (\textit{left}) and CI Tau (\textit{right}). Middle panels show DS Tau (\textit{left}) and IP Tau (\textit{right}). Bottom panel is IM Lup. The MUSE outflow/jet images are averaged from the different emission lines that are used to estimate the $\overline{\mathrm{PA}}_{\mathrm{outflow/jet}}$. We adopt the geometrical configuration visually demonstrated in Figure 3 from \citet{Pietu_2007} for the disks with positive and negative inclination. CI Tau has positive disk inclination, and the $\overline{\mathrm{PA}}_{\mathrm{outflow/jet}}$ is measured from north to west. The rest of the sources have negative disk inclination and the $\overline{\mathrm{PA}}_{\mathrm{outflow/jet}}$ is measured from north to east.}
    \label{DLTau_and_CITau_diagram}
    \end{figure*}

   \subsection{MUSE and ALMA image registration}
    
     When comparing the MUSE images with ALMA data, we focus on comparing the geometry of the dusty outer disk plane, as measured with ALMA, and the position angle of the outflows/jets, as measured with MUSE. With the assumption that the disk is circular and the star is at its center, the ALMA observations can be used to recover the inclination and the position angle relative to the fitted disk center. In our sample, the systems have their disk geometry calculated through visibilities \citep{Long_2018} or image fitting \citep{Huang_2018}, and we adopted those published values for our inclination and position angle for the outer disk plane. As both works fit the center of the disk under the assumptions mentioned above, the PA$_\mathrm{dust}$ already includes the observational and instrumental uncertainties, which remain fixed in this work.
     
     With MUSE, our assumption to recover the PA$_\mathrm{outflow/jet}$ is that the star is centered in the image frame. In order to confirm this, we fit a 2D circular Gaussian to the specific channels we used to analyze the PA$_\mathrm{outflow/jet}$. We do this to the reduced data cube without PSF subtraction, assuming that the outflow/jet  emission is negligible compared to the stellar emission. We found that for each channel, the center of the Gaussians is less than 1 pixel (or smaller than 25mas) from the center of the images for all sources. Therefore, we measured the PA$_\mathrm{outflow/jet}$ relative to the center of the image where the star is. 
     
     When comparing ALMA with MUSE, the assumptions are that the outer disk geometry (incl, PA$_\mathrm{dust}$) was measured relative to the same point as the PA$_\mathrm{outflow/jet}$, which is the stellar position. Additionally, the outer disk plane did not change even when both images were taken at different epochs. The comparison of the geometry of the disk with different instruments is similar to the work done by \citet{Bohn_2022}, where they combined the K-band from GRAVITY/VLTI to constrain the geometry of the inner disk by finding the best parametric SED solution to the squared visibilities. These authors  used ALMA $^{12}$CO and $^{13}$CO data to constrain the geometry of the outer disk. Their determination of potential misalignment between the inner and outer disks is narrowed down by comparing the inclinations and position angles defined by the planes of the inner and outer disks, respectively. The \citet{Gravity_2020} describes the position angle constrained by ALMA to be a well-fixed parameter in order to describe the geometry of the disk. We include the rotational accuracy as 0.2$^{\circ}$ when propagating the uncertainty for each PA$_\mathrm{outflow/jet}$ value in Table \ref{tab:jets_properties} (see also Appendix \ref{sec:rotation_residual_analysis}).

    \section{Results}
    \label{sect_3}

    We identified seven forbidden emission lines: [\ion{O}{i}]~$\lambda\lambda$6300, 6363, [\ion{N}{ii}]~$\lambda\lambda$6548, 6583, H$\mathrm{\alpha}$, [\ion{S}{ii}]\,$\lambda\lambda$6716, 6730, which are seen to be originating from the innermost part of the disk as outflow-like or jet-like structures (Fig.\ref{fig:supra}, Fig.\ref{fig:DLTau_jet_dust}, Figure \ref{spectrum1}, Figure \ref{spectrum2}, and Figure \ref{spectrum3}). Some of these lines have been reported in some of the sources we analyzed here (i.e., \citet{2Simon_2016, Fang2018, Banzatti2019}). We used \citet{Natta_2014}, \citet{Podio_2006}, and the NIST atomic spectra database for the line identification and line uncertainty. The PA$_\mathrm{outflow/jet}$ is defined as the position angle of the outflow/jet  measured in degrees that lie approximately in the direction perpendicular to the rotational axis of the disk. In the next section, we detail how the geometrical derivation of the PA$_\mathrm{outflow/jet}$ is conducted. It is very likely that we are not solely seeing jets, but also outflows and (perhaps) winds that cannot be revealed due to the low spectral resolution of MUSE. Any description and analysis of winds are out of the scope of the present work, as it is necessary to assess the few km s$^{-1}$ velocity components that characterize them. We performed a Gaussian fit to the emission lines to study their velocity components and, finally, we measured the length of the outflow/jet  to estimate the mass loss from the [\ion{O}{i}]~$\lambda$6300 line. 



 \begin{table*}[htp!]
   \caption{Outflow/jet properties.}
   \label{tab:jets_properties}
   \small 
   \centering 
   \begin{tabular}{lccccccccccr} 
   \noalign{\smallskip} \hline \hline
   \noalign{\smallskip} & \multicolumn{4}{c}{[\ion{O}{i}] $\lambda6300.30 ~\pm$~0.010~\AA} &  & \multicolumn{4}{c}{[\ion{N}{ii}] $\lambda6548.05 ~\pm$~0.10~\AA} \\
   \cline{2-5}
   \cline{7-10}
   \noalign{\smallskip} Name & FWHM & $v_{\mathrm{outflow/jet}}$ & $F_{\lambda}$ & PA$_{\mathrm{outflow/jet}}$ &  & FWHM & $v_{\mathrm{outflow/jet}}$ & $F_{\lambda}$ & PA$_{\mathrm{outflow/jet}}$ \\ 
     & (km~s$^{-1}$) &  (km~s$^{-1}$) & (erg~s$^{-1}$~cm$^{-2}$~\AA$^{-1}$) & ($^{\circ}$) & & (km~s$^{-1}$) &  (km~s$^{-1}$) & (erg~s$^{-1}$~cm$^{-2}$~\AA$^{-1}$) & ($^{\circ}$) \\
   \hline
   DL Tau & 135.22 & -217.8$\pm$2.0 & 7.96$\times$10$^{-15}$ & 143.41$\pm$0.25 & & 130.89 & -216.3$\pm$2.0  & 1.93$\times$10$^{-15}$ & -- \\ 
   CIDA9 & -- &-- & -- & -- & & -- & -- & -- & -- \\
   CI Tau & 160.90 & -209.7$\pm$1.3 & 1.18$\times$10$^{-15}$ & 78.12$\pm$0.61 & & -- & -- & -- & --\\ 
   DS Tau & 143.70 & 161.9$\pm$1.6 & 1.21$\times$10$^{-15}$ & 69.79$\pm$0.68 & & -- & -- & -- & -- \\ 
   GO Tau & -- & -- & -- & -- & & -- & -- & -- & --\\
   IP Tau & 204.58 & -83.0$\pm$2.0 & 2.87$\times$10$^{-16}$ & 81.35$\pm$1.53 & & -- & -- & -- & -- \\ 
   IM Lup & 262.19 & -95.4$\pm$1.3 & 4.91$\times$10$^{-16}$ & 59.08$\pm$1.67 & & -- & -- & -- & -- \\ 
   GW Lup & -- & -- & -- & -- & & -- & -- & -- & -- \\
   \hline
\end{tabular}
\end{table*}

 \begin{table*}[htb]
   \label{tab:jets_properties_cont}
   \small 
   \centering 
   \begin{tabular}{lccccccccccr} 
   \noalign{\smallskip} \hline \hline
   \noalign{\smallskip} & \multicolumn{4}{c}{H$\mathrm{\alpha}$ $\lambda$6562.80 $\pm~10^{-5}$~\AA} &  &\multicolumn{4}{c}{[\ion{S}{ii}] $\lambda$6716.44 $\pm$~0.010~\AA} \\
   \cline{2-5}
   \cline{7-10}
    \noalign{\smallskip} Name & FWHM & $v_{\mathrm{outflow/jet}}$ & $F_{\lambda}$ & PA$_{\mathrm{outflow/jet}}$ &  & FWHM & $v_{\mathrm{outflow/jet}}$ & $F_{\lambda}$ & PA$_{\mathrm{outflow/jet}}$ \\ 
     & (km~s$^{-1}$) &  (km~s$^{-1}$) & (erg~s$^{-1}$~cm$^{-2}$~\AA$^{-1}$) & ($^{\circ}$) & & (km~s$^{-1}$) &  (km~s$^{-1}$) & (erg~s$^{-1}$~cm$^{-2}$~\AA$^{-1}$) & ($^{\circ}$) \\
   \hline
   DL Tau & 121.73 & -222.9$\pm$2.0 & 2.28$\times$10$^{-14}$ & 142.79$\pm$0.23 & & 124.87 & -208.0$\pm$2.0 & 7.31$\times$10$^{-15}$ & 143.22$\pm$0.23 \\ 
   CIDA9 & -- & -- & -- & -- & & -- & -- & -- & --\\
   CI Tau & 154.65 & -184.8$\pm$1.3 & 5.51$\times$10$^{-15}$ & -- & & 150.17 & -195.6$\pm$1.3 & 3.74$\times$10$^{-16}$ & --\\
   DS Tau & -- & -- & -- & -- & & 115.56 & 151.2$\pm$1.6 & 6.48$\times$10$^{-16}$ & -- \\
   GO Tau &-- &-- &-- & -- & & -- & -- & -- & --\\
   IP Tau & -- & -- & -- & -- & & -- & -- & -- & -- \\
   IM Lup & -- & -- & -- & -- & & -- & -- & -- & -- \\
   GW Lup & -- & -- & -- & -- & & -- & -- & -- & -- \\
   \hline
\end{tabular}
\end{table*}

\begin{table*}[htb]
   \small 
   \centering 
   \begin{tabular}{lccccccccccr} 
   \noalign{\smallskip} \hline \hline
   \noalign{\smallskip} & \multicolumn{4}{c}{[\ion{O}{i}] $\lambda$6363.78 $\pm$~0.010~\AA} &  &\multicolumn{4}{c}{[\ion{N}{ii}] $\lambda$6583.45 $\pm$~0.10~\AA} \\
   \cline{2-5}
   \cline{7-10}
    \noalign{\smallskip} Name & FWHM & $v_{\mathrm{outflow/jet}}$ & $F_{\lambda}$ & PA$_{\mathrm{outflow/jet}}$ &  & FWHM & $v_{\mathrm{outflow/jet}}$ & $F_{\lambda}$ & PA$_{\mathrm{outflow/jet}}$ \\ 
     & (km~s$^{-1}$) &  (km~s$^{-1}$) & (erg~s$^{-1}$~cm$^{-2}$~\AA$^{-1}$) & ($^{\circ}$) & & (km~s$^{-1}$) &  (km~s$^{-1}$) & (erg~s$^{-1}$~cm$^{-2}$~\AA$^{-1}$) & ($^{\circ}$) \\
   \hline
   DL Tau & 135.67 & -219.5$\pm$2.0 & 2.64$\times$10$^{-15}$ & 143.77$\pm$0.31  & & 116.65 & -217.8$\pm$2.0 & 4.93$\times$10$^{-15}$ & 143.51$\pm$0.23 \\  
   CIDA9 & -- & -- & -- & -- & & -- & -- & -- & --\\
   CI Tau & -- & -- & -- & -- & & 158.72 & -225.4$\pm$1.3 & 1.20$\times$10$^{-15}$ & 79.68$\pm$0.92 \\ 
   DS Tau & -- & -- & -- & -- & & 129.32 & 194.8$\pm$1.6 & 2.18$\times$10$^{-15}$ & 71.59$\pm$0.39 \\ 
   GO Tau & -- &-- &-- & -- & & -- & -- & -- & --\\
   IP Tau & -- & -- & -- & -- & & -- & -- & -- & -- \\
   IM Lup & -- & -- & -- & -- & & 135.32 & -124.1$\pm$1.3 & 3.77$\times$10$^{-16}$ & 52.96$\pm$0.90 \\ 
   GW Lup & -- & -- & -- & -- & & -- & -- & -- & -- \\
   \hline
\end{tabular}
\end{table*}

\begin{table*}[htb!]
   \small 
   \centering 
   \begin{tabular}{lcccccr} 
   \noalign{\smallskip} \hline \hline
   \noalign{\smallskip} & \multicolumn{4}{c}{[\ion{S}{ii}] $\lambda$6730.82 $\pm$~0.010~\AA} \\
   \cline{2-5}
    \noalign{\smallskip} Name & FWHM & $v_{\mathrm{outflow/jet}}$ & $F_{\lambda}$ & PA$_{\mathrm{outflow/jet}}$\\ 
     & (km~s$^{-1}$) &  (km~s$^{-1}$) & (erg~s$^{-1}$~cm$^{-2}$~\AA$^{-1}$) & ($^{\circ}$)\\
   \hline
   DL Tau & 122.54 & -236.6$\pm$2.0 & 1.16$\times$10$^{-14}$ & 142.81$\pm$0.23 \\  
   CIDA9 & -- & -- & -- & -- \\
   CI Tau & 152.89 & -189.7$\pm$1.3 & 7.35$\times$10$^{-16}$ & 79.57$\pm$0.79 \\ 
   DS Tau & 130.87 & 167.9$\pm$1.6 & 1.11$\times$10$^{-15}$ & -- \\
   GO Tau & -- & -- & -- & -- \\
   IP Tau & -- & -- & -- & -- \\
   IM Lup & -- & -- & -- & -- \\
   GW Lup & -- & -- & -- & -- \\
   \hline
\end{tabular}
\caption*{\footnotesize{\textbf{Note:} The outflow/jet is obtained by de-projecting the line centroids from Gaussian fits, as explained in Sect. \ref{sec:outflow_jet_velocities}; the errors reported in this table do not include the uncertainty in wavelength calibration. Symbol '--' means no detection. All PA$_{\mathrm{outflow/jet}}$ uncertainties includes the 0.2$^{\circ}$ MUSE rotational uncertainty.}}
\end{table*}

\begin{figure*} [!htbp]
    \centering
    \includegraphics[width=12cm]{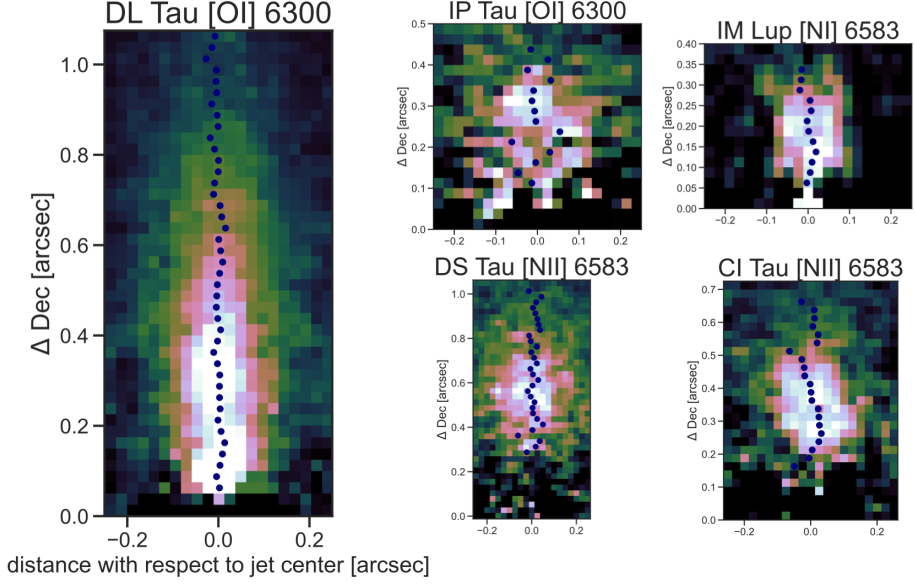} 
    \caption{Successive Gaussian centers shown as navy-colored dots. Wiggles are seen in DL Tau and IP Tau at [\ion{O}{i}]~$\lambda$6300, and for CI Tau, DS Tau, and IM Lup at [\ion{N}{ii}]~$\lambda$6583.}
    \label{fig:joydivision}
    \end{figure*}

    \subsection{Geometrical fitting of the outflows/jets}
    \label{geometrical_fitting}
    
    In order to derive the PA$_{\mathrm{outflow/jet}}$ for each spectral line in all disks, we fit the jet intensity peaks with a linear function (after stellar subtraction). We read the maximum value across the outflow/jet for the forbidden emission lines detected on each source and store the x and y positions corresponding to the jet emission into a new array. These are then converted into polar coordinates, where $r = \sqrt{x^{2}+y^{2}}$ and $\theta = \arctan(y/x)$. 
    Jet intensity peaks are only considered if they are above 5$\sigma$ of the background. For other imaging parameters and wavelengths (see Table \ref{tab:jet_lengths} and \ref{tab:spectral_widths}). The retrieved positions contain the angular information to calculate the PA$_{\mathrm{outflow/jet}}$. 
    In order to fit a line along these retrieved positions of the jet, we employed a fitting method called the orthogonal distance regression (ODR; \citet{brown1990statistical}). This routine gathers the retrieved data and the linear model function to produce the best linear parameters for the jet. The linear model function follows the simple ordinary prescription of a line, $f(x_{i}; \vec{\beta}) = \beta_{1}x + \beta_{0}$, where, $x_{i}$ is the position in pixels, and $\beta_{0}$ and $\beta_{1}$ are the unknown parameters to be found by the ODR routine. We assign an initial standard deviation for each jet peak position to be 1 pixel. Choosing a value of 1 pixel is a conservative selection for the initial standard error as we found that when estimating the PA$_{\mathrm{outflow/jet}}$, the corresponding error of the Gaussian center is smaller than 1 pixel. As the function we used for the fit, $f(x_{i}; \vec{\beta)}$, is said to be linear in variables, $x_{i}$, and linear in parameters, $\vec{\beta}$, the total error in the routine is estimated by applying the linear least squares method. That is, by finding the set of parameters for which the sum of the squares of the $n$ orthogonal distances from the $f(x_{i}; \vec{\beta})$ curve to the $n$ data points is minimized. This error associated at the end of the fit follows: min $\sum_{i=1}^{n} [y_{i} - f(x_{i}+\delta_{i};\vec{\beta)}]^{2} + \delta_{i}^{2}$, where $\delta_{i}$ is the error associated to each $n$ data point.

    For the iterative routine to estimate the PA$_{\mathrm{outflow/jet}}$ for each emission line, we used (as an initial estimate) the PA$_{\mathrm{outflow/jet}}$ from the locations of the jet intensity peaks from the first part. Then, we used this first value for the  PA$_{\mathrm{outflow/jet}}$ as a seed for a second iteration, where we fit Gaussians to profiles that are orthogonal to the jet axis defined by the first PA$_{\mathrm{outflow/jet}}$. Depending on the source, the first three to five pixels were not used to avoid artifacts from the stellar subtraction at the image center. For example, a comparison between the peak intensity locations and the Gaussian centers is shown for DL Tau in Fig. \ref{fig:linear_jet_DLTau_diagram2}. Those Gaussian centers were then used to calculate the final PA$_{\mathrm{outflow/jet}}$ for each emission line. The final PA$_{\mathrm{outflow/jet}}$ of the disks for which we detect bright emission lines are shown in Table \ref{tab:jets_properties}. Then, we took the average, $\overline{\mathrm{PA}}_{\mathrm{outflow/jet}}$, from the PA$_{\mathrm{outflow/jet}}$ estimates of different emission lines in Table \ref{tab:jets_properties} and we added the corresponding errors in quadrature. This $\overline{\mathrm{PA}}_{\mathrm{outflow/jet}}$ value for each source was then used to compared with the PA$_\mathrm{dust}$ (see Table \ref{tab:disk_properties}). We note that the CIDA9, GO Tau, and GW Lup jets were too compact at the center, making it difficult to properly estimate any PA$_{\mathrm{outflow/jet}}$. Before making the comparison between the $\overline{\mathrm{PA}}_{\mathrm{outflow/jet}}$ and the PA$_{\mathrm{dust}}$, it is important to understand the side of the disk plane which the emission line is coming from. We used the same geometrical convention as in Figure 3 from \citet{Pietu_2007} to indicate the real disk plane configuration through the inclination for each disk system.
    
    In Fig.\ref{DLTau_and_CITau_diagram}, we portray this convention by superimposing the outflows/jets to the mm dust continuum of the protoplanetary disk. It shows a better visualization of the measured $\overline{\mathrm{PA}}_{\mathrm{outflow/jet}}$, in comparison with the PA$_\mathrm{dust}$, for DL Tau (\textit{left panel}) for the negative outer disk inclination case, and CI Tau (\textit{right panel}) for the positive outer disk inclination case. DS Tau, IP Tau, and IM Lup have negative outer disk inclinations. The PA$_\mathrm{dust}$ follows the usual standard definition of the position angle of the major axis of the outer disk, from 0$^{\circ}$ to 180$^{\circ}$, from north to east. Since DL Tau has a negative outer disk inclination, the PA$_\mathrm{outflow/jet}$ values are measured counterclockwise from the north, and it is similar to saying that the disk is flipped. On the other hand, CI Tau has a positive outer disk inclination then the PA$_\mathrm{outflow/jet}$ values are measured clockwise from the north. Having identified the orientation of the outer disk inclination, we determined the rotation of the protoplanetary disk-outflow/jet system (see Fig.\ref{fig:DLTau_jet_dust}). 
    
    We took into account the difference between the $\overline{\mathrm{PA}}_{\mathrm{outflow/jet}}$, which is related to the PA of the innermost disk, and the PA$_\mathrm{dust}$. For this, we followed:

    \begin{equation}
    \text{difference} =
    \begin{cases}
    \lvert \lvert \overline{\mathrm{PA}}_{\mathrm{outflow/jet}}-\text{PA}_{\mathrm{dust}}\lvert-90^{\circ} \lvert, & \text{if incl.<0} \\
     \lvert \lvert \overline{\mathrm{PA}}_{\mathrm{outflow/jet}}+\text{PA}_{\mathrm{dust}}  \lvert -90^{\circ} \lvert, & \text{if incl.>0}
    \end{cases}
    \end{equation}

    We found that the difference between the $\overline{\mathrm{PA}}_\mathrm{outflow/jet}$ and the PA$_\mathrm{dust}$ to be small in general (see Table \ref{tab:disk_properties}). For the sources analyzed here, the estimation of the PA$_\mathrm{outflow/jet}$ and their associated errors lie within the estimation of the PA$_\mathrm{dust}$ and its associated error range. Under the assumption that the outflow (or jet) axis is perpendicular to the inner disk plane, we do not find any misalignment between the inner and outer disk. The spread of the PA$_\mathrm{outflow/jet}$ values in Table \ref{tab:jets_properties} depend on the uncertainties and values coming from the Gaussian fit centers. 
    The symmetric Gaussian fit model is not well suited to capture possible asymmetries in outflows/jets produced by low signal-to-noise variations, especially close to the center of the images where the rms is not constant (see Figure \ref{fig:joydivision}). 
    Our rms values are estimated in a region free from stellar and outflow/jet emission; however, the rms value close to the image center is not constant due to contamination when subtracting the stellar spectrum. Those effects are not systematically considered in our PA$_\mathrm{outflow/jet}$ estimates, therefore, our uncertainties should be considered as lower limits. 
    It is the case that the estimation of the PA$_\mathrm{outflow/jet}$ for IM Lup was more difficult because of its compact extent and low signal-to-noise level. 
    The intensity variations in the outflow/jet are comparable to noise variation level, resulting in a PA$_\mathrm{outflow/jet}$ difference of 6$^{\circ}$ between the [\ion{O}{i}] $\lambda6300$ and [\ion{N}{ii}] $\lambda6583$ lines. 
    By calculating the difference between PA$_\mathrm{outflow/jet}$ and PA$_\mathrm{dust}$ we get 5.18$^{\circ}\pm1.77^{\circ}$ and 0.94$^{\circ}\pm1.08^{\circ}$ for [\ion{O}{i}] $\lambda6300$ and [\ion{N}{ii}] $\lambda6583$ line, respectively. 
    The $\overline{\mathrm{PA}}_{\mathrm{outflow/jet}}$ and PA$_\mathrm{dust}$ are different by $2.12^{\circ}\pm1.99^{\circ}$, therefore, the difference is within the uncertainty. This is further discussed in \S\ref{sect_4}. 
    
    \begin{figure}[htbp]
    \centering
    \includegraphics[width=7cm]{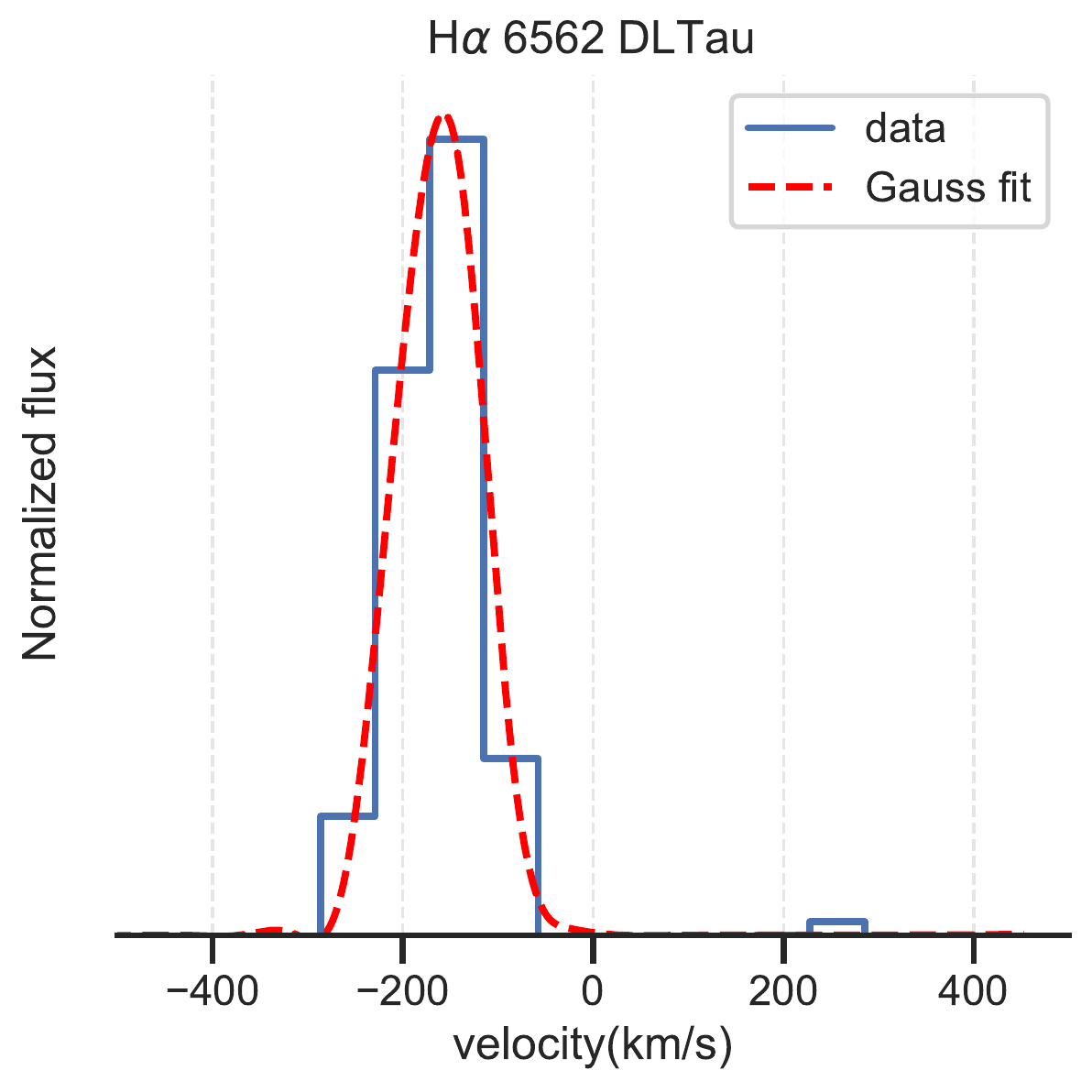} 
    \caption{H${\alpha}$ line profile identified for DL Tau probing the high line-of-sight velocity component coming from the strong outflow/jet. The red dashed line represent the line-of-sight maximum velocity. }
    \label{line_profile_Halpha}
    \end{figure}

\begin{figure*}[htp!]
    \centering
    \includegraphics[width=12cm]{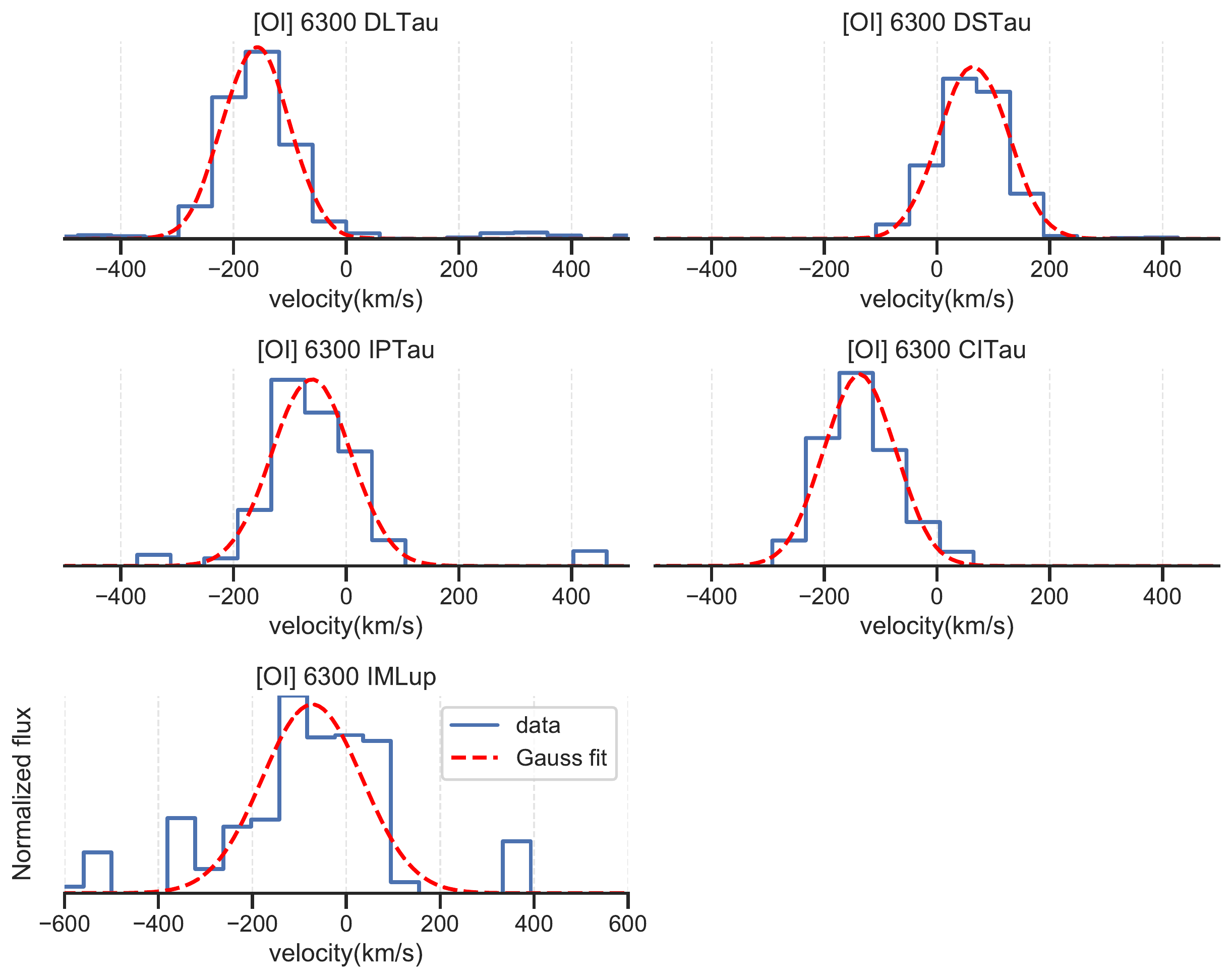} 
    \caption{[\ion{O}{i}]$\lambda$6300 line profiles identified in five sources from our sample. DL Tau and CI Tau have their line centered at velocities greater than 100 km s$^{-1}$, which is characteristic of jets. Unlike DL Tau and CI Tau, the disks of DS Tau, IP Tau, and IM Lup show smaller shifts of less than 100 km s$^{-1}$ and broader line widths significantly overlapping with 0 km s$^{-1}$, possibly including the LVC emission as a result. DS Tau has the most centered line.}
    \label{line_profile_OI6300}
    \end{figure*}
    
    Overall, we detected two essential geometrical features. One is a broadening of the respective intensity distribution with distance, indicating that the jet width increases with distance from the star. The other feature is the variation of the location of the intensity maximum, indicating a wiggling of the jet axis. A periodic pattern in these changes may hint at a precession or orbiting jet \citep{Fendt_1998}. In Fig. \ref{fig:joydivision}, the wiggle patterns are marked by the Gaussian centers as navy color dots. We further investigate how these Gaussian centers vary in position with respect to their distance from the jet axis (see Fig. \ref{linear_jet_DLTau_diagram} and Fig. \ref{fig:linear_jet_DLTau_diagram2}). In order to check whether or not these successive Gaussian centers have periodicity, we calculate their deviation of them from the jet axis. Applying a simple sinusoidal function in a periodogram does not provide significance, even if essentially done by eye -- as there seems to be some quasi-periodic behavior in the Gaussian centers. In \S\ref{jet_precession}, we further discuss how the patterns of the Gaussian centers could possibly hint at a sign of jet precession.

\begin{figure*}[!htbp]
    \centering
    \includegraphics[width=12cm]{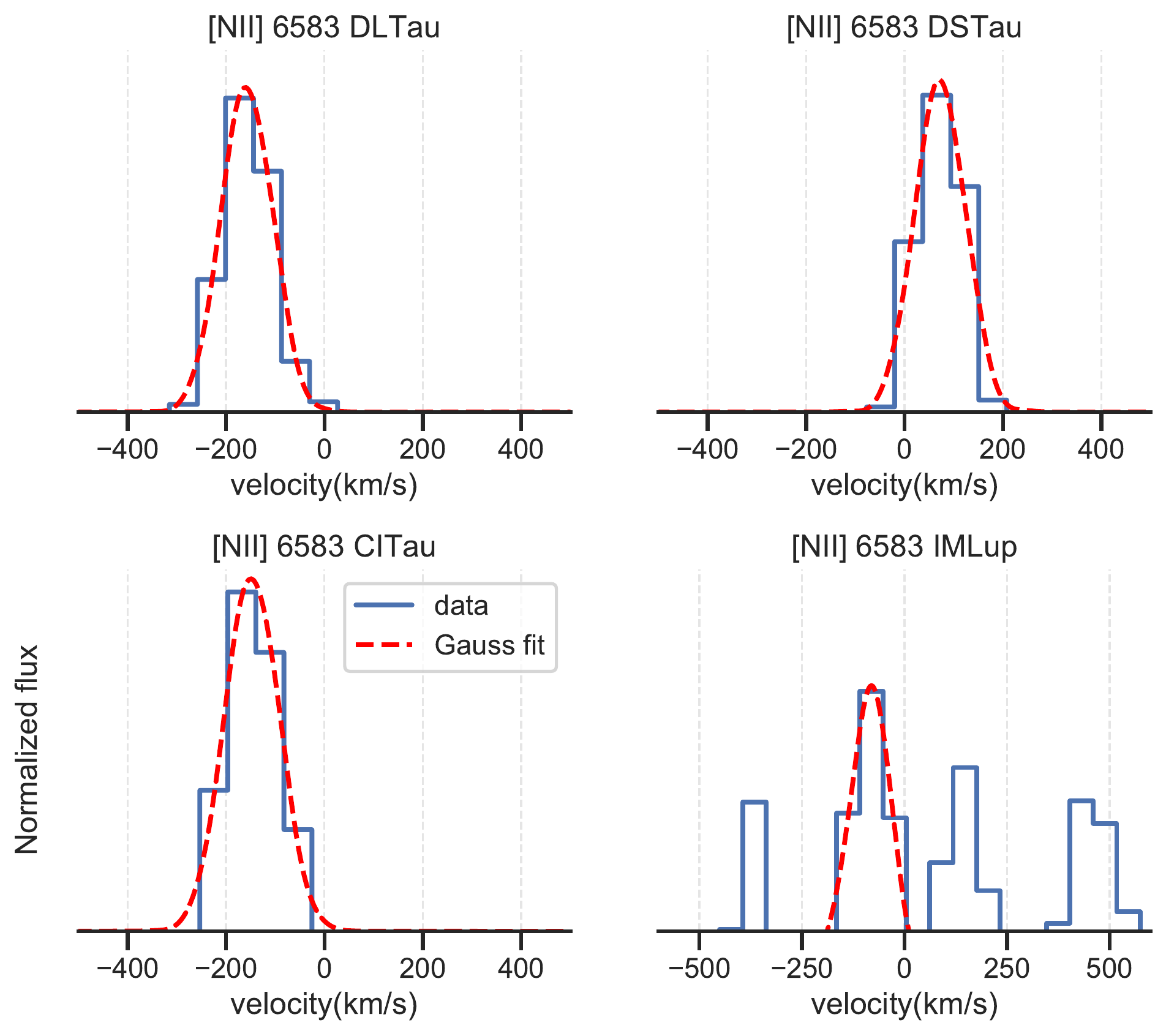} 
    \caption{[NII]$\lambda$6583 line profiles  identified in four sources from our sample. Same case as in Fig.\ref{line_profile_OI6300}, DL Tau and CI Tau probe high-velocity components, while DS Tau and IM Lup are more centered to zero.}
    \label{line_profile_NII6583}
    \end{figure*}
    
\begin{figure*}[!htbp]
    \centering \includegraphics[width=12cm]{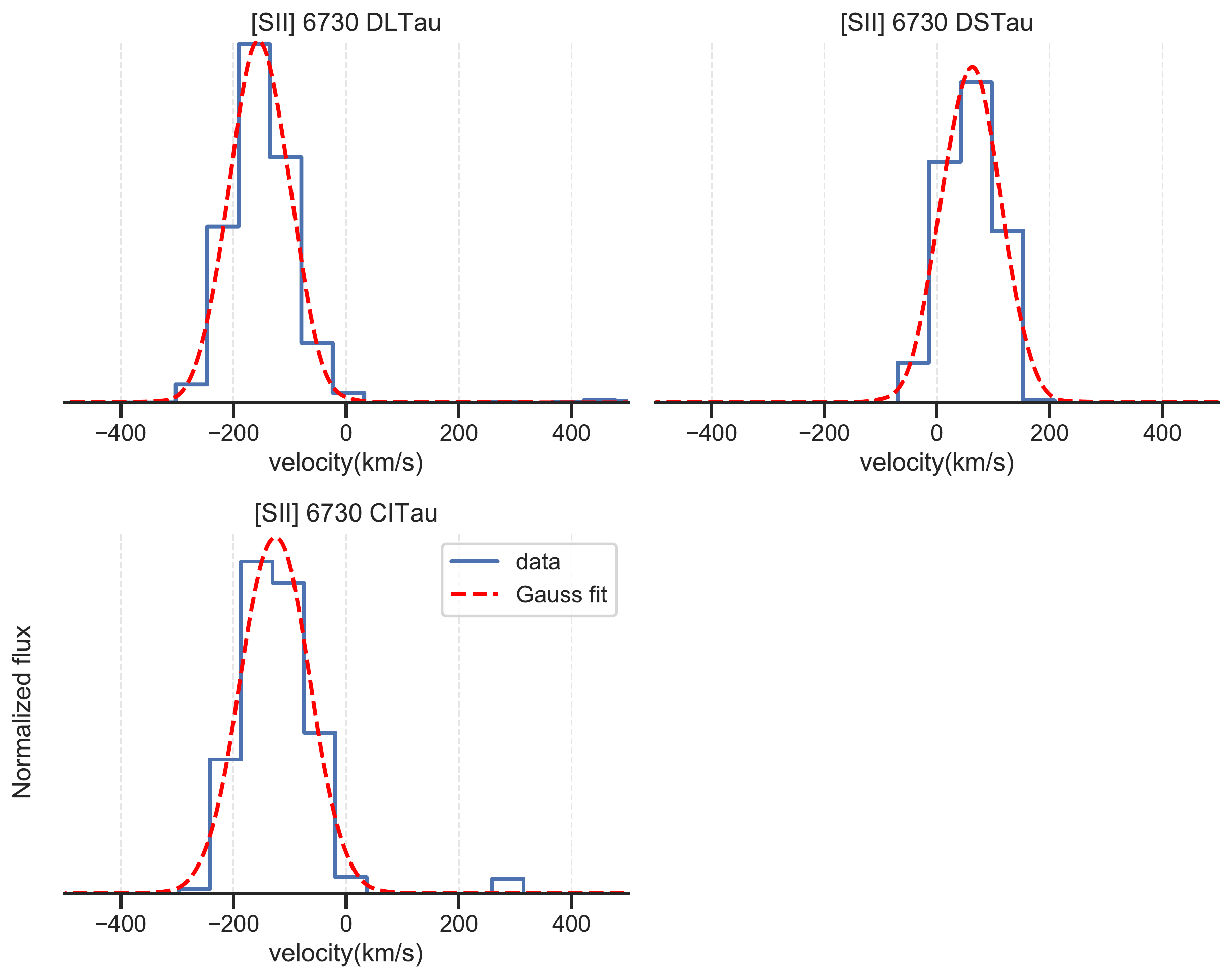} 
    \caption{[SII]$\lambda$6730 line profile identified for DL Tau, CI Tau, and DS Tau.}
    \label{line_profile_SII6730}
    \end{figure*}

    \subsection{Outflow/jet velocities}
    \label{sec:outflow_jet_velocities}
    
    The emission of these forbidden optical emission lines has been established to trace outflows/jets and winds in T Tauri stars. From their line velocity, we can characterize whether it is a high-velocity or a low-velocity component \citep{Edwards_1987, Hartigan_1995}. For each spectral frame in the data cube, we summed pixel fluxes over the region defined by the dashed yellow rectangles in Fig.\ref{fig:supra} to obtain a spectrum of the emission. To measure the emission line centroids to sub-pixel precision, we take the line peak from a Gaussian fit to the data (see Fig. \ref{line_profile_Halpha}, Fig. \ref{line_profile_OI6300}, Fig. \ref{line_profile_NII6583}, and Fig. \ref{line_profile_SII6730}). In Table \ref{tab:jets_properties}, we report the de-projected outflow/jet velocities estimated from the Gaussian fits as follows:
     $v_{\mathrm{outflow/jet}} = c \left( \left(\lambda_\mathrm{gauss}-\lambda_\mathrm{ref}\right) / \lambda_\mathrm{ref}\right)/ \cos(\mathrm{incl}), $ 
    where $c$ is the speed of light, incl is the inclination of the disk (Table \ref{tab:disk_properties}), $\lambda_\mathrm{gauss}$ is the Gaussian fit centroid, and $\lambda_\mathrm{ref}$ is the rest wavelength of each forbidden line in the air.
    The errors on the de-projected velocities, $v_{\mathrm{outflow/jet}}$, in Table \ref{tab:jets_properties} are propagated from the Gaussian fit and the disk inclination, but do not include the uncertainty in wavelength calibration that is likely between 5 km~s$^{-1}$ up to 50 km~s$^{-1}$ (see \citep{Xie_2020}). Most of the $v_{\mathrm{outflow/jet}}$ that we could estimate from the emission lines are blue-shifted and their values correspond mostly (but are not limited to) to the high-velocity components. We determine that for each source, we only see one side of the outflow/jet and the dust from their protoplanetary disk obscures the receding part of these outflow/jet velocities. Unlike the one-sided blue-shifted sources, we report the red-shifted velocities for DS Tau. 

     The detection of H$\mathrm{\alpha}$ line seems to stand out in the spectra (Fig.\ref{spectrum1},  Fig.\ref{spectrum2}, and Fig.\ref{spectrum3}). However, the H$\mathrm{\alpha}$ emission is seen spread in an area surrounding the line spread function, similar to a high noise intensity that H$\mathrm{\alpha}$ introduces from a high photon count level. This effect is not necessarily seen prominently in DL Tau (Fig. \ref{line_profile_Halpha}). An instrumental artifact causes this spread and the reason behind this effect is unknown. Similar effects have been seen and analyzed by \citet{Xie_2020}, where the H$\alpha$ line-to-continuum ratio varies across the field. These variations are so high that the residuals due to this instrument issue are stronger than those from photon noise. The noise reduction is taken into account when applying the principal component analysis (PCA; \citet{Soummer_2012}) method, but the error scale is so high that quantifying the real emission of H$\mathrm{\alpha}$ over the area is difficult.
     
     The structured spectral line profiles reported here show the intrinsic outflow/jet velocity peaks or the line-of-sight velocity for the lines that we were able to capture in the spectra (Fig. \ref{line_profile_Halpha}, Fig. \ref{line_profile_OI6300}, Fig. \ref{line_profile_NII6583}, and Fig. \ref{line_profile_SII6730}). The velocity values reported here are corrected with the stellar radial velocity that is reported in Table \ref{tab:disk_properties} from both \citet{Banzatti2019} and \citet{Fang2018}. The [\ion{O}{i}]$\lambda$6300, H$\alpha$, [\ion{N}{ii}]$\lambda$6583, and [\ion{S}{ii}]$\lambda$6730 line profiles (Fig. \ref{line_profile_OI6300}, Fig. \ref{line_profile_NII6583}, and Fig. \ref{line_profile_SII6730}) for DL Tau and CI Tau are considerably blue-shifted with line center velocities much greater than 100 km~s$^{-1}$. The deprojection velocities show that the velocity of the outflow/jet, $v_{\mathrm{outflow/jet}}$, for DL Tau and CI Tau are greater or very close to 200 km~s$^{-1}$ tracing strong outflows/jets, as shown in see Table \ref{tab:jets_properties}. High-resolution spectra for DL Tau have shown a single low-velocity and a high-velocity component  \citep{Banzatti2019}. The high-velocity component dominates the MUSE image. In addition, DL Tau hosts the most extended and collimated outflow/jet, reaching approximately 180 AU. For CI Tau, a strong outflow/jet is reported for the first time: its morphology is different from the DL Tau one, as seen in Fig. \ref{DLTau_and_CITau_diagram} (right panel). By taking 200 km~s$^{-1}$ as an approximate velocity of the jets in DL Tau and CI Tau, and a distance of 159 pc and 160 pc, respectively, then the proper motion of their extended outflow/jet is about 0$\farcs$26 per year.
     
     The emission lines of DS Tau, IP Tau, and IM Lup have rather minor line-of-sight velocity shifts of less than 100 km~s$^{-1}$ (except for DS Tau) and broader line widths considerably overlapping 0 km~s$^{-1}$, which could hide some LVC emission. DS Tau is the only source in this sample where the line emission appears red-shifted rather than blue-shifted. This red-shifted emission is consistent with a high disk inclination (see Table \ref{tab:disk_properties}) observed with ALMA, where the outflow/jet appears only from the back side of the disk (see Fig.\ref{fig:DLTau_jet_dust}). The IP Tau disk is a transition disk as seen in mm continuum emission \citep{Long_2018}. The MUSE data for IP Tau shows a blue-shifted component in the [\ion{O}{i}]~$\lambda$6300 line with properties in between the HVC and the LVC, which is in good agreement with what has been observed in high-resolution spectral analysis \citep{Banzatti2019}. Interestingly, this line has been reported to vary over time \citep{2Simon_2016}. A recent work by \citet{Bohn_2022} found no potential misalignment between the inner and outer disk in IP Tau when analyzing the VLT/GRAVITY and ALMA data. IM Lup is known to have a blue-shifted LVC- BC measured from the [\ion{O}{i}]~$\lambda$6300 line \citep{Fang2018}.
    
    \subsection{Are emission lines resolved?}

We have classified whether or not the emission lines we report are resolved, marginally resolved, or unresolved (see Table \ref{tab:spectral_widths}), depending on the resolution limit that MUSE has at the wavelength of a given emission line (Fig. \ref{fig:muse_spectral_resolution}). Resolving the line width implies that the emission line width is larger than the line-broadening of MUSE. The [\ion{S}{ii}]~$\lambda$6716 line, for example, is marginally equal to the line-broadening FWHM of MUSE (115 km~s$^{-1}$) in DL Tau (Fig.\ref{fig:DLTau_widths}); but in CI Tau, it is resolved given that the measured width is approximately 150 km~s$^{-1}$ (see Fig.\ref{fig:CITau_widths}), which is greater than the instrumental resolution. Marginally resolved lines and unresolved lines, in particular, should be considered to upper limits rather than precise estimates of the emission line widths \citep[i.e.,][]{Eriksson_2020}. The H$\mathrm{\alpha}$ line in almost all samples, except in CI Tau, is mostly unresolved because their measured width is less than the line-broadening FWHM of MUSE at H$\alpha$ (119 km~s$^{-1}$). The spectral capability of MUSE has enabled to solve for the velocity components of the forbidden emission lines, leading to a good venue for the exploration of the momentum transport of the disk as the disk losses mass, in which such estimate depends on the line profile and the resolution limit of the instrument.

    \subsection{Mass-loss rate}    
    \label{mass_loss_rate}

     The mass-loss rate considers the velocity and the length from the bright [\ion{O}{i}]$\lambda$6300 line luminosity of the densest bulk of the outflow/jet. The [\ion{O}{i}]$\lambda$6300 line is optically thin and used to trace the total mass in the outflow/jet \citep{Hartigan_1994, Giannini_2015, Nisini_2018, Fang2018}. The mass-loss rate is:

    \begin{equation}
    \label{eq:mass_loss}
    \begin{split}
    \dot{M}_{\mathrm{loss}} = C(T,n_{e}) \left( \frac{V_{\perp}}{100~\mathrm{km~s^{-1}}} \right)
    \left( \frac{l_{\perp}}{100~\mathrm{au}} \right)^{-1} \\
    \left( \frac{L_{\mathrm{6300}}}{L_{\odot}} \right)~ M_{\odot}~yr^{-1}
    \end{split}
    ,\end{equation}
    where $V_{\perp}$ is the outflow/jet velocity that is deprojected from the LoS (see Table \ref{tab:jets_properties}). The length of the outflow/jet is $l_{\perp}=$d$_\mathrm{pc}~\theta,$ where d$_\mathrm{pc}$ is the distance of the source given in Table \ref{tab:disk_properties}, and $\theta$ is the length of the outflow/jet measured in arsec (see Table \ref{tab:jet_lengths} and \ref{tab:mass_loss}). The length of the outflow/jet is measured until the emission flux drops below 10$^{-17}$~erg~s$^{-1}$~cm$^{-2}$~\AA$^{-1}$. $C(T,n_{e})$ is a coefficient of the [\ion{O}{i}]~$\lambda$6300 line transition from the energy state 2$\,\to\,$1, that depends on the gas temperature via thermally excited collisions of electrons. At $T=10,000$ K, we used $C(T,n_{e})$ = 9.0$\times10^{-5}$ for a $n_{e}$ = 5.0$\times10^{4}$~cm$^{-3}$, both values provided by \citet{Fang2018}. The resulting mass-loss rates of the disk-outflow/jet sources are listed in Table.\ref{tab:mass_loss}. These mass-loss rates range (1.1-6.5) $\times$10$^{-7}$-10$^{-8}$~$M_{\odot}~yr^{-1}$, which is in agreement with values reported by \citet{Hartigan_1994, Hartigan_1995, Nisini_2018, Fang2018}. A good advantage of the MUSE data is that the spatial extension of the outflow/jet can be measured. We use the distances of the sources specified in Table \ref{tab:disk_properties} to convert from angular length (arcsecs) to physical length (au). We adopted the same scaling distance as in \citet{Fang2018} at 100 au as a normalization factor of the extension of the outflow/jet. As noted, the mass-loss rate is proportional to the line luminosity of [\ion{O}{i}]~$\lambda$6300, which assumes a low electron number density that is much lower than the critical density ($n_{\mathrm{H}}\sim$10$^{6}$~cm$^{-3}$) and an atomic abundance similar to the standard interstellar medium ($\sim$10$^{-4}$).

    \begin{table}[pht!]
    \caption{Mass-loss rates for the disk-outflow/jet systems.} 
    \label{tab:mass_loss}
    \small 
    \centering 
    \begin{tabular}{lcccccr} 
    \hline \hline
    \toprule
        Name & log $L_{
        \mathrm{6300}}$  & $\theta$   & $l_{\perp}$ & $\dot{M}_{\mathrm{loss}}$ \\
        &  ($L_{\odot}$) & (arcsec)  & (AU) & ($M_{\odot}~yr^{-1}$) \\
   \hline
    DL Tau & -2.20 & 1.1 & 180.5 & 4.6$\times10^{-7}$ \\  
    CI Tau  & -2.99 & 0.7 & 112.2 & 1.1$\times10^{-7}$\\
    DS Tau  & -3.02 &  0.9  & 142.6 & 6.5$\times10^{-8}$ \\ 
    IP Tau  & -3.82 & 0.5 & 64.7 & 1.2$\times10^{-8}$\\  
    IM Lup  & -3.43 & 0.4  & 62.3 & 3.4$\times10^{-8}$ \\  
    \hline 
    \end{tabular}
    \end{table}

\noindent

\subsection{Outflow/jet width in DL Tau}
\label{line_width}

Figure \ref{jet_width} shows the width of six emission lines of the outflow/jet in DL Tau. The intensity profiles are compiled by stacking the Gaussian fits of the intensity slices across the outflow/jet as a function of the distance from the outflow/jet axis. The outflow/jet widths in Fig. \ref{jet_width} represent the average length, namely $\sim$0$\farcs$5, of the extension of the outflow/jet in DL Tau. We converted the width of the outflow/jet) from pixels using the MUSE spatial resolution of 0$\farcs$025 per pixel to angular distance: [\ion{O}{i}]~$\lambda$6300 = 0$\farcs$16, [\ion{O}{i}]~$\lambda$6363 = 0$\farcs$12, [\ion{N}{ii}]~$\lambda$6583 = 0$\farcs$11, [\ion{S}{ii}]~$\lambda$6716 = 0$\farcs$31, [\ion{S}{ii}]~$\lambda$6730 = 0$\farcs$21, and H$\alpha$ = 0$\farcs$17. Each emission line width in au is shown in the legend box. The only emission line that is not included in the [\ion{N}{ii}]~$\lambda$6548 line because it is very faint.

Each line profile traces different layers of the outflow/jet. The [\ion{O}{i}]~$\lambda$6363 line and the [\ion{N}{ii}]~$\lambda$6583 line share very similar narrow widths and, compared to the other lines, they both seem to trace the deepest layer of the outflow (jet). These two lines also have similar FWHM (see Table \ref{tab:jets_properties}), encouraging further studies to see whether [\ion{O}{i}]~$\lambda$6363 and the [\ion{N}{ii}]~$\lambda$6583 could perhaps share the same physical properties of the emitting area. Interestingly, we see similar widths for H$\alpha$ and [\ion{O}{i}]~$\lambda$6300. On the other hand, we found [\ion{S}{ii}]~$\lambda$6716 to be the broadest line, followed by [\ion{S}{ii}]~$\lambda$6730 line. 

The brightest and most detected lines are [\ion{O}{i}]~$\lambda$6300 and [\ion{N}{ii}]~$\lambda$6583 as a result of collisional excitation of electrons in shock fronts. The [\ion{S}{ii}]~$\lambda$6716/[\ion{S}{ii}]~$\lambda$6730 and  [\ion{N}{ii}]~$\lambda$6583/[\ion{O}{i}]~$\lambda$6300 line ratio can help to estimate the physical conditions concurring in these scenarios. Empirical fittings \citep{Profaux_2014, Ellerbroeck_2014} and shock model analysis \citep{Hartigan_1994} suggest that these two line ratios are coming from regions with electron densities, $n_{e}$, ranging from 10$^{3}$ to $10^{4}$~cm$^{-3}$, ionization fractions ranging from 0.2 to 10, shock velocities ranging from 60 to 80~km~s$^{-1}$, and magnetic field strengths ranging from 10 to $10^{4}~\mu$G. Studies of other line ratios also provide useful information on the physical conditions. Previous analysis from the [\ion{O}{i}] at $\lambda$5577.3 and $\lambda$6300.3 line ratio in the LVC emitting region have determined that these are emitted in rather dense gas ($n_{e}\geq10^{7}$ cm$^{-3}$), where the gas density relative to hydrogen, $n_{H}$, is $\geq10^{10}$ cm$^{-3}$. In temperatures between 5000 and 10$^{4}$ K, these estimations are based on models where the excitation energies are due to collisions of electrons in regions dominated by neutral Hydrogen \citep{Natta_2014, 2Simon_2016, Fang2018}. \citet{Lavalley_2000} mention that from the [\ion{N}{ii}]~$\lambda$6583/[\ion{O}{i}]~$\lambda$6300 the line ratio diagnostic, the ratio increases with ionization fraction, in regions where $n_{e}\geq n_{H}$. \citet{Lavalley_2000} also stated that the [\ion{S}{ii}]~$\lambda$6716 and [\ion{S}{ii}]~$\lambda$6730 line ratio is a well-known decreasing function of the electronic density until $n_{e}\geq n_{cr} \sim 10^{4}$ cm$^{-3}$, the critical density for collision with electrons of [\ion{S}{ii}]. We encourage further analysis of the line ratio of these forbidden emission lines to the gas temperature, gas density, electron density, and ionization fraction that is characteristic depending on the region of the emitting area.

    \begin{figure}[!htbp]
    \centering
    \includegraphics[width=9cm]{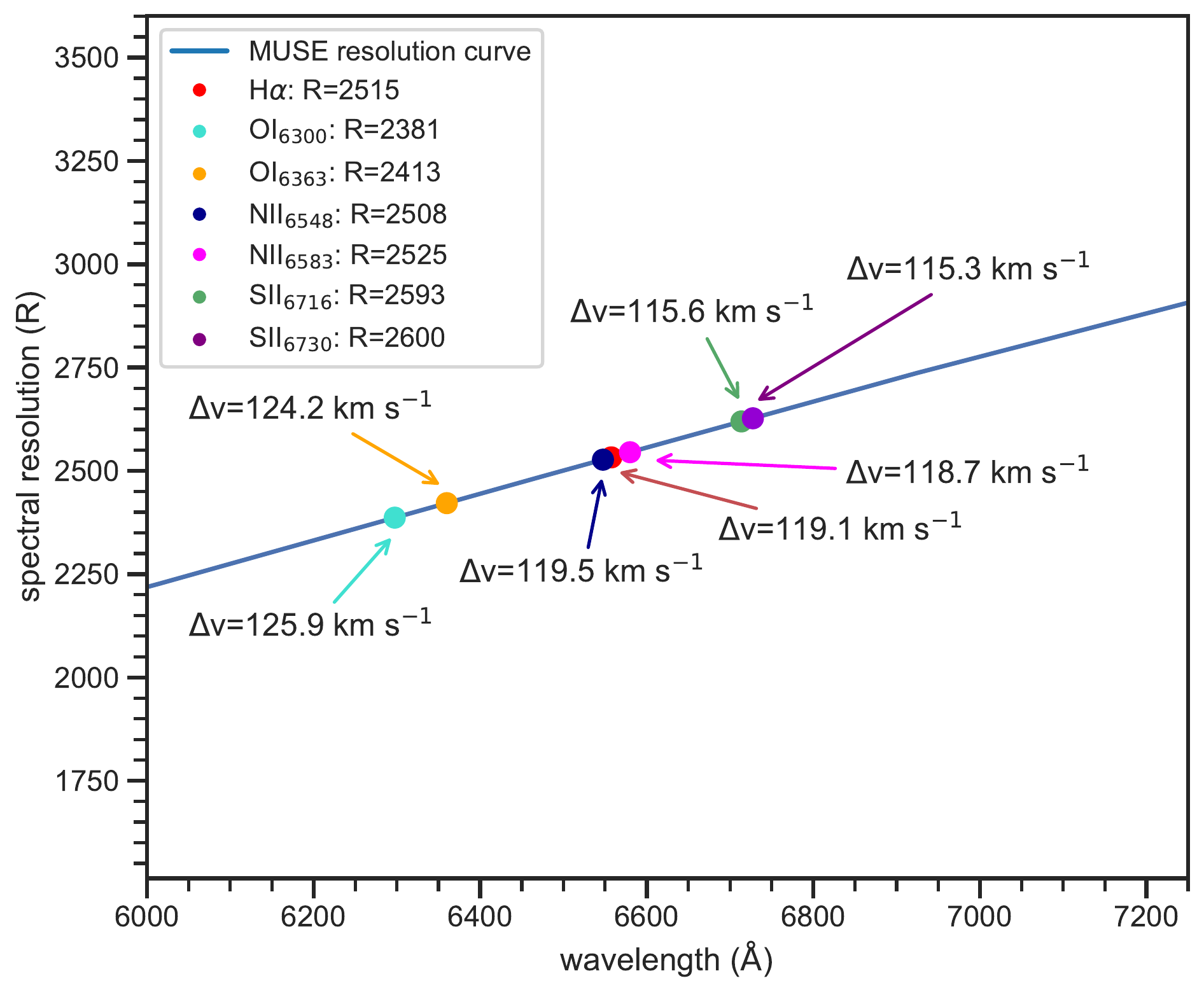} 
    \caption{Spectral resolution, R=$\lambda/\Delta \lambda$, versus wavelength of the seven forbidden emission lines used in the analysis here. The blue curve shows the resolution curve of MUSE at increments of 500~\AA. Each R value for each line is linearly interpolated from the MUSE resolution curve. The values were obtained from the MUSE User Manual (version 11.4, Fig.18).}
    \label{fig:muse_spectral_resolution}
    \end{figure}

\section{Discussion}
\label{sect_4}
The forbidden emission lines analyzed here originate from the center of the protoplanetary disks. Their shape indicates the presence of outflow/jet systems tracing the ongoing process of mass accretion to the star and mass extraction from the inner disk. A full analysis of the mass accretion from the lines is out of the scope of the current work. However, we are motivated to continue this analysis and compare the kinematics with the relative flux between HVC and LVC, with high-resolution spectra from \citet{Banzatti2019}. 

From the eight T Tauri systems, the averaged position angle of the outflow/jet, $\overline{\mathrm{PA}}_\mathrm{outflow/jet}$, for DL Tau, DS Tau, CI Tau, IP Tau, and IM Lup (see Table \ref{tab:disk_properties}) show no tendency of misalignment between the inner disk and the outer disk. The outflows/jets with low signal-to-noise levels and compact extensions close to the image center are challenging to the Gaussian fit method. Due to these effects, the orthogonal profiles of the outflow/jet can take non-Gaussian shapes which, combined with the variable rms in the image center, produce a systematic underestimation of the PA$_{\mathrm{outflow/jet}}$ uncertainties. For IM Lup, determining a PA$_{\mathrm{outflow/jet}}$ from the [\ion{O}{i}]~$\lambda$6300 and [\ion{N}{ii}]~$\lambda$6583 lines was challenging due to the systematic issues mentioned resulting in a $\Delta$PA$_{\mathrm{outflow/jet}}=6^{\circ}$ between the two emission lines. However, the PA$_{\mathrm{outflow/jet}}$ values from  [\ion{O}{i}]~$\lambda$6300 and from [\ion{N}{ii}]~$\lambda$6583 are tracing the same physical outflow/jet axis. The PA average of both emission lines for IM Lup are within the uncertainty range, therefore, no misalignment is detected between the inner and outer disk. However, we encourage follow-up observations with higher signal-to-noise to improve the PA$_{\mathrm{outflow/jet}}$ estimates for IM Lup. If there was a misalignment between the inner and outer disk, it would be an order of magnitude smaller than the reported values on other sources: $72^{\circ}$ for HD 100453 \citep{Benisty_2017}, $30^{\circ}$ for HD14306  \citep{Benisty_2018}, $30^{\circ}$ for DoAR 44 \citep{Casassus_2018}, and $70^{\circ}$ for HD 142527 \citep{Marino_2015}. One piece of observational evidence supporting the determination of no misalignment in IM Lup is the lack of shadows in scattered light \citep{Avenhaus_2018}. For IP Tau, our result of no misalignment agrees with what has been found by \citet{Bohn_2022} when performing a parametric model fit to the squared visibilities. The estimation of the PA$_{\mathrm{outflow/jet}}$ for different emission lines is a reliable way to illustrate the orientation of the inner disk and compare it to the PA$_{\mathrm{dust}}$. The two most detected lines in our sample are [\ion{O}{i}]~$\lambda$6300 and [\ion{N}{ii}]~$\lambda$6583 (see Table \ref{tab:jets_properties}). An interesting future question would be how these emission lines observed in the inner parts of the disk are aligned to the stellar rotation axis, finding that spectro-polarimetric observations could help provide an answer \citep[i.e.,][]{Donati_2019}. Our understanding of the nature of the emission lines and the stellar properties dependence is limited, as it is unclear what the strength of the magnetic field is in the inner disk of T Tauri sources.

\begin{figure} [htp!]
    \centering
    \includegraphics[width=9cm]{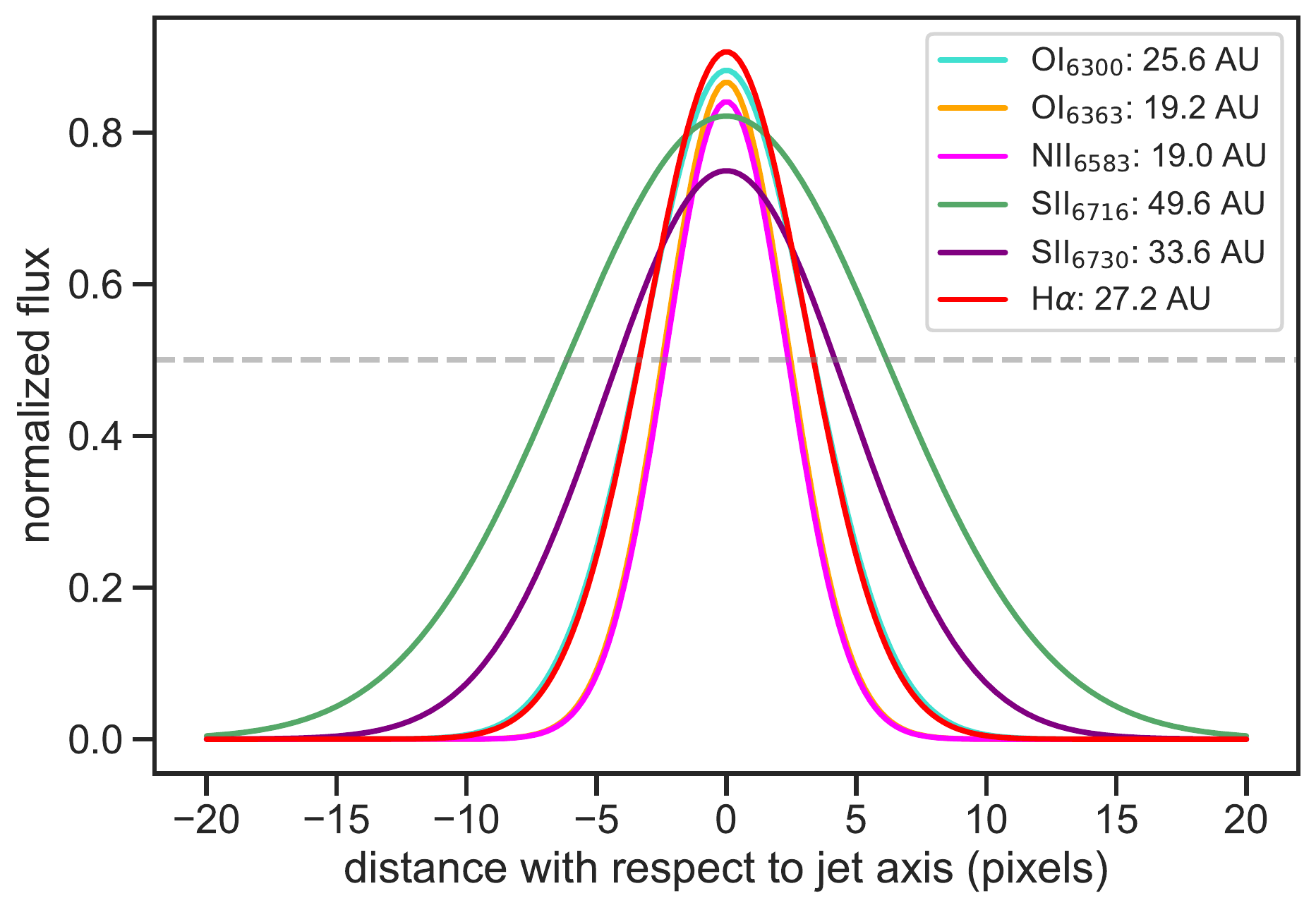}
    \caption{Line-width profile of the emission lines  analyzed in DL Tau. Each line profile is the result of a compilation of the Gaussian fits to intensity slices across the outflow/jet and their maximum intensity peaks are centered at the jet axis. The width of the outflow/jet, in au, is shown in the legend box.}
    \label{jet_width}
    \end{figure}

The line profiles help characterize the velocity components that describe the outflow/jet emission. We confirm a strong blue-shifted outflow (jet) in DL Tau and CI Tau with a $v_{\mathrm{outflow/jet}}$ greater or approximately 200~km~s$^{-1}$ reported from their emission lines. DS Tau is also a source showing HVC, but because of its high inclination (i=65$^{\circ}$) close to the edge-on orientation disk, an LVC can be embedded into the HVC \citep[see also ][]{Banzatti2019}. The line profile of IP Tau and IM Lup show low HVC and are also seen to encompass LVC parts as their profile also approaches the 0 km~s$^{-1}$. Although we are limited by the resolution of MUSE to analyze the LVC, high-resolution studies have confirmed that DL Tau, DS Tau, and IM Lup show the presence of disk winds when analyzing the [\ion{O}{i}]~$\lambda$6300 line \citep{Banzatti2019}. 

Furthermore, the emission line profile of the maximum intensity peaks for DL Tau probe different layers in the outflow/jet suggesting that the temperature, the gas density, the electron density, and the ionization fraction is different depending on the proximity to the jet center. The H$\mathrm{\alpha}$ emission in all disks is seen throughout the disk surface, except in DL Tau. Although it has been assumed to be a systematic issue (i.e., imperfections on the PSF part of the H$\alpha$), at the same time, we do not discard the possibility of a highly ionized atmosphere above the surface of the disk.
Despite the low signal-to-noise level in IP Tau, we found some emission from both, an extremely weak signal in the red-shifted component and the blue-shifted one; with the characteristic of a wide inner gap in the dust continuum emission. We stress that the H$\mathrm{\alpha}$ emission should be interpreted with caution before drawing a definitive conclusion that these are real velocity shifts. 

The mass-loss rate calculated for the outflows/jets are on the order of 10$^{-7}$-10$^{-8}$~$M_{\odot}~yr^{-1}$ (see Table \ref{tab:mass_loss}) -- a result that has also been found in previous works \citep[i.e.,][]{Hartigan_1995, Nisini_2018,Fang2018}. In the outer parts, where the emission of the outflow/jet is weaker, the length could be better defined for longer time integration that would lead to a higher sensitivity. As a result, deeper observations could potentially detect longer outflow/jet lengths. The mass-loss rate formula is also used, assuming that the peak of the emission line happens in the shock front, where the velocity of the shock is used instead of the intrinsic velocity of the outflow/jet. \citet{Hartigan_1994} showed different mass-loss rate calculations in both the postshock and the shock front and stressed that these values should not differ greatly from those based on electron densities and ionization fractions. The [\ion{O}{i}]~$\lambda$6300 line is primarily seen in most sources, probing gas slightly closer to the shock front than [\ion{S}{ii}]~$\lambda$6730, where the critical density is about 100 times lower \citep{Fang2018}. Determining the true angular momentum transport due to the outflow mass loss is very difficult as it requires the knowledge of the full 3D velocity vector and mass content in the outflow. The next step will be to improve the modeling of the velocity, density, and temperature structure of the outflows to generate synthetic line observations and compare them with actual observations. With the help of Fig. \ref{jet_width}, further analysis must be considered to compare the environment where these lines are emitting from. Physical parameters such as ionization fraction, electron density, gas temperature, and the rotational velocity are key to modeling the production of winds in disks.

\subsection{Jet wiggles as a potential sign of precession}
\label{jet_precession}

The outflows/jets show different morphologies. When plotting the Gaussian centers in Fig. \ref{fig:joydivision}, it seems that in jets that are not too collimated, the amplitude of the wiggles is higher and asymmetric. As the wiggles do not show any clear periodic pattern, these are not associated with any inner disk misalignment. Instead, these are distinctive to the jet launching and evolution mechanism. Investigating whether there are any photometric variations in the inner disk of these sources could help us better understand whether there might be a body orbiting close to the central star \citep{Herbst_2001, Cody_2018}.
Fully 3D simulations of a MHD jet launching in young binaries, considering tidal forces \citep{2015ApJ...814..113S,2018ApJ...861...11S}, have found that the inner disks will be warped and that the jet axis, which remains perpendicular to the disk, is subsequently re-aligned as a consequence of the disk precession.

The angle of the jet precession cone derived is slight and about $8\degr$, but it could be much less -- that is, perhaps approaching the differences between position angles identified here for T Tauri systems. However, their simulation was run only for one binary orbit and thus could not follow a whole precession period. Similar simulations, run in hydrodynamics but with forced precession of the jet nozzle, were applied for the precession jet source SS\,433 \citep{2014A&A...561A..30M}, for example. These simulations show that the expected sinusoidal pattern of jet propagation varies along the jet, with larger amplitude and wavelength for more considerable distances, similar to our data indicating a jet cone opening up with distance. Furthermore, a more complex model fitting may be necessary to derive a conclusive statement about precession.

As we do not know the magnetic field structure and strength, defining a statement about the viability of kink modes in jets is not feasible. As an alternative to a precession of the jet axis, we might assume that the jet kink instability could cause the wiggling jet structure. Depending on the jet magnetization, this instability limits the expansion of the jet \citep{2008A&A...492..621M}. 
   
\section{Conclusion}
\label{sect:sect_5}

We analyzed spatially resolved emission lines, [\ion{O}{i}]~$\lambda\lambda$6300, 6363, [\ion{N}{ii}]~$\lambda\lambda$6548, 6583, H$\mathrm{\alpha}$, and [\ion{S}{ii}]\,$\lambda\lambda$6716, 6730,  in five T Tauris: DL Tau, CI Tau, DS Tau, IP Tau, and IM Lup hosting outflows/jets. We conducted an analysis to estimate the PA$_{\mathrm{outflow/jet}}$ of the emission lines coming from the innermost region of the disk. The average of the position angles for different emission lines was used to compare with the PA$_{\mathrm{dust}}$ from the outer disk obtained in a previous work. We also performed a simple kinematic analysis to describe the velocity components of the line profiles and the line width of the outflow/jet in DL Tau. Finally, we calculated the mass-loss rate of the outflow/jet by using the [\ion{O}{i}]~$\lambda$6300 line based on physical parameters derived in this work. Our main conclusions are summarized as follows:

\begin{enumerate}
    \item  The PA$_{\mathrm{outflow/jet}}$ values are in good agreement, with differences of about 1$^{\circ}$, except for IM Lup that is 2.1$^{\circ}$, with the previously determined PA$_{\mathrm{dust}}$. Therefore, we do not find any evidence for a potential misalignment of the inner disk with regard to the outer disk.
    \item The DL Tau and CI Tau emission lines are strongly blue-shifted, showing a velocity profile greater than 200 km~s$^{-1}$ associated with strong outflows/jets. The IP Tau, DS Tau, and IM Lup emission lines are less shifted and closely probing low-velocity components more associated with outflows and winds. The velocity components exhibit different outflows/jets and wind morphology in their systems. 
    \item The line width of the emission lines in DL Tau probes different layers in the outflow/jet, with the [\ion{O}{i}]~$\lambda6363$ line and the [\ion{N}{ii}]~$\lambda6583$ line probing the deepest layer and the [\ion{S}{ii}]~$\lambda6716$ line and the [\ion{S}{ii}]~$\lambda6730$ line probing the widest layer. The [\ion{O}{i}]~$\lambda6363$ and [\ion{N}{ii}]~$\lambda6583$ lines have very similar widths as well as H$\alpha$ and [\ion{O}{i}]~$\lambda6363$ lines hinting to certain correlation in the ionization fraction and electronic density functions as mentioned in \citet{Lavalley_2000}.
    \item Our estimated values for the mass loss are in agreement as being on the order of (1.1-6.5) $\times$10$^{-7}$-10$^{-8}$~$M_{\odot}~yr^{-1}$, including the measurement of the length of the outflow/jet. This value is comparable to previous works for other sources and instruments.
\end{enumerate}

\begin{acknowledgements}
      We would like to thank the referee for providing useful comments in order to improve the quality and understanding of this work. This work is supported by the \emph{European Research Council (ERC)} project under the European Union's Horizon 2020 research and innovation programme number 757957. Thanks to Paola Pinilla for providing the dust continuum data. Also, thanks to Matthias Samland and Sebastiaan Haffert for useful discussions about the MUSE instrument and data analysis. This paper make use of the ALMA data: ADS/JAO.ALMA\#2016.1.01164.S,ADS/JAO.ALMA\#2016.1.00484.L, ADS/JAO.ALMA\#2015.1.00888.S,ADS/JAO.ALMA\#2017.A.00006.S. ALMA is a partnership of ESO (representing its member states), NSF (USA) and NINS (Japan), together with NRC (Canada), MOST and ASIAA (Taiwan), and KASI (Republic of Korea), in cooperation with the Republic of Chile. The Joint ALMA Observatory is operated by ESO, AUI/NRAO and NAOJ. N.K. acknowledges support provided by the Alexander von Humboldt Foundation in the framework of the Sofja Kovalevskaja Award endowed by the Federal Ministry of Education and Research. Th.H. acknowledges support from the European Research Council under the Horizon 2020 Framework Program via the ERC Advanced Grat Origins 83 24 28. G-DM acknowledges the support of the DFG priority program SPP 1992 ``Exploring the Diversity of Extrasolar Planets'' (KU~2849/7-1 and MA~9185/1-1). G-DM also acknowledges the support from the Swiss National Science Foundation under grant BSSGI0$\_$155816 ``PlanetsInTime''. Parts of this work have been carried out within the framework of the NCCR PlanetS supported by the Swiss National Science Foundation. SK acknowledges funding from UKRI in the form of a Future Leaders Fellowship (grant no. MR/T022868/1). We would like to thank the matplotlib team \citep{Hunter_2007} for the good quality tool in order to better visualize the data. To \citet{Piqueras_2017} for the mpdaf software to analyze the MUSE data.
    
\end{acknowledgements}

\bibliography{muse}
\begin{appendix}

\section{Example of linear model function}
\label{linear_model_function}

\begin{figure} [htp!]
    \centering
    \includegraphics[width=8cm]{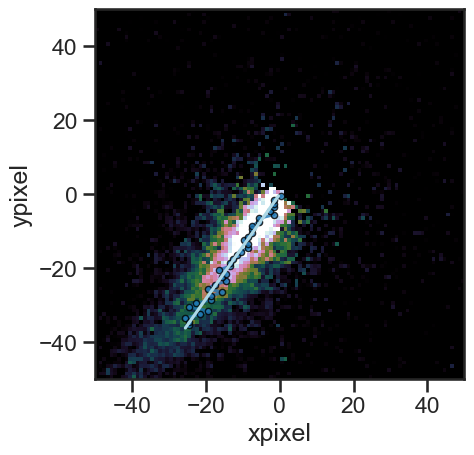} 
    \includegraphics[width=8cm]{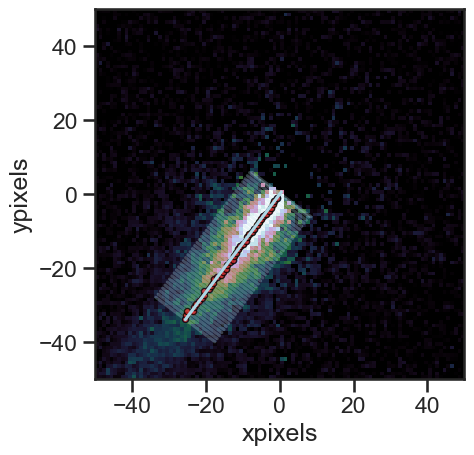} 
    \caption{DL Tau [\ion{O}{i}]~$\lambda$6300 outflow/jet as an example of the linear model function (light blue) based on the Gaussian centers (grey dots, \textit{top}). We fit 100 crossing lines (\textit{bottom}) in order to estimate the width across the outflow/jet.}
    \label{linear_jet_DLTau_diagram}
    \end{figure}

\begin{figure} [htp!]
    \centering
    \includegraphics[width=8cm]{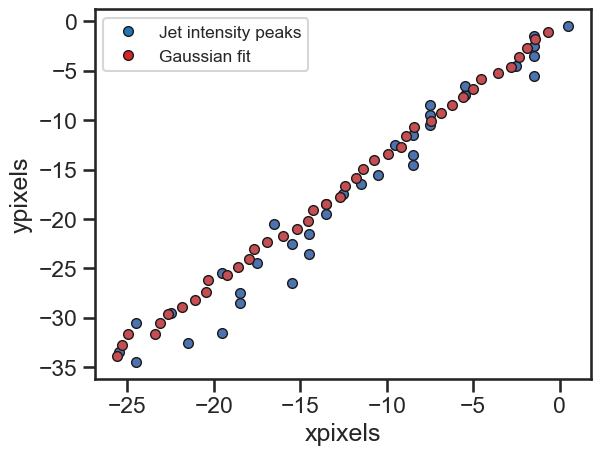} 
    \caption{[\ion{O}{i}]~$\lambda$6300 successive outflow/jet intensity peaks across the jet axis in blue. We overplot the Gaussian centers from the fit in red, which is then used to estimate the PA$_\mathrm{outflow/jet}$. The deviations of the outflow/jet intensity peaks could be caused by noise in the data.}
    \label{fig:linear_jet_DLTau_diagram2}
    \end{figure}

Figure \ref{linear_jet_DLTau_diagram} shows an example of the linear function model for [\ion{O}{i}]~$\lambda$6300 in DL Tau to estimate the PA$_{\mathrm{outflow/jet}}$. As mentioned in Sect.~\ref{geometrical_fitting}, we performed a Gaussian fit to maximum jet intensity peaks locations in each row in order to avoid any bias by noise in the data. These Gaussian centers are orthogonal with respect to the line model function in light blue. We applied the same methodology to the other sources in our sample.
Figure  \ref{fig:linear_jet_DLTau_diagram2} shows the comparison between
the the maximum intensity peaks across the outflow/jet in blue and the
intensity peaks fitted by a Gaussian in red.

    \section{Outflow/jet lengths}
    
    To determine the length of the outflow/jet, the linear model function goes across the outflow/jet until it reaches the rms of the image. The number of elements or pixels representing the length are in Table \ref{tab:jet_lengths}. The starting point of the outflow/jet length comes from the center of the image, where the star is located. We consider these lengths when calculating the mass-loss rate using the [\ion{O}{i}]~$\lambda$6300 in \S\ref{mass_loss_rate}.
    
    \begin{table*}[pht!]
    \caption{Outflow/jet length estimates.} 
    \label{tab:jet_lengths}
    \centering
    \begin{tabular}{lcccccr} 
    \hline \hline
    \addlinespace 
    Source & Line & rms &  Length & Length \\
    &  &  &  (pixels) & (arcsec) & \\
    \addlinespace 
    \hline
    \multirow{7}{*}{DL Tau} & [\ion{O}{i}]~$\lambda$6300 & 11.34 & 43  & 1.07 & \\
                         & [\ion{O}{i}]~$\lambda$6363 & 10.93 & 32  & 1.10 & \\ 
                         & [\ion{N}{ii}]~$\lambda$6548 & 11.47 & 44 & 1.10 & \\ 
                         & [\ion{N}{ii}]~$\lambda$6583 & 12.45 & 65  & 1.63 &  \\ 
                         & H$\mathrm{\alpha}$ & 11.84 & 83 & 2.07 & \\ 
                         & [\ion{S}{ii}]~$\lambda$6716 & 11.0 & 79  & 1.98 &  \\ 
                         & [\ion{S}{ii}]~$\lambda$6730 & 15.22 & 78 & 1.95 &  \\  
                         \midrule \addlinespace 
    \multirow{3}{*}{CI Tau}  &  [\ion{O}{i}]~$\lambda$6300 & 14.0 & 29 & 0.73 & \\
                         & [\ion{N}{ii}]~$\lambda$6583 & 13.76 & 30 & 0.75 &  \\
                         & [\ion{S}{ii}]~$\lambda$6730 & 10.26 & 31 & 0.78 &  \\
                         \midrule \addlinespace 
    \multirow{2}{*}{DS Tau}   & [\ion{O}{i}]~$\lambda$6300 & 24.65 & 36 & 0.90 & \\ 
                         & [\ion{N}{ii}]~$\lambda$6583 & 19.75 & 42 & 1.05 & \\ 
                         \midrule \addlinespace 
    \multirow{1}{*}{IP Tau} & [\ion{O}{i}]~$\lambda$6300 & 15.03 & 20 & 0.50 & \\
                         \midrule \addlinespace 
    \multirow{2}{*}{IM Lup}  &  [\ion{O}{i}]~$\lambda$6300 & 13.17 & 16 & 0.40 & \\ 
                         & [\ion{N}{ii}]~$\lambda$6583 & 15.27 & 16 & 0.40 & \\  
                         \midrule \addlinespace 
    \end{tabular}
    \end{table*}
    
    \section{Disk inclination and rotation}
    
    In Figure \ref{fig:DLTau_jet_dust}, we identify whether these disk-outflow (disk-jet) systems have positive or negative inclination to determine the direction to where the disk is rotating. We look at the rotation of the system based on the $^{12}$CO channel maps. For IM Lup, we use \citet{Pinte_2020}. Also, the inclination of IM Lup can easily be seen by looking at the bright side of the disk in near-infrared image from \citet{Avenhaus_2018}. For CI Tau, we use \citet{Rosotti_2021}. For IP Tau and DS Tau, we use \citet{Simon_2017}. For DL Tau, there is no evidence of the $^{12}$CO channel maps in the literature nor any observations taken and posted in the ALMA archive. By looking at the blue-shifted outflow/jet in the MUSE image, it is intuitive that we are looking at the back side of the disk in DL Tau. For PDS 70, we use \citet{Keppler_2019} and \citet{Isella_2019}.
    
    \begin{figure}[htp!]
    \centering
    \includegraphics[width=8cm]{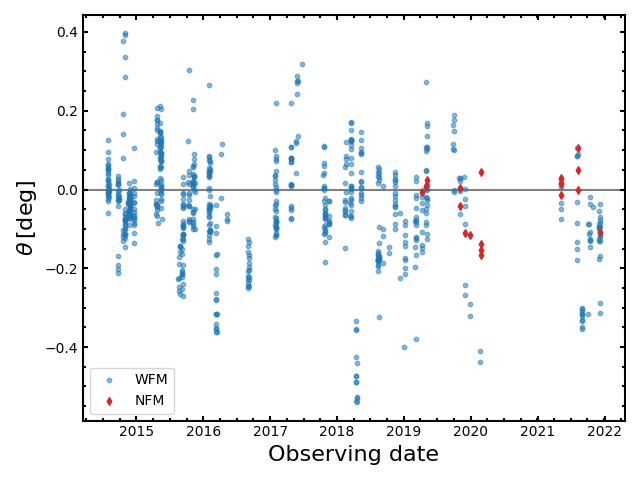}
    \caption{Residual rotation of the WFM observations (blue) taken until late 2021 and the NFM observations (red). Note: the NFM residuals are less than 0.2 degrees.}
    \label{fig:rotation_residual_analysis}
    \end{figure}
    
    \begin{figure*}[htp!]
    \centering
    \includegraphics[width=20cm]{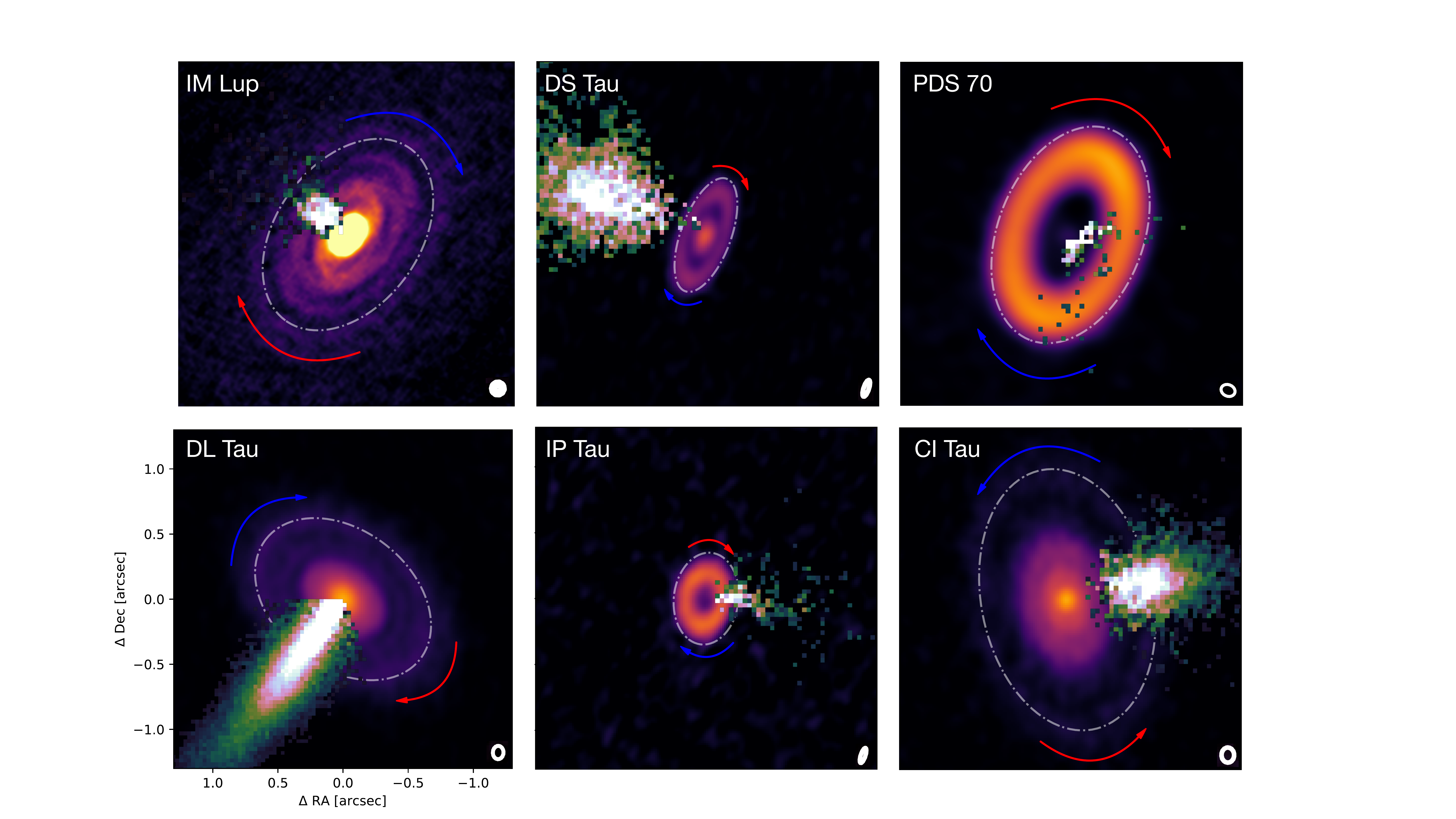} 
    \caption{Composite images of disk-outflow (disk-jet) systems. The outflows/jets are superimposed to the dust continuum disk detected by ALMA Cycle 4 at 1.33 mm \citep{Long_2018}. The arrows, representing the redshifted and blueshifted velocity components, shows the disk rotation. With this determination, IM Lup, DS Tau, PDS 70 (the top three panels) as well as  DL Tau and IP Tau have a negative disk inclination, whereas, CI Tau have a positive inclination. The outflow/jet in CI Tau, IP tau, and DL Tau is comprised of the emission from [\ion{O}{i}]~$\lambda$6300. For IM Lup, and DS Tau are [\ion{N}{ii}]~$\lambda$6583. Lastly, we decided to try out the data set of PDS 70 already published by \citet{Haffert_2019}, as there is some emission at the center seen in H$\mathrm{\alpha}$. The disk inclination is -51.7$\pm$0.1  and the PA$_{\mathrm{dust}}$ is 156.7$\pm$0.1$^{\circ}$ as derived from ALMA Cycle 5 continuum observations at 0.855 mm by \citep[see also][]{Isella_2019,Keppler_2019}. We estimated the PA$_{\mathrm{outflow/jet}}$ for PDS 70 as 154.82$\pm$5.92$^{\circ}$.}
    \label{fig:DLTau_jet_dust}
    \end{figure*}

    \section{Rotational offset analysis in MUSE}
    \label{sec:rotation_residual_analysis}
    
    We carried out a residual rotation analysis for the NFM data of globular clusters in order to obtain the accuracy of the MUSE field-of-view orientation (see \citet{Kamann_2018}). This analysis was done with \textsc{PampelMuse} software \citep{Kamann_2013}, which uses the centers of dense star clusters, which have been pre-imaged with Hubble Space Telescope (HST) -- and then the stars available are identified in the astrometric reference catalogues in the MUSE cubes. This is done using a coordinate transformation with six parameters that takes into account shifts along the x- and y-axes, while other parameters are measure of distortions and rotations. Figure \ref{fig:rotation_residual_analysis} shows the residual rotation for all the wide-field mode (WFM) observations (blue) taken until late 2021 and the NFM observations (red). From the first sight, there is no systematic difference between the results derived from WFM and NFM cubes, which is reassuring and the resulting uncertainty is less than 1 degree for both modes -- and less than 0.2 degrees for the NFM in particular. The same analysis was used by \citet{Emsellem_2022}, where the MUSE field of view orientation is of the order of a few tenths  of a degree. Based on this residual analysis, we consider the rotational accuracy of MUSE NFM to be 0.2 degrees, which is a conservative selection (see column 8 in Table \ref{tab:disk_properties}).  As expected, this uncertainty value does not change the actual uncertainty estimates for our PA$_{\mathrm{outflow/jet}}$ reported in Table \ref{tab:disk_properties}.

    \section{Spectrum}
    
\begin{figure*}[ht]
     \centering
     \begin{subfigure}[a]{0.9\textwidth}
         \centering
         \includegraphics[width=\textwidth]{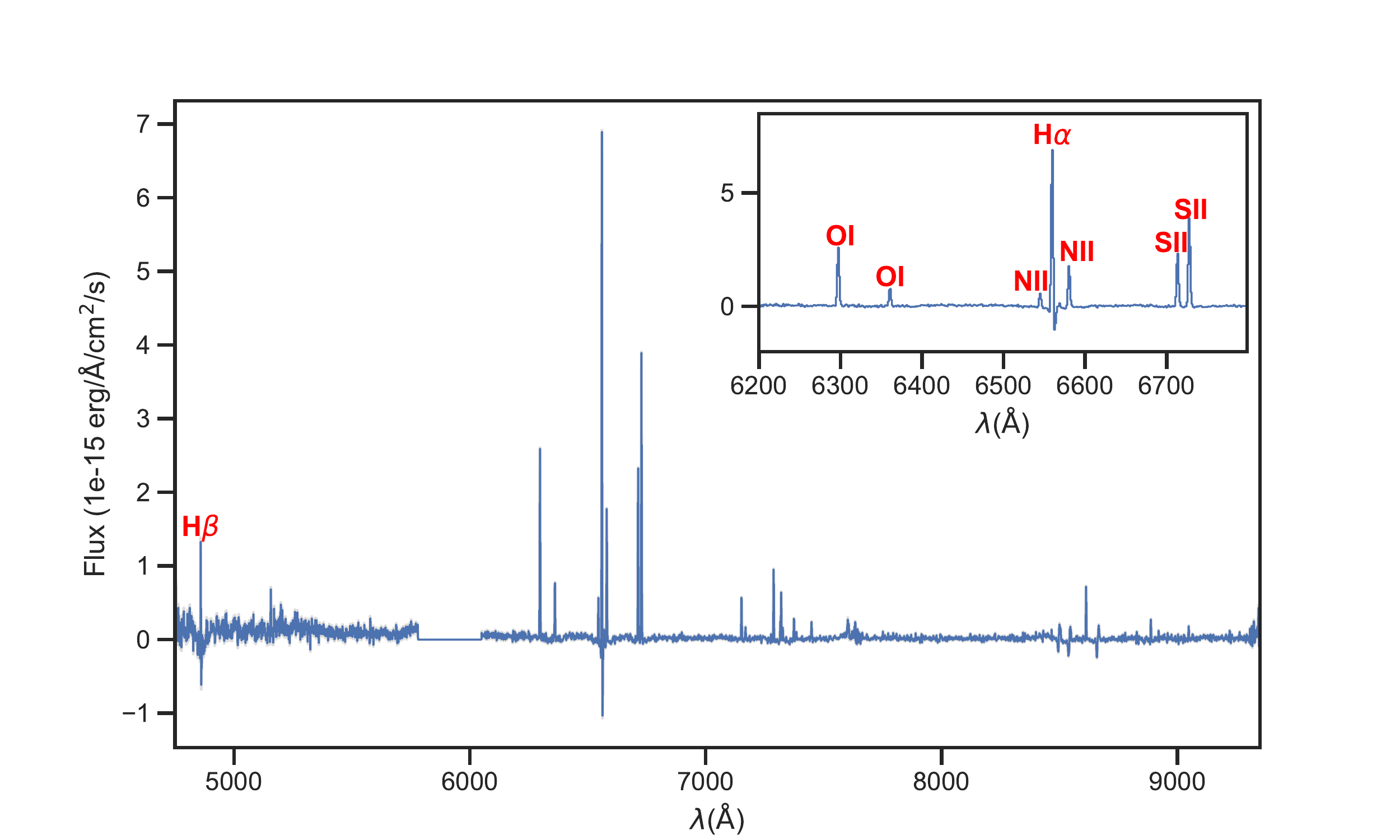}
         \caption{DL Tau}
         \label{fig:dltau}
     \end{subfigure}
     \hfill
     \begin{subfigure}[b]{0.9\textwidth}
         \centering
         \includegraphics[width=\textwidth]{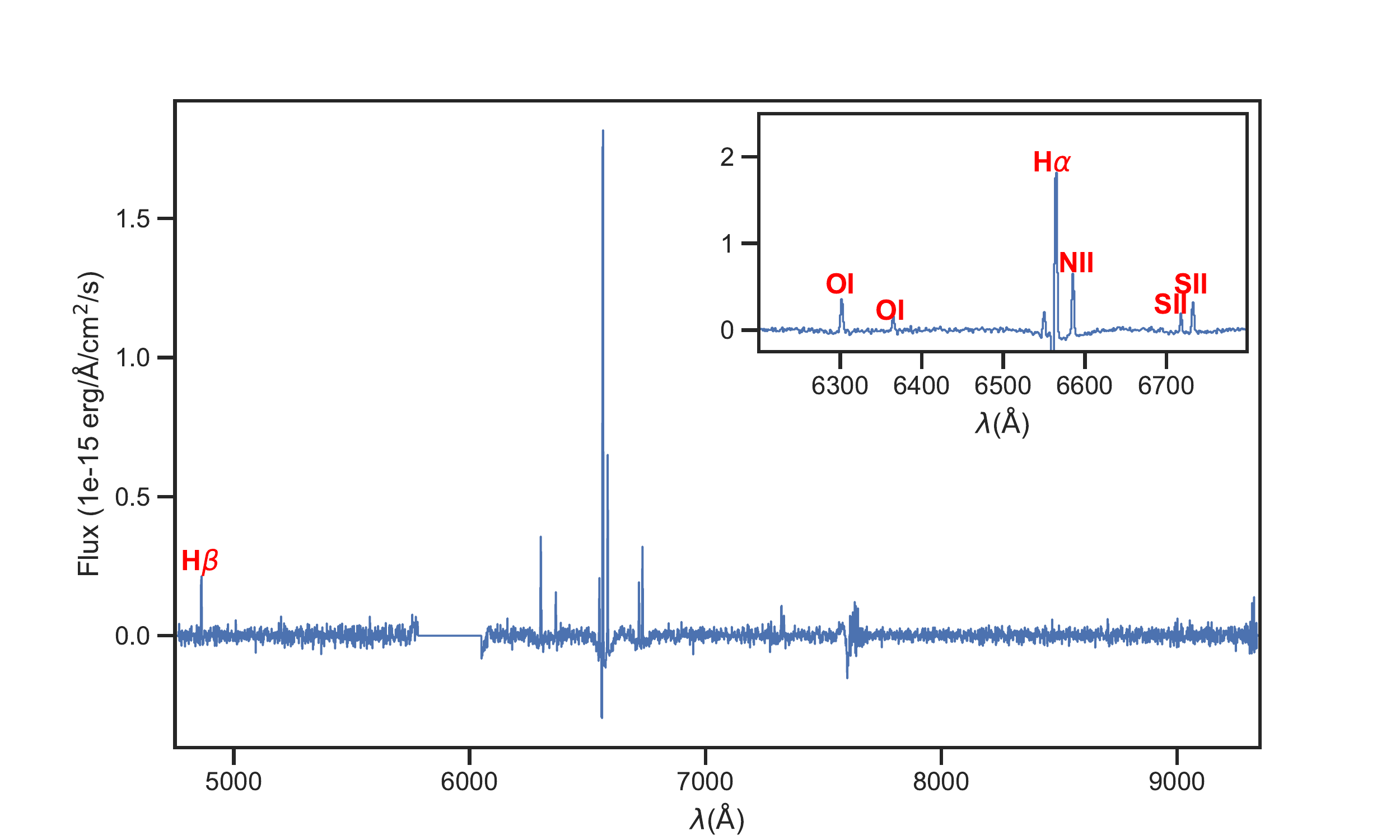}
         \caption{DS Tau}
         \label{fig:dstau}
     \end{subfigure}
     \hfill
     \caption{Spectrum in the area enclosed to the jets of DL Tau and DS Tau.}
     \label{spectrum1}
    \end{figure*}

\begin{figure*}[ht]
     \centering
     \begin{subfigure}[c]{0.9\textwidth}
         \centering
         \includegraphics[width=\textwidth]{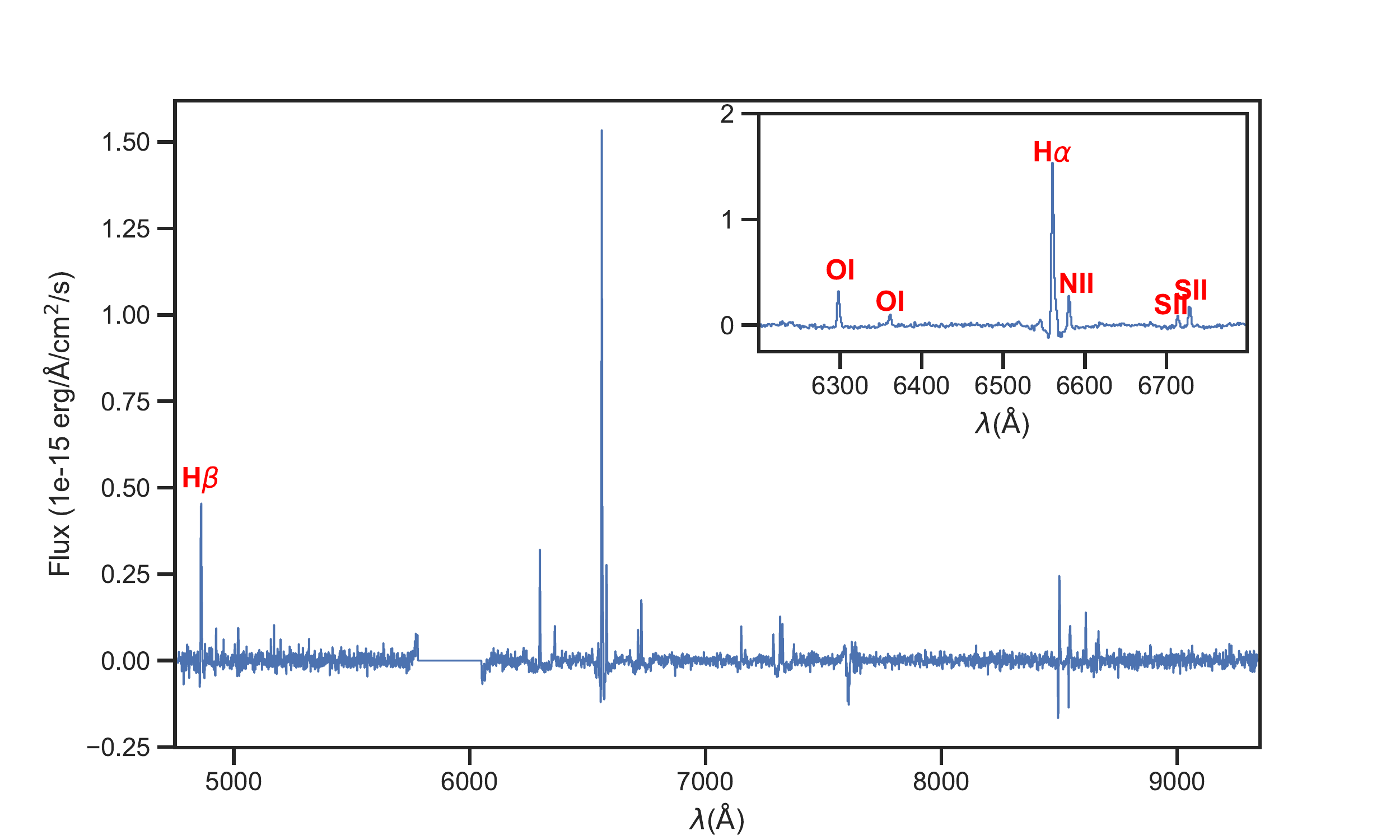}
         \caption{CI Tau}
         \label{fig:citau}
     \end{subfigure}
     \hfill
    \begin{subfigure}[d]{0.9\textwidth}
    \centering
    \includegraphics[width=\textwidth]{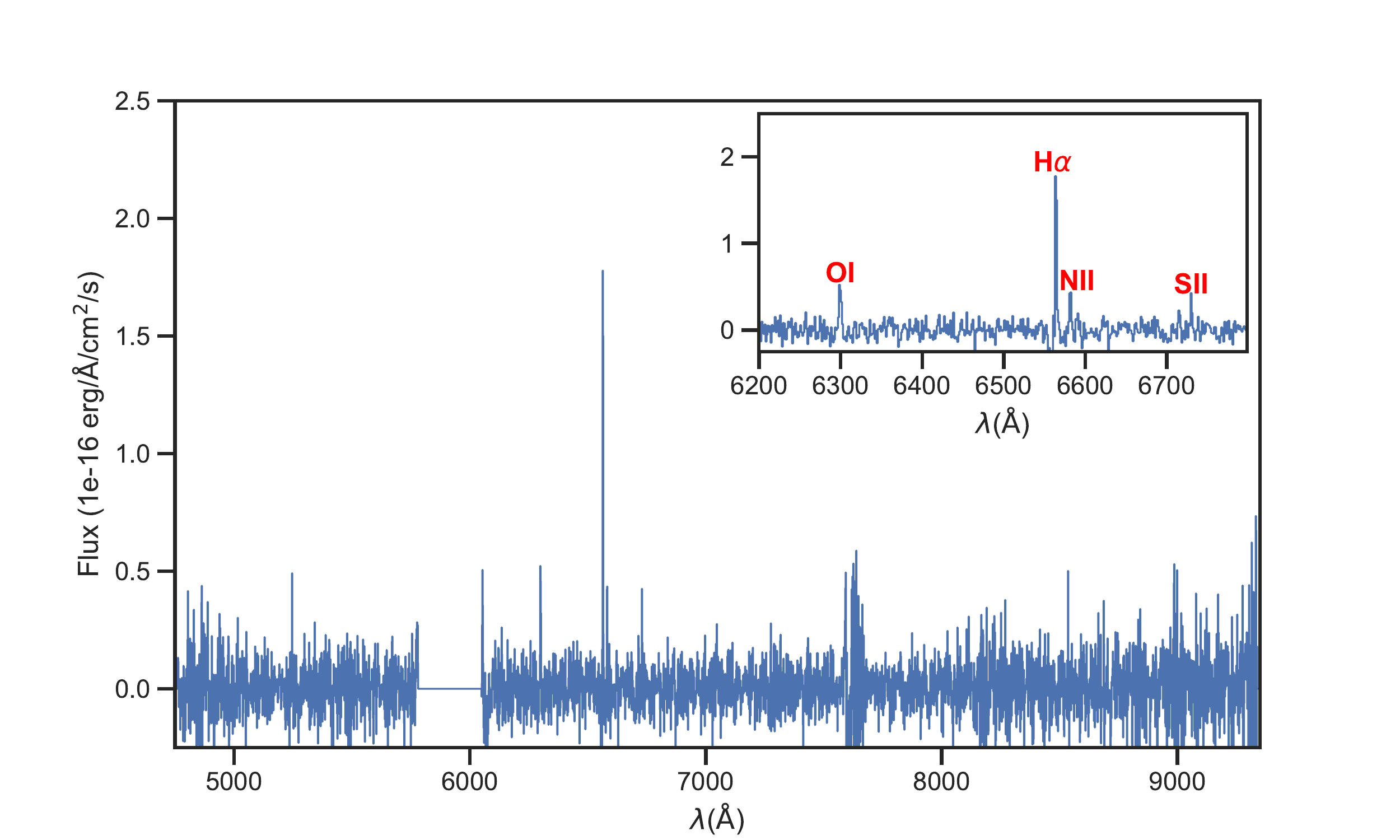} 
    \caption{IP Tau}
    \label{fig:iptau}
    \end{subfigure}
    \hfill
    \caption{Spectrum in the area enclosed to the jets of CI Tau and IP Tau.}
    \label{spectrum2}
    \end{figure*}

\begin{figure*}[ht]
    \centering
    \begin{subfigure}[b]{0.9\textwidth}
    \centering
    \includegraphics[width=\textwidth]{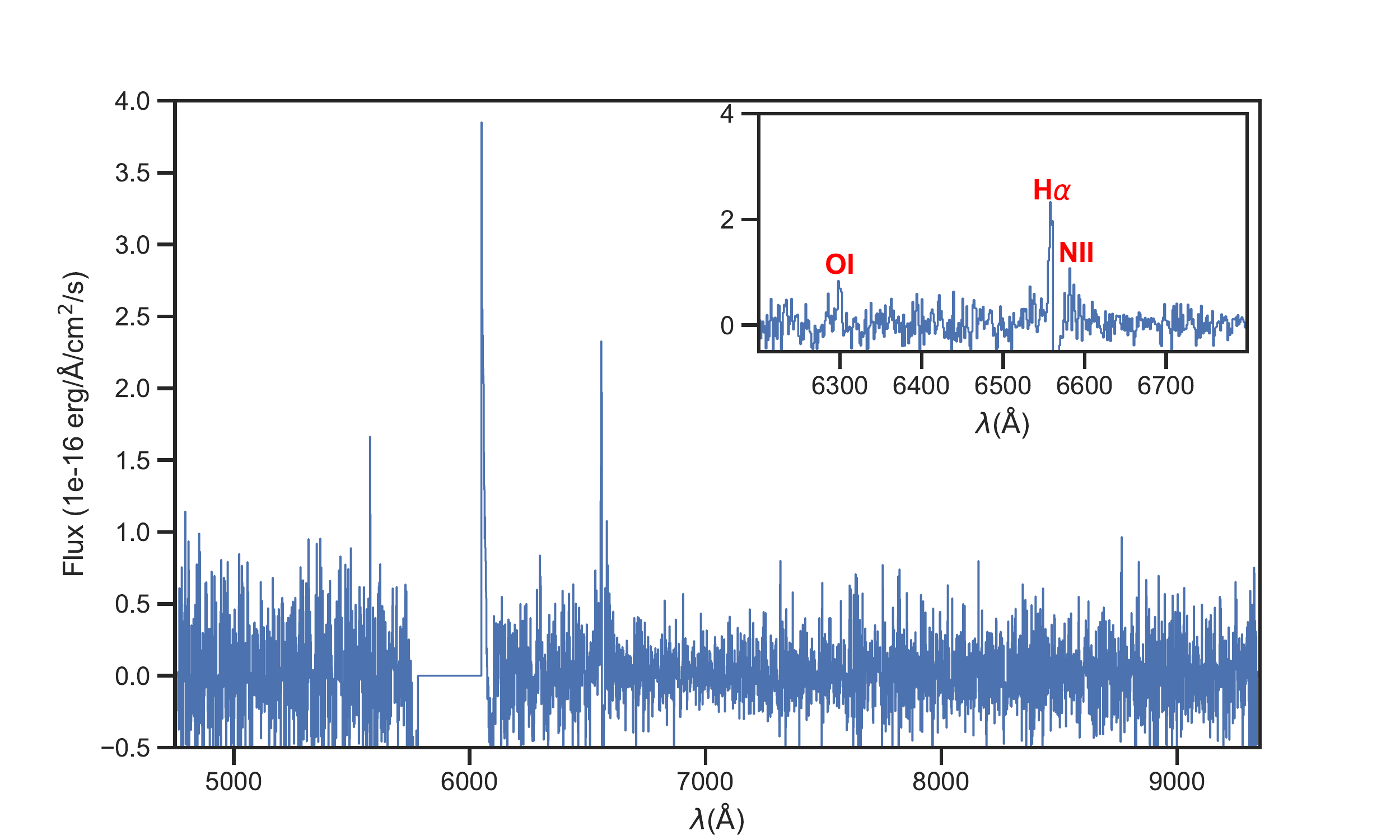}
    \caption{IM Lup}
    \label{fig:imlup}
    \end{subfigure}
    \hfill
    \caption{Spectrum in the area enclosed to the jet of IM Lup.}
    \label{spectrum3}
    \end{figure*}

    Figure \ref{spectrum1}, Figure \ref{spectrum2}, and Figure \ref{spectrum3} show the spectrum by spatially summing the area enclosed to the jet over all channels for all sources presented in this work. In the upper-right, we zoom onto the wavelength range where the seven forbidden emission lines are identified and analyzed here. DL Tau shows the best signal-to-noise emission in all of the seven lines analyzed here. The DL Tau spectrum shows the presence of just a couple of more emission lines in the blue regime, such as H$\beta$ $\lambda$4861 as well as a weak one at $\sim$5156~\AA~(see Fig.~\ref{fig:dltau}).
    We also found very faint emission of H$\beta$ in DS Tau and CI Tau. We report emission lines in the red optical regime as well at around [\ion{Ca}{ii}] $\lambda$7288, [\ion{O}{ii}] $\lambda\lambda$7320,7329, 7373, and [\ion{Fe}{ii}] $\lambda\lambda$ 8619, 8892 for DL Tau and CI Tau, which host strong outflows/jets. We encourage further analysis of these other lines that lie out of the scope of this paper.
    The strong outflow/jet is seen very clear and collimated in all seven emission lines in DL Tau, reaching close to 200 au in extension. As a comparison, the extension of the outflow/jet in CI Tau only reach around 60 au.

    \section{Resolution profile}
    
\begin{figure*}
     \centering
     \begin{subfigure}[b]{0.3\textwidth}
         \centering
         \includegraphics[width=\textwidth]{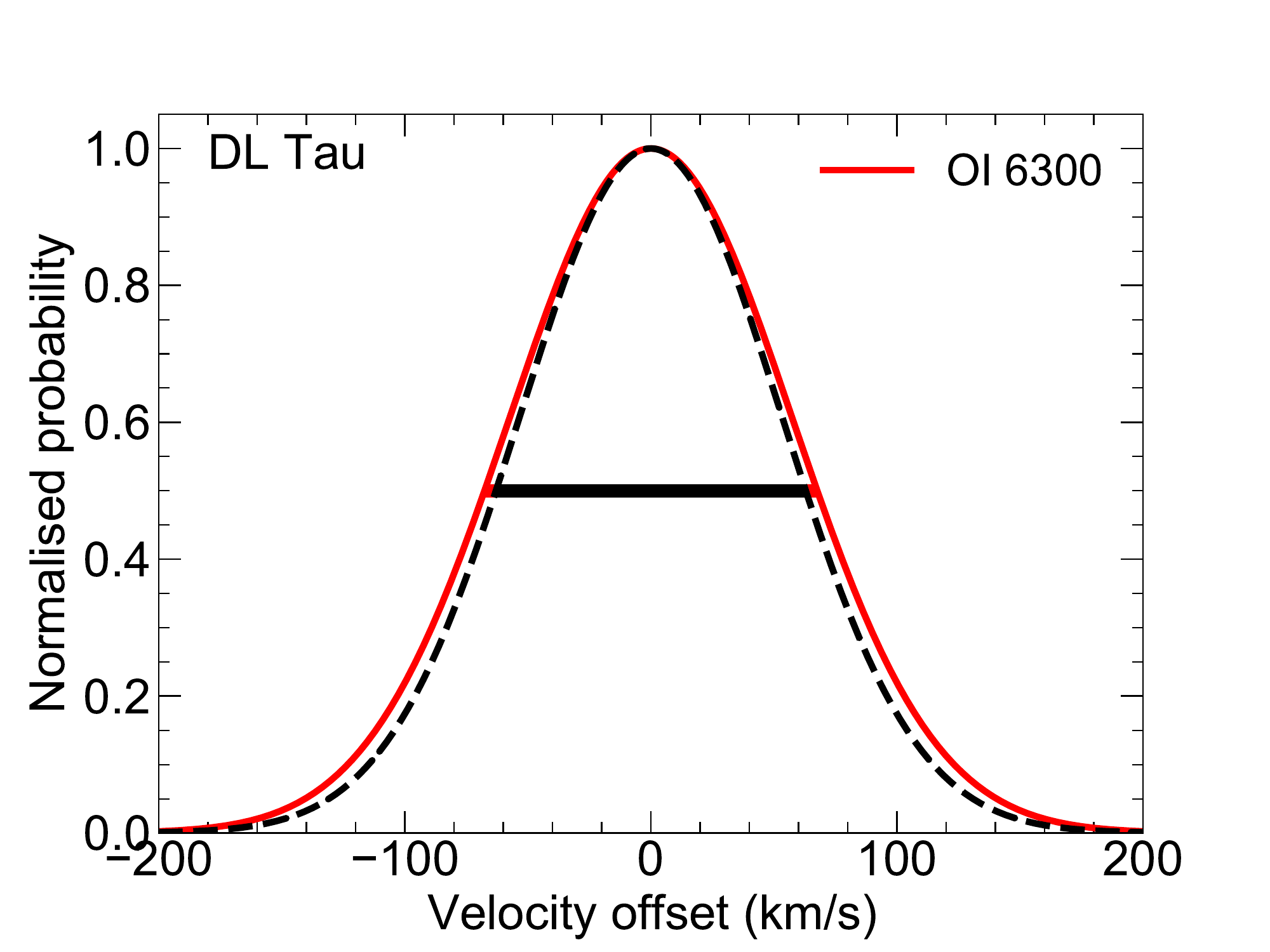}
     \end{subfigure}
     \begin{subfigure}[b]{0.3\textwidth}
         \centering
         \includegraphics[width=\textwidth]{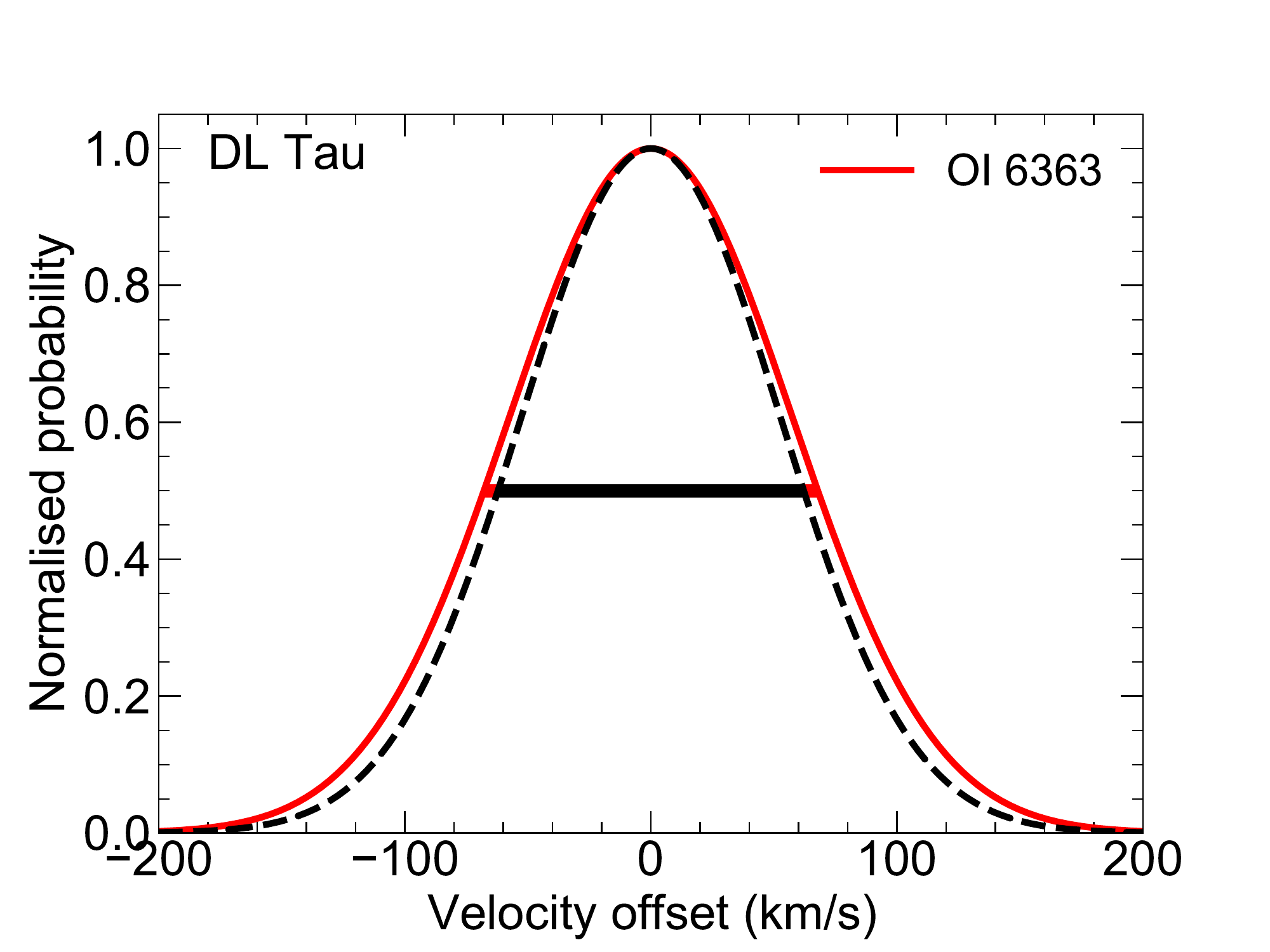}
     \end{subfigure}
     \begin{subfigure}[b]{0.3\textwidth}
         \centering
         \includegraphics[width=\textwidth]{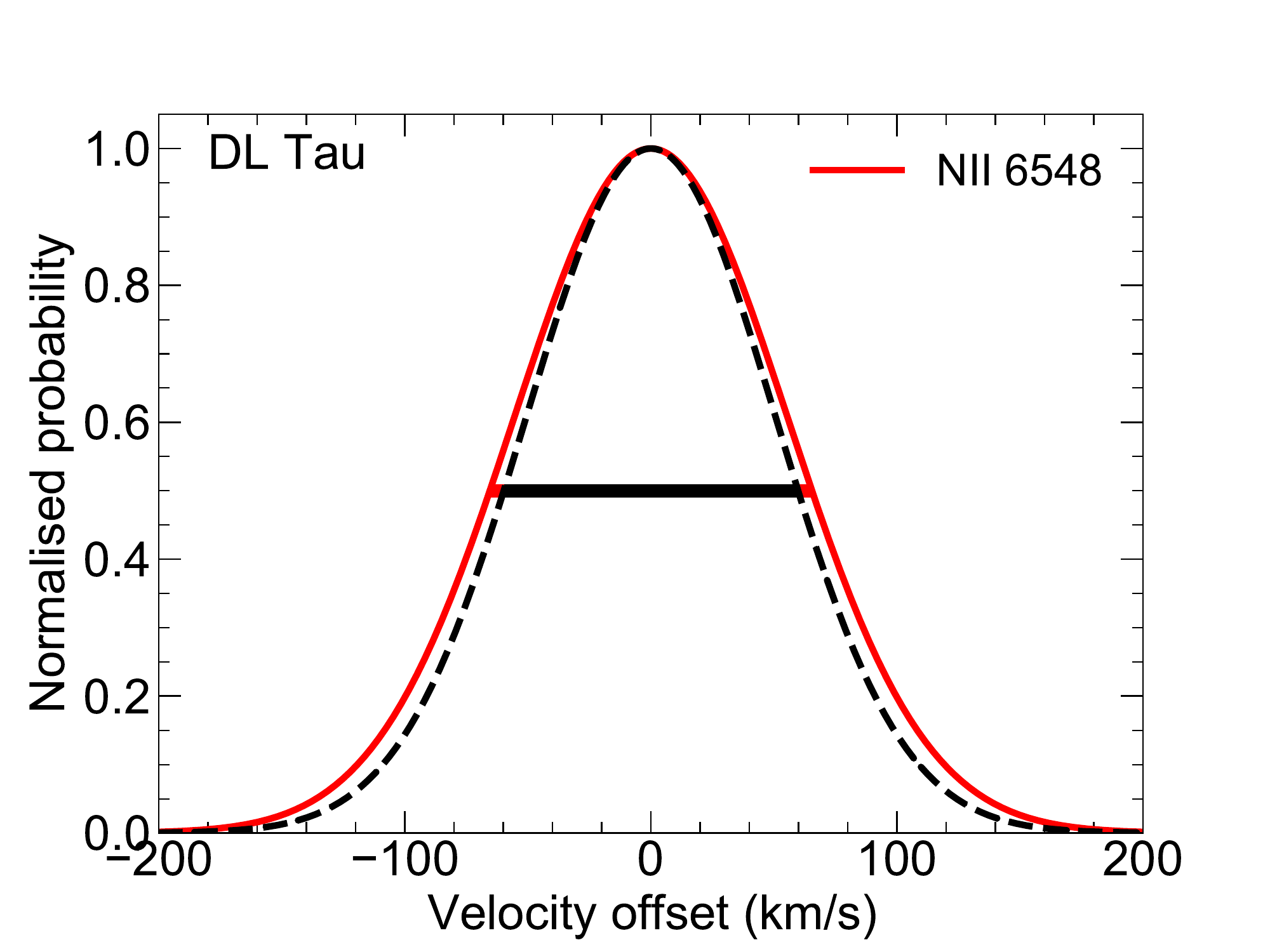}
     \end{subfigure}
     \begin{subfigure}[b]{0.3\textwidth}
         \centering
         \includegraphics[width=\textwidth]{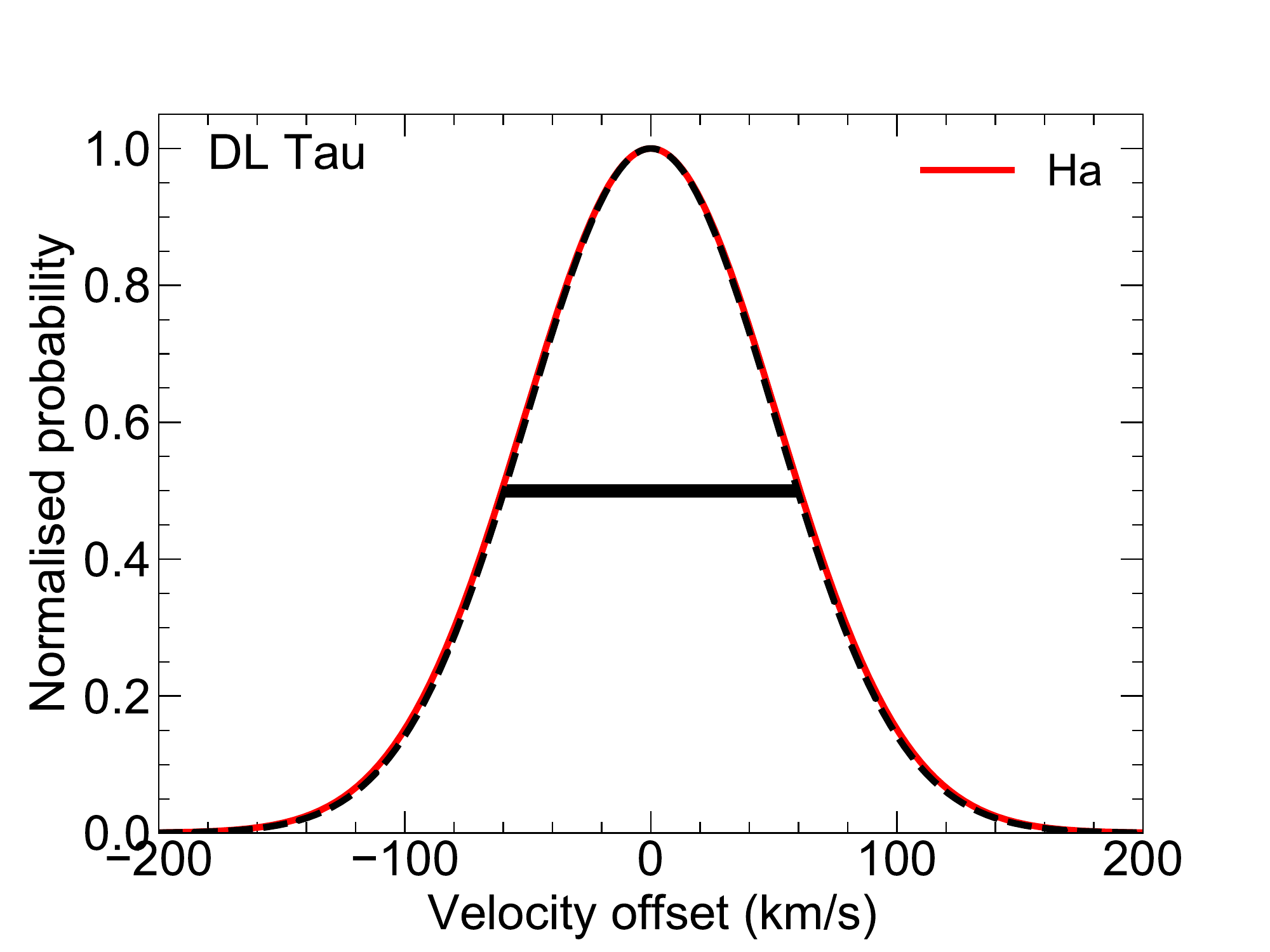}
     \end{subfigure}
     \begin{subfigure}[b]{0.3\textwidth}
         \centering
         \includegraphics[width=\textwidth]{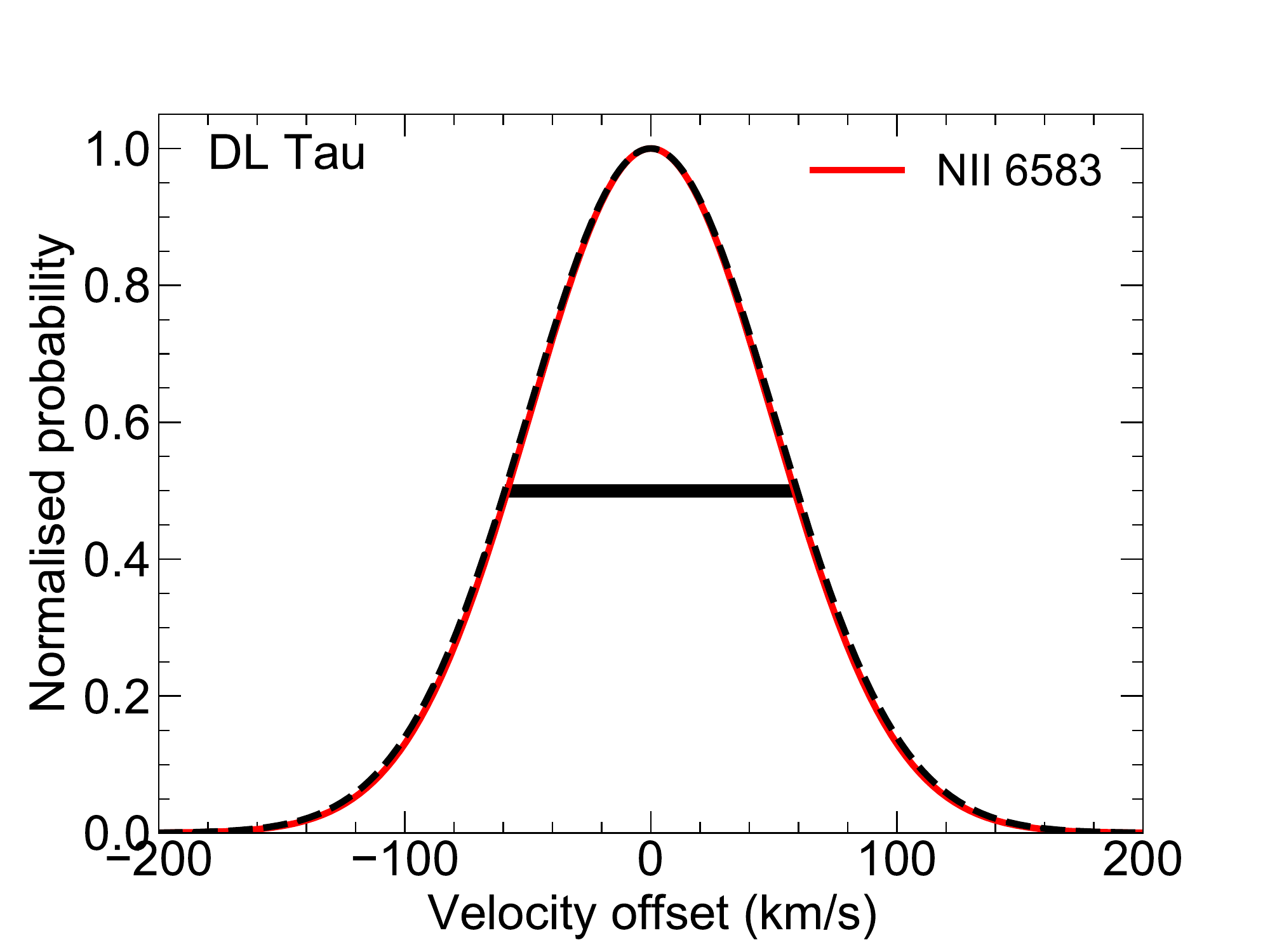}
     \end{subfigure}
     \begin{subfigure}[b]{0.3\textwidth}
         \centering
         \includegraphics[width=\textwidth]{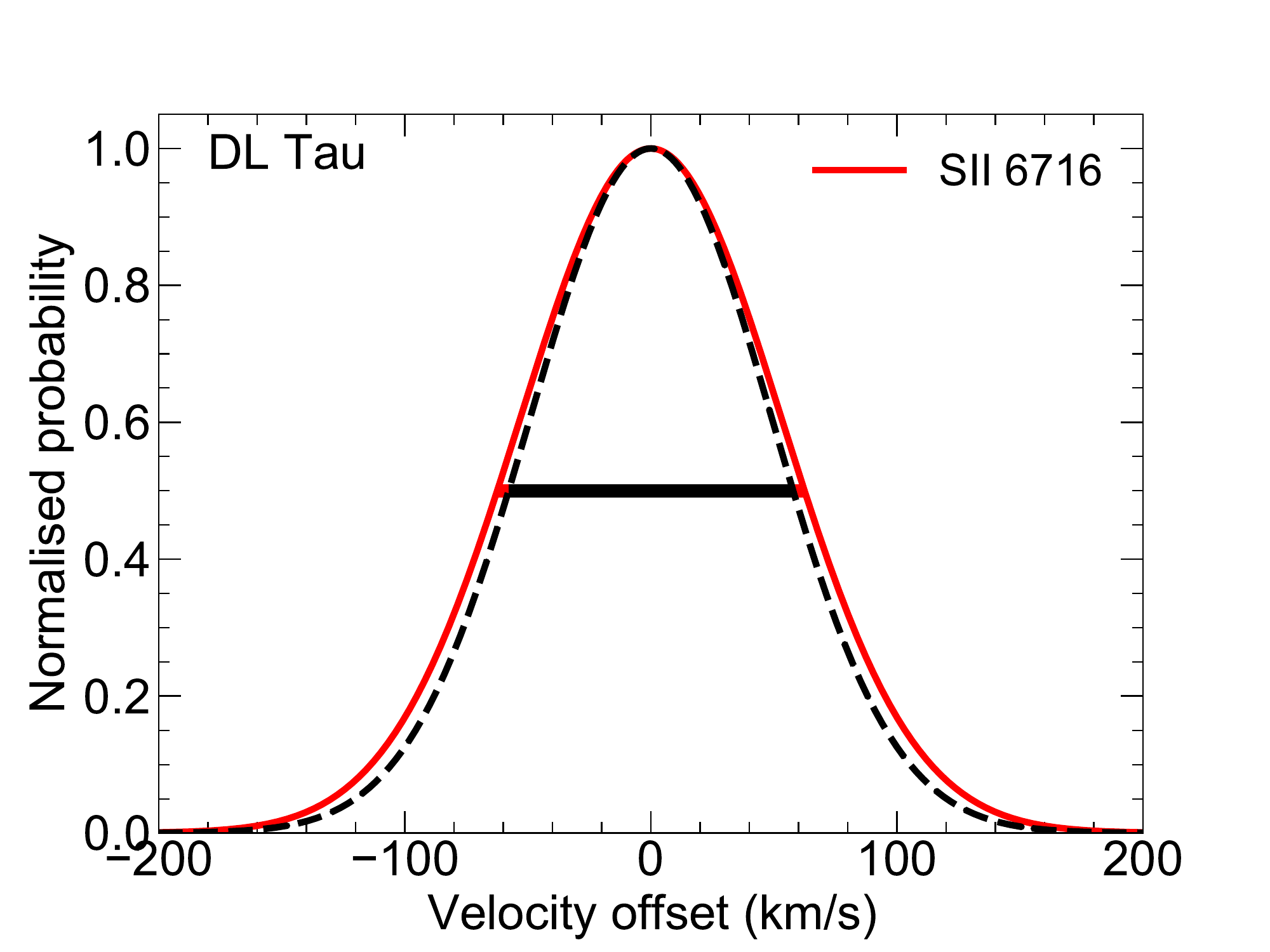}
     \end{subfigure}
     \begin{subfigure}[b]{0.3\textwidth}
         \centering
         \includegraphics[width=\textwidth]{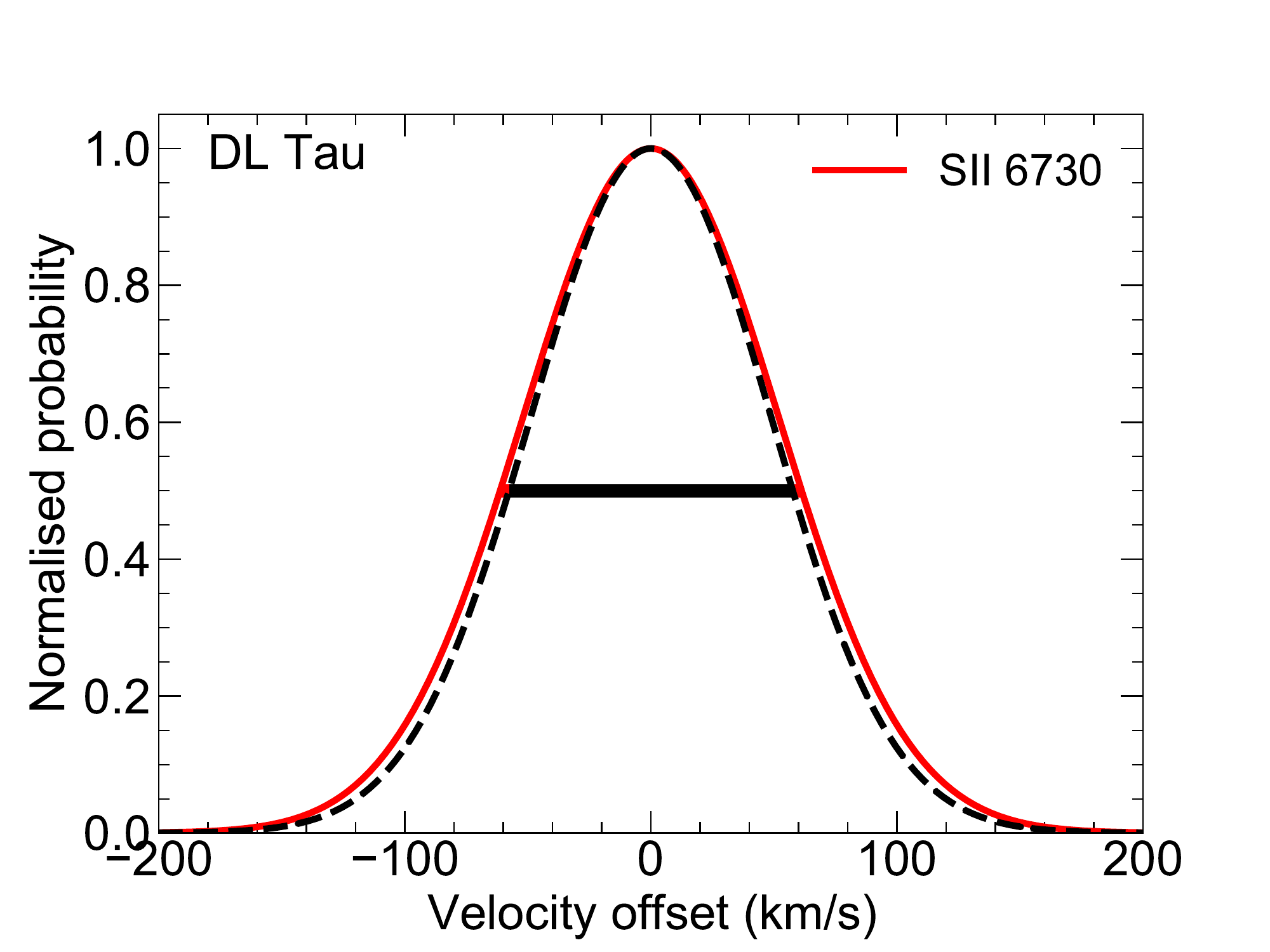}
     \end{subfigure}
        \caption{DL Tau line widths in red compared to the line-broadening of MUSE in black-dashed line.}
        \label{fig:DLTau_widths}
\end{figure*}

\begin{figure*}
     \centering
     \begin{subfigure}[b]{0.3\textwidth}
         \centering
         \includegraphics[width=\textwidth]{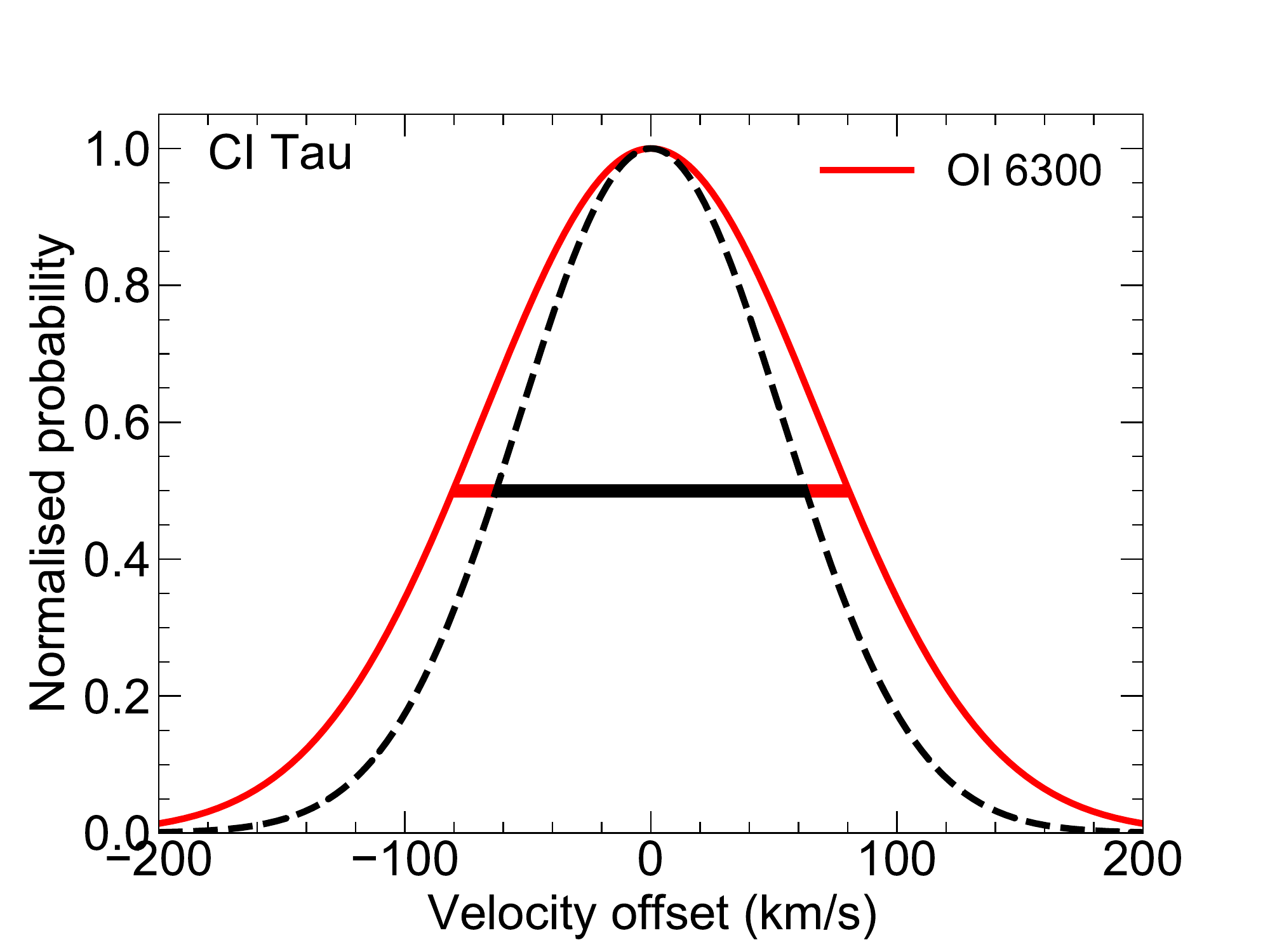} 
     \end{subfigure}
     \begin{subfigure}[b]{0.3\textwidth}
         \centering
         \includegraphics[width=\textwidth]{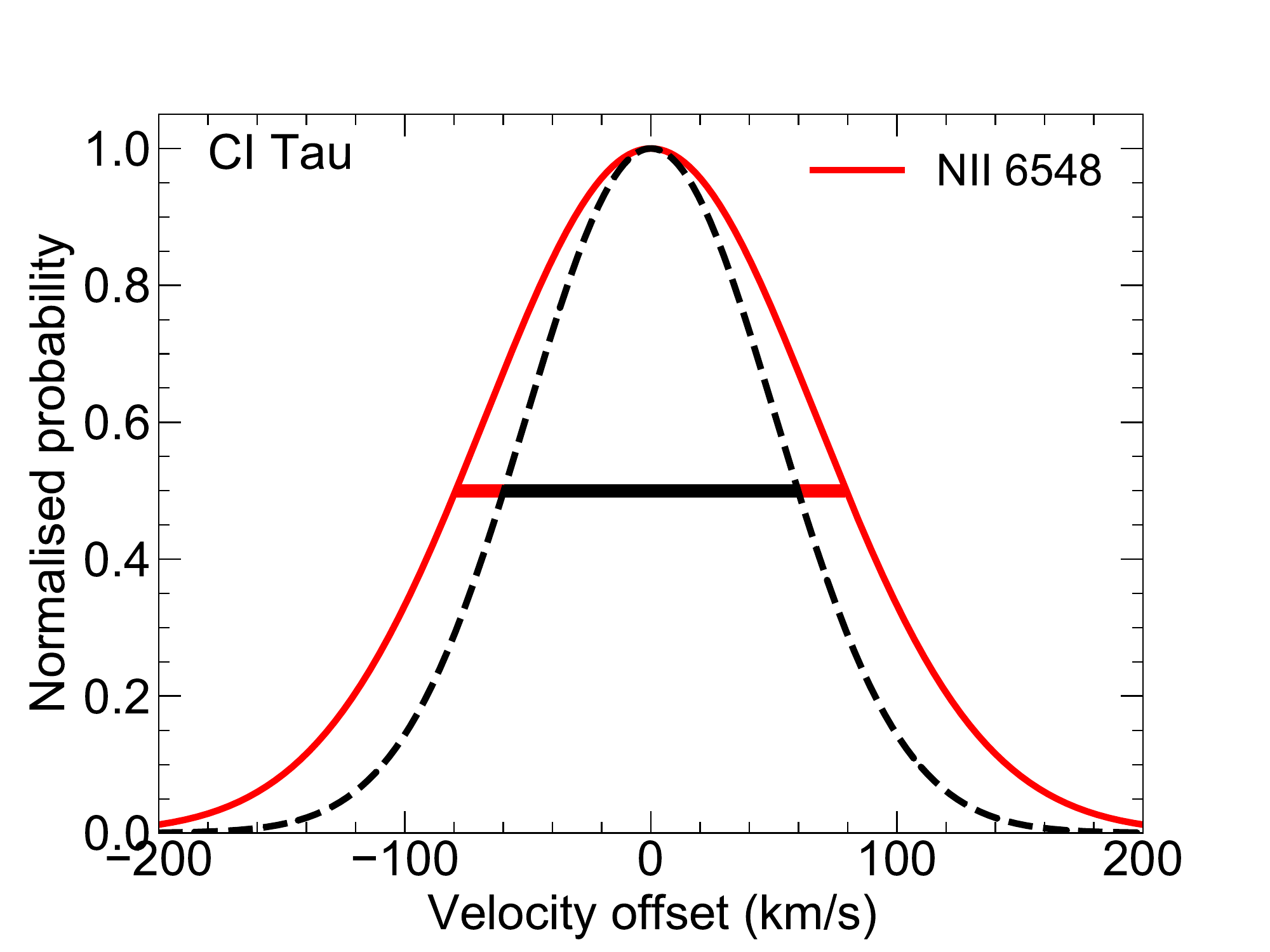}
     \end{subfigure}
     \begin{subfigure}[b]{0.3\textwidth}
         \centering
         \includegraphics[width=\textwidth]{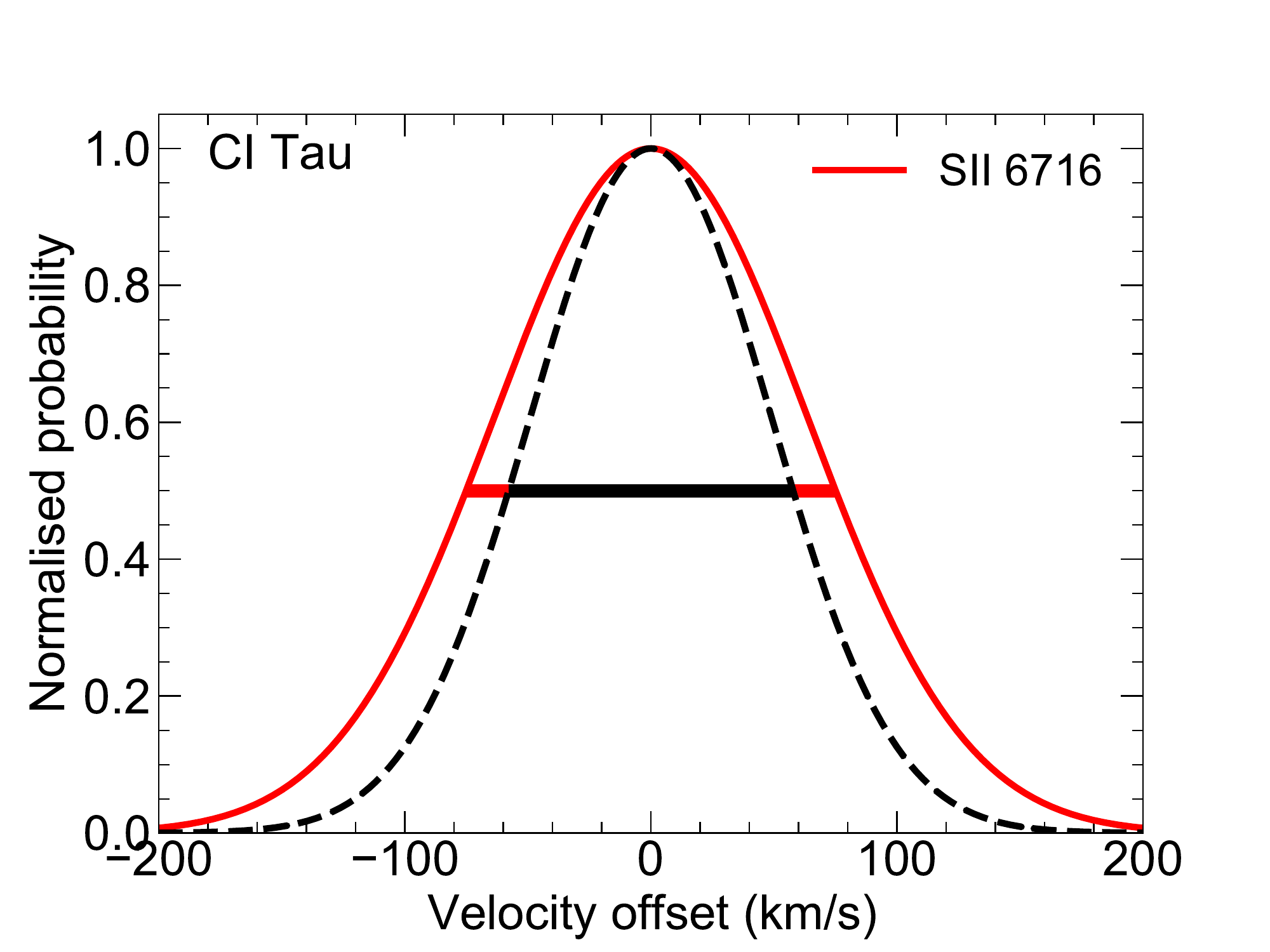}
     \end{subfigure}
     \begin{subfigure}[b]{0.3\textwidth}
         \centering
         \includegraphics[width=\textwidth]{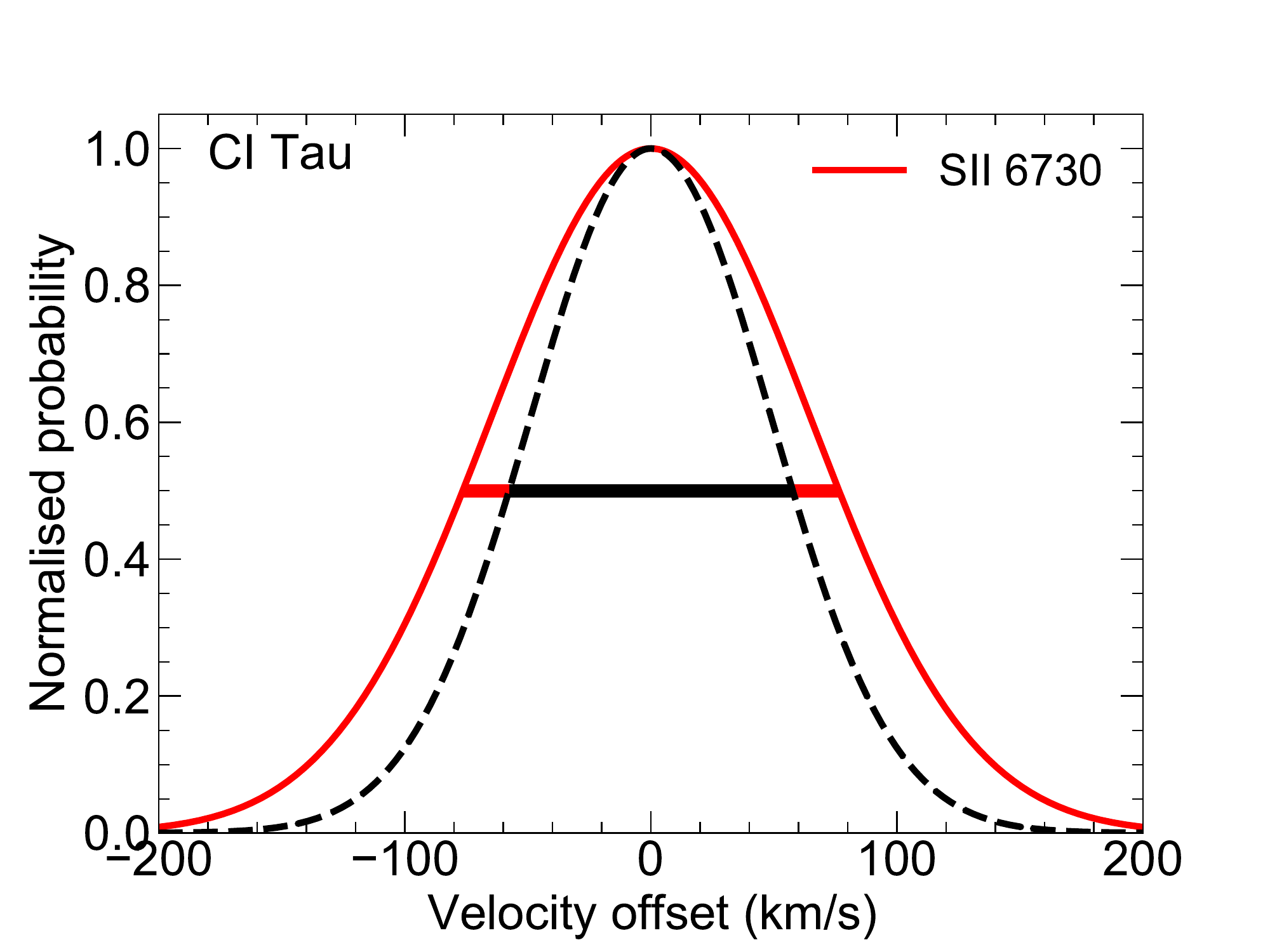}
     \end{subfigure}
        \caption{CI Tau line widths in red compared to the line-broadening of MUSE in black-dashed line.}
        \label{fig:CITau_widths}
\end{figure*}

\begin{figure*}
     \centering
     \begin{subfigure}[b]{0.3\textwidth}
         \centering
         \includegraphics[width=\textwidth]{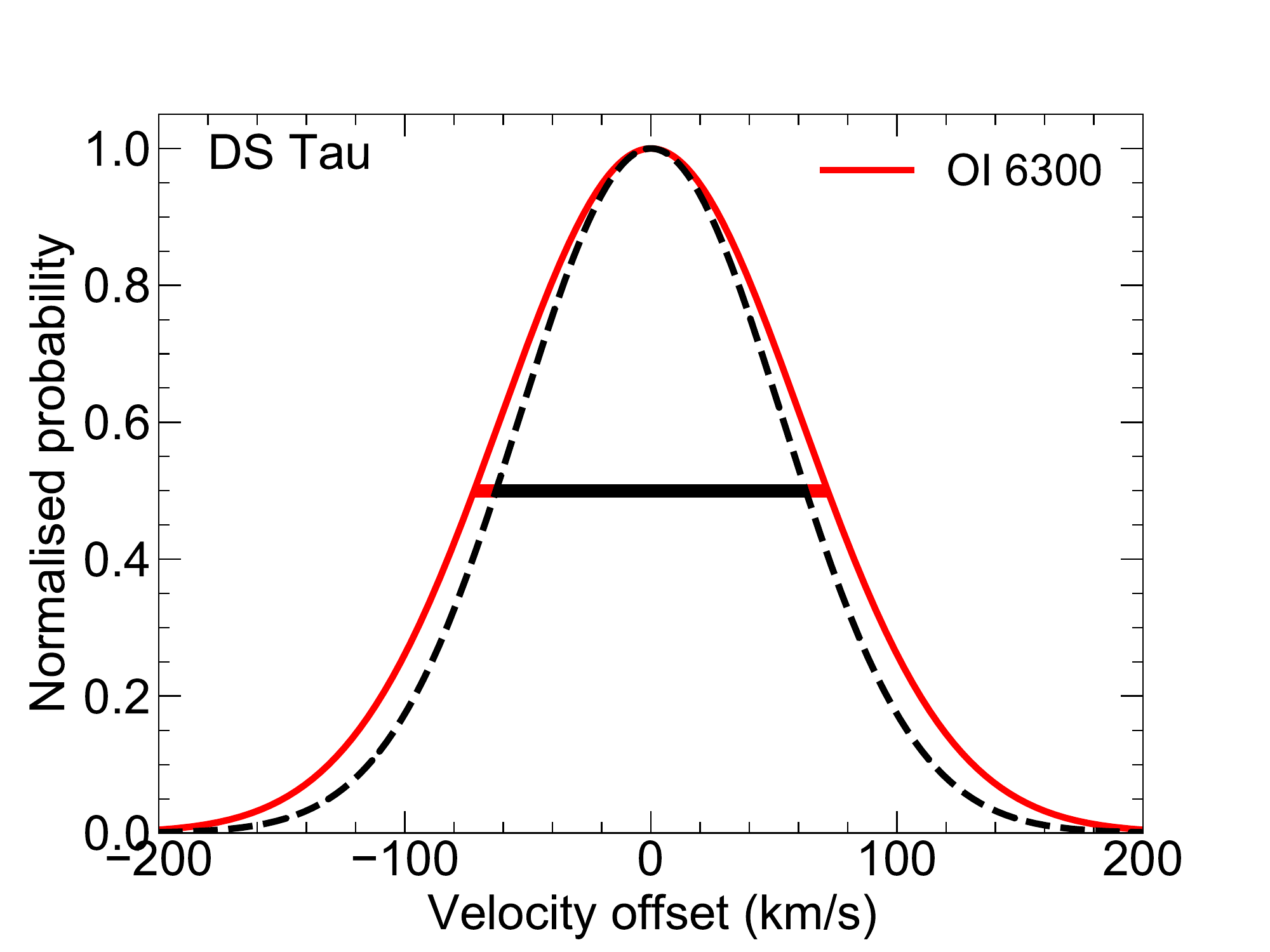} 
     \end{subfigure}
     \begin{subfigure}[b]{0.3\textwidth}
         \centering
         \includegraphics[width=\textwidth]{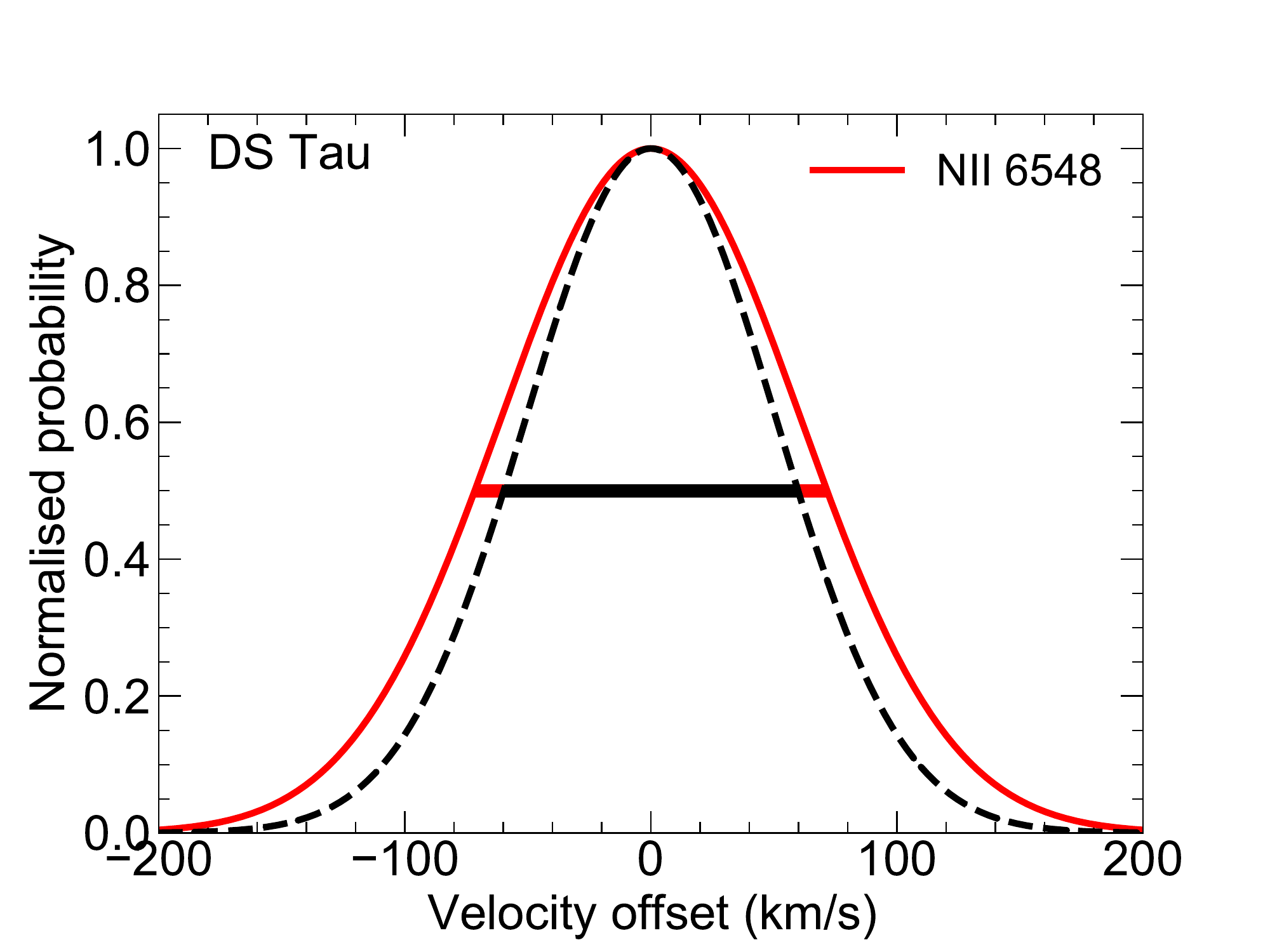}
     \end{subfigure}
     \begin{subfigure}[b]{0.3\textwidth}
         \centering
         \includegraphics[width=\textwidth]{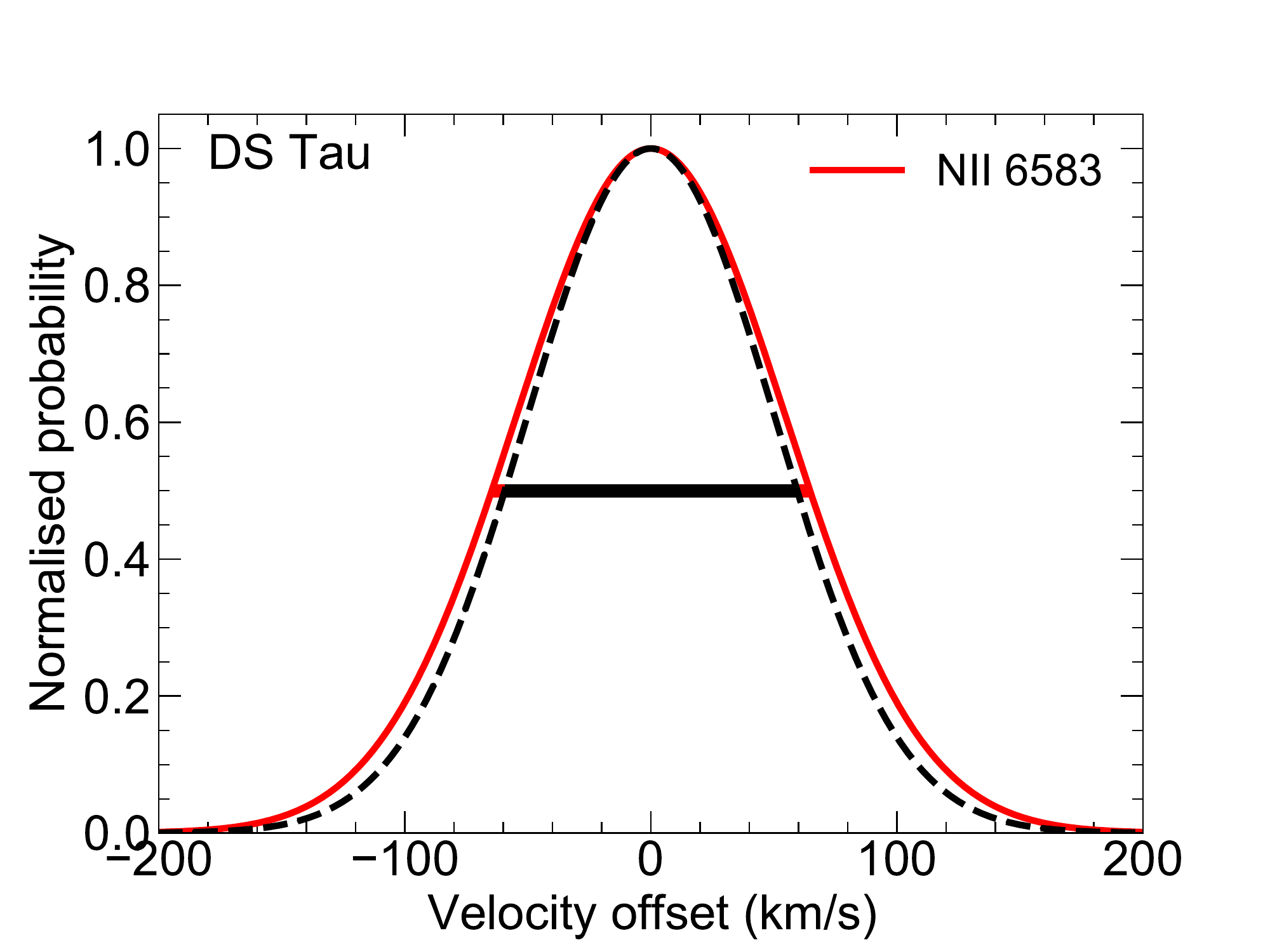}
     \end{subfigure}
     \begin{subfigure}[b]{0.3\textwidth}
         \centering
         \includegraphics[width=\textwidth]{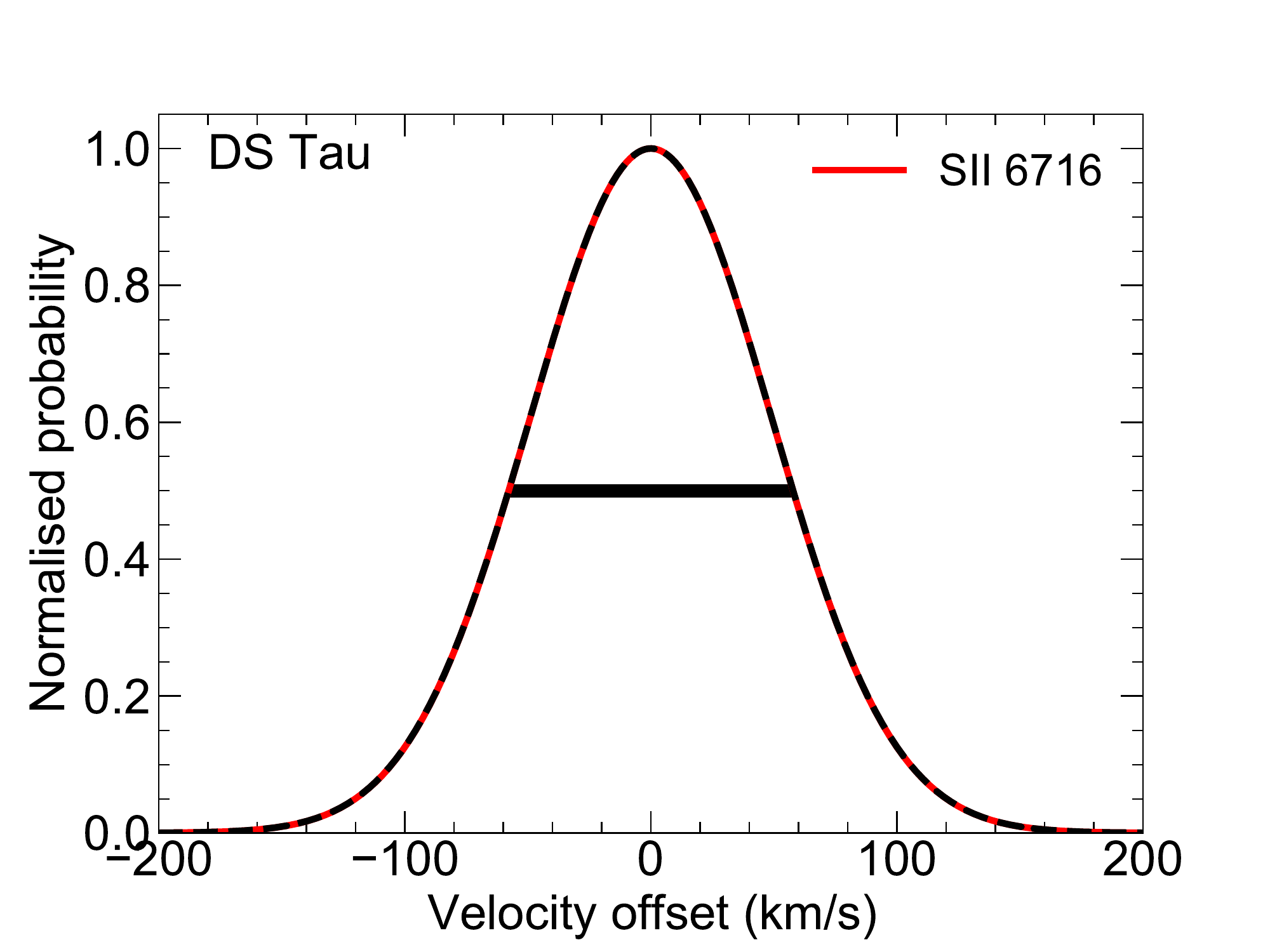}
     \end{subfigure}
     \begin{subfigure}[b]{0.3\textwidth}
         \centering
         \includegraphics[width=\textwidth]{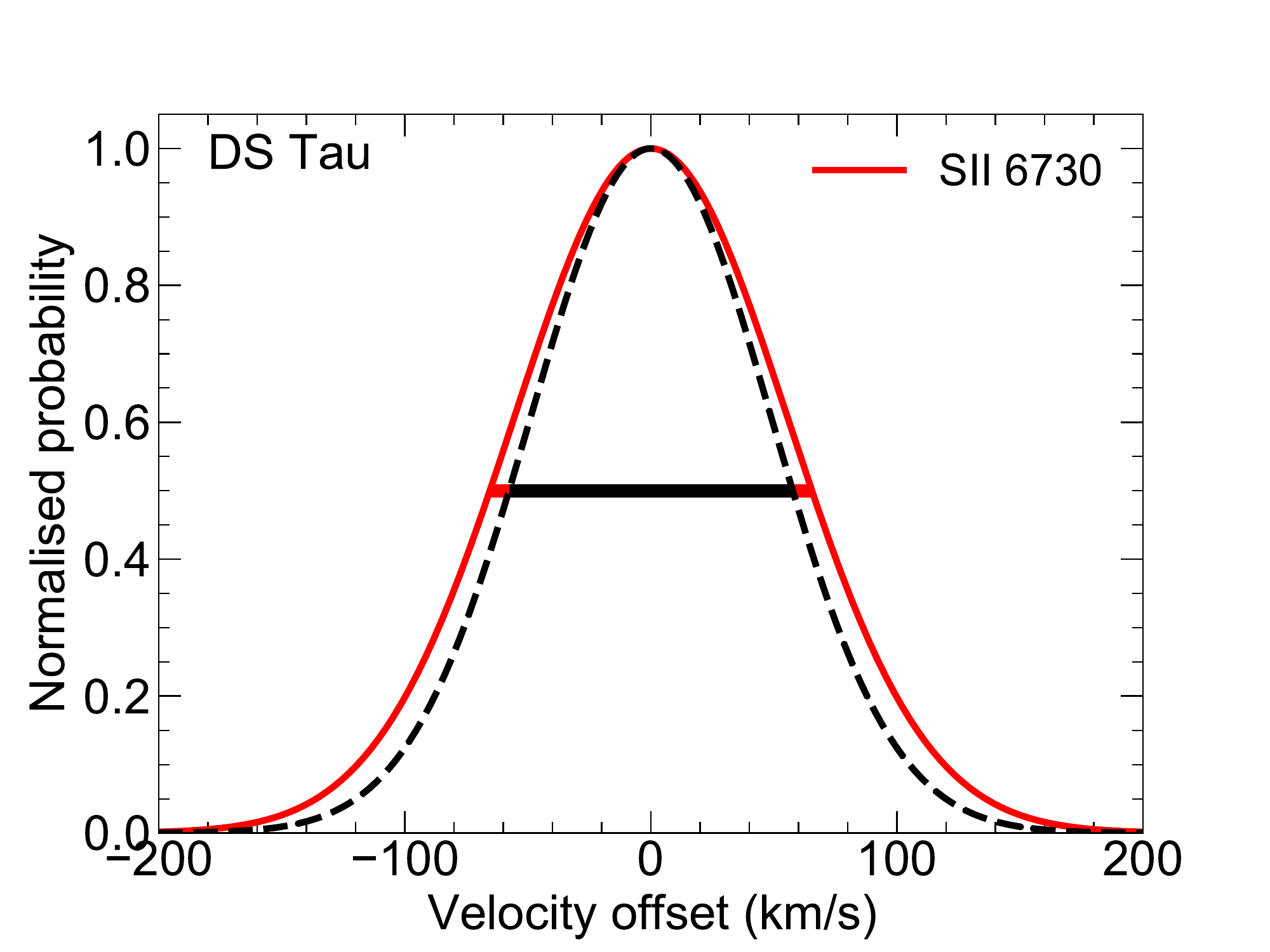}
     \end{subfigure}
        \caption{DS Tau line widths in red compared to the line-broadening of MUSE in black-dashed line.}
        \label{fig:DSTau_widths}
\end{figure*}

\begin{figure*}
     \centering
     \begin{subfigure}[b]{0.3\textwidth}
         \centering
         \includegraphics[width=\textwidth]{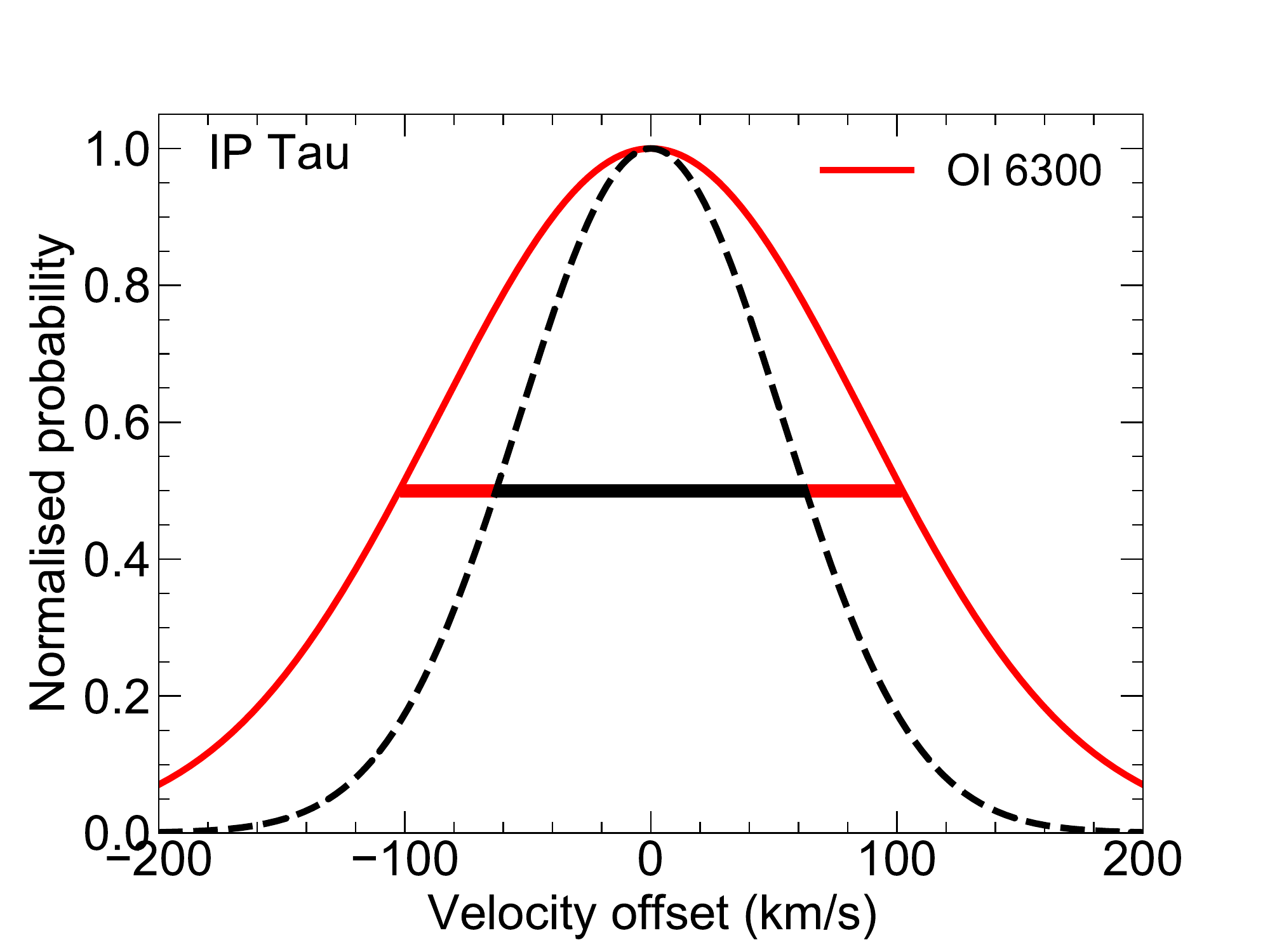} 
     \end{subfigure}
        \caption{IP Tau line widths in red compared to the line-broadening of MUSE in black-dashed line.}
        \label{fig:IPTau_widths}
\end{figure*}

\begin{figure*}
     \centering
     \begin{subfigure}[b]{0.3\textwidth}
         \centering
         \includegraphics[width=\textwidth]{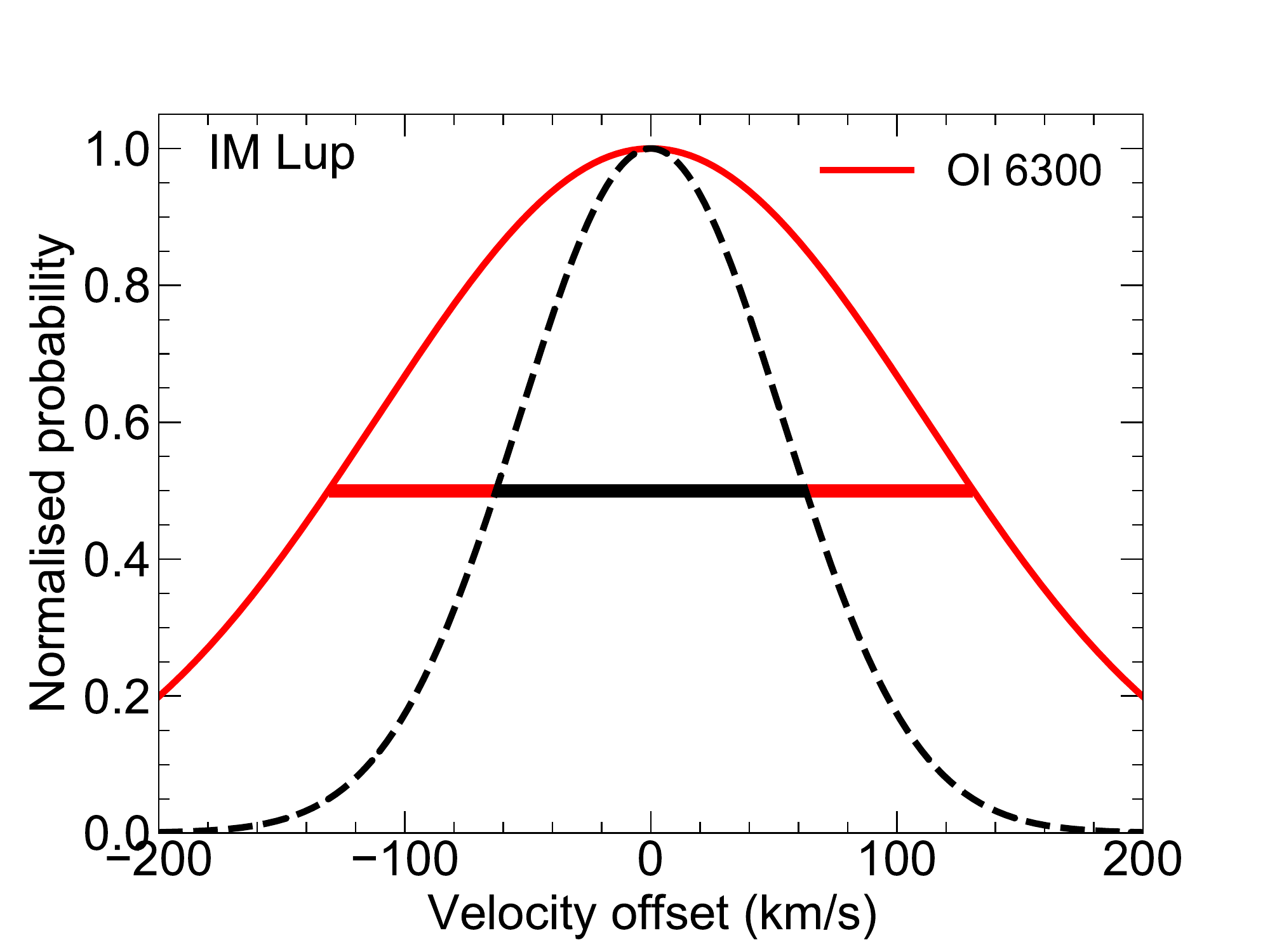} 
     \end{subfigure}
     \begin{subfigure}[b]{0.3\textwidth}
         \centering
         \includegraphics[width=\textwidth]{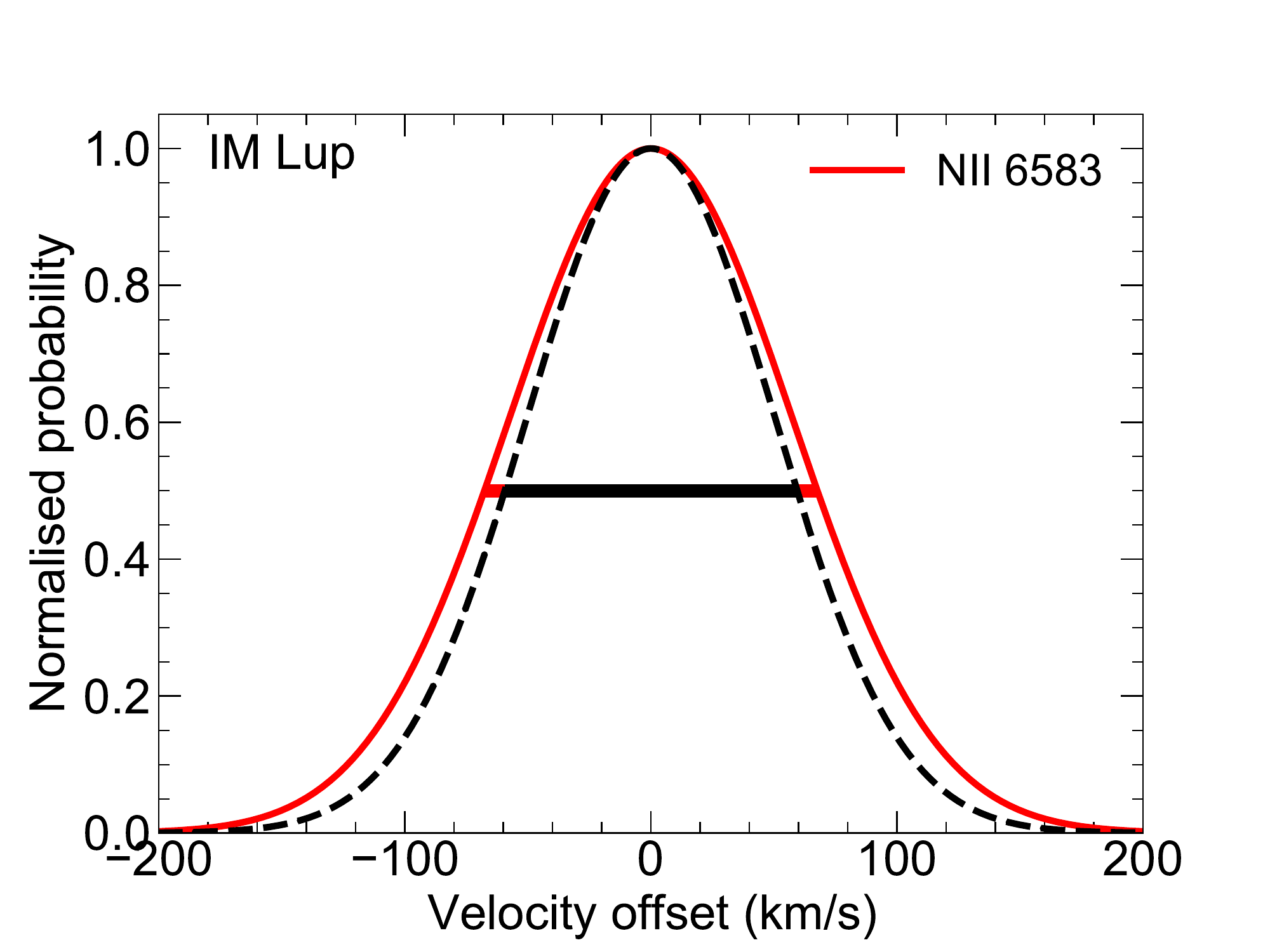}
     \end{subfigure}
        \caption{IM Lup line widths in red compared to the line-broadening of MUSE in black-dashed line.}
        \label{fig:IMLup_widths}
\end{figure*}

    Table \ref{tab:spectral_widths} shows the summary of the spectral resolution properties to be compared to the line-broadening of MUSE (see Figure \ref{fig:muse_spectral_resolution}). Figure \ref{fig:DLTau_widths}, Figure \ref{fig:CITau_widths}, Figure \ref{fig:DSTau_widths}, Figure \ref{fig:IPTau_widths}, and Figure \ref{fig:IMLup_widths} show the actual line profile for all five disk-outflow/jet systems. We note that most of the lines analyzed here are marginally resolved, meaning the observed line width (red curve) is very close to the line-broadening of MUSE (black-dashed line).

    \begin{table*}[pht!]
    \caption{Summary of the spectral width of the disk-outflow/jet systems compared to the line-broadening of MUSE from Fig. \ref{fig:muse_spectral_resolution}.} 
    \label{tab:spectral_widths}
    \centering
    \begin{tabular}{lccccccccr} 
    \hline \hline
    \addlinespace 
    Source & Line & FWHM = 2$\sqrt{2\mathrm{log}(2)}\sigma$ &  Spatial resolution  & $\lambda_\mathrm{center}$ & Line width & Resolved? \\
    &  &  &  (mas) &  (\AA) & (km s$^{-1}$) & \\
    \addlinespace 
    \hline
    \multirow{7}{*}{DL Tau} & [\ion{O}{i}]~$\lambda$6300 & 2.84 & 71.0  & 6297.47 & 135.22 & Marginally \\ 
                         & [\ion{O}{i}]~$\lambda$6363 & 2.87 & 71.8  & 6360.69 & 135.67 & Marginally \\ 
                         & [\ion{N}{ii}]~$\lambda$6548 & 2.86 & 71.5  & 6544.92 & 130.89 &  Marginally\\ 
                         & [\ion{N}{ii}]~$\lambda$6583 & 2.56 & 64.0  & 6579.97 & 116.65 & No \\ 
                         & H$\mathrm{\alpha}$ & 2.66 & 66.5 & 6559.97 & 121.22 &  No \\ 
                         & [\ion{S}{ii}]~$\lambda$6716 & 2.79 & 69.8  & 6713.38 & 124.87 & Marginally \\ 
                         & [\ion{S}{ii}]~$\lambda$6730 & 2.75 & 68.8 & 6727.47 & 122.54 & Marginally \\ 
                         \midrule \addlinespace 
    \multirow{5}{*}{CI Tau}  &  [\ion{O}{i}]~$\lambda$6300 & 3.38 & 84.5 & 6297.71 & 160.90 & Yes \\ 
                         & [\ion{N}{ii}]~$\lambda$6583 & 3.49 & 87.3 & 6580.21 & 158.72 & Yes \\
                         & [\ion{S}{ii}]~$\lambda$6716 & 3.36 & 84.0 & 6713.89 & 150.17 & Yes\\ 
                         & [\ion{S}{ii}]~$\lambda$6730 & 3.43 & 85.8  & 6727.71 & 152.89 & Yes \\ 
                         \midrule \addlinespace 
    \multirow{5}{*}{DS Tau}   & [\ion{O}{i}]~$\lambda$6300 & 3.02 & 75.5 & 6301.51 & 143.70 & Marginally \\ 
                         & [\ion{N}{ii}]~$\lambda$6548 & 3.12 & 78.0  & 6550.06 & 143.18 & Marginally \\ 
                         & [\ion{N}{ii}]~$\lambda$6583 & 2.84 & 71.0 & 6585.26 & 129.32 & Marginally \\ 
                         & [\ion{S}{ii}]~$\lambda$6716 & 2.59 & 64.8 & 6718.01 & 115.56 & No\\ 
                         & [\ion{S}{ii}]~$\lambda$6730 & 2.94 & 73.5 & 6732.76 & 130.87 & Marginally \\ 
                         \midrule \addlinespace 
    \multirow{1}{*}{IP Tau} & [\ion{O}{i}]~$\lambda$6300 & 4.30 & 107.5  & 6298.50 & 204.58 & Yes \\ 
                         \midrule \addlinespace 
    \multirow{3}{*}{IM Lup}  &  [\ion{O}{i}]~$\lambda$6300 & 5.51 & 137.8  & 6299.17 & 262.19 & Yes \\ 
                         & [\ion{N}{ii}]~$\lambda$6583 & 2.97 & 74.3  & 6581.67 & 135.32 & Marginally \\  
                         \midrule \addlinespace 

    \end{tabular}
    \caption*{\footnotesize{\textbf{Note:} The $\lambda_\mathrm{center}$ refers to the center of the jet area where the flux is brightest.}}
    \end{table*}
\end{appendix}
\end{document}